\documentclass[fleqn,onecolumn,usenatbib]{mnras}
\usepackage{newtxtext,newtxmath}

\usepackage[T1]{fontenc}

\DeclareRobustCommand{\VAN}[3]{#2}
\let\VANthebibliography\thebibliography
\def\thebibliography{\DeclareRobustCommand{\VAN}[3]{##3}\VANthebibliography}

\usepackage{graphicx}	
\usepackage{amsmath}	
\usepackage{amssymb}	
\usepackage{multicol}        
\usepackage{bm}		
\usepackage{pdflscape}	
\usepackage{natbib}
\usepackage[font=Large]{caption} 
\usepackage{adjustbox}
\usepackage{deluxetable}
\usepackage{longtable}
\usepackage[caption=false]{subfig}
\usepackage{hyperref}
\usepackage[left]{lineno}
\graphicspath{{./}{figures/}}
\newcommand{\aoptfullev}{a_{\text{OPT,ev}}}
\newcommand{\aopttrimev}{a_{\text{OPT,trim,ev}}}
\newcommand{\bopttrimev}{b_{\text{OPT,trim,ev}}}
\newcommand{\boptfullev}{b_{\text{OPT,ev}}}
\newcommand{\coptfullev}{c_{\text{OPT,ev}}}
\newcommand{\copttrimev}{c_{\text{OPT,trim, ev}}}
\newcommand{\sigmaintoptfullev}{\sigma_{\text{int}_{\text{OPT,ev}}}}
\newcommand{\sigmaintopttrimev}{\sigma_{\text{int}_{\text{OPT,trim,ev}}}}
\newcommand{\sigmaintopttrim}{\sigma_{\text{int}_{\text{OPT,trim}}}}
\newcommand{\aoptfull}{a_{\text{OPT}}}
\newcommand{\aopttrim}{a_{\text{OPT,trim}}}
\newcommand{\bopttrim}{b_{\text{OPT,trim}}}
\newcommand{\boptfull}{b_{\text{OPT}}}
\newcommand{\coptfull}{c_{\text{OPT}}}
\newcommand{\copttrim}{c_{\text{OPT,trim}}}
\newcommand{\sigmaintoptfull}{\sigma_{\text{int}_{\text{OPT}}}}
\newcommand{\axfullev}{a_{\text{X,PLAT,ev}}}
\newcommand{\axtrimev}{a_{\text{X,trim,ev}}}
\newcommand{\bxtrimev}{b_{\text{X,trim,ev}}}
\newcommand{\bxfullev}{b_{\text{X,PLAT,ev}}}
\newcommand{\cxfullev}{c_{\text{X,PLAT,ev}}}
\newcommand{\cxtrimev}{c_{\text{X,trim,ev}}}
\newcommand{\sigmaintxfullev}{\sigma_{\text{int}_{\text{X,PLAT,ev}}}}
\newcommand{\sigmaintxtrimev}{\sigma_{\text{int}_{\text{X,trim,ev}}}}
\newcommand{\axfull}{a_{\text{X,PLAT}}}
\newcommand{\axtrim}{a_{\text{X,trim}}}
\newcommand{\bxtrim}{b_{\text{X, trim}}}
\newcommand{\bxfull}{b_{\text{X,PLAT}}}
\newcommand{\cxfull}{c_{\text{X,PLAT}}}
\newcommand{\cxtrim}{c_{\text{X, trim}}}
\newcommand{\sigmaintxfull}{\sigma_{\text{int}_{\text{X,PLAT}}}}
\newcommand{\sigmaintxtrim}{\sigma_{\text{int}_{\text{X, trim}}}}
\newcommand{\sigmaint}{\sigma_{\text{int}}}

\title{Optical and X-ray GRB Fundamental Planes as Cosmological Distance Indicators}

\author[Dainotti et al.]{
Dainotti, M. G.,$^{1,2,3}$\thanks{E-mail: mdainotti@stanford.edu}
Nielson, V.$^{4,5}$
Sarracino, G.,$^{6,7}$
Rinaldi, E.,$^{8,9,10}$
Nagataki, S.,$^{11,10}$
\newauthor
Capozziello, S.,$^{6,7,12}$
Gnedin, O. Y.$^{5}$
and Bargiacchi, G., $^{7,12}$
\\
$^{1}$National Astronomical Observatory of Japan, 2 Chome-21-1 Osawa, Mitaka, Tokyo 181-8588, Japan\\
$^{2}$The Graduate University for Advanced Studies, SOKENDAI, Shonankokusaimura, Hayama, Miura District, Kanagawa 240-0193, Japan\\
$^{3}$Space Science Institute, Boulder, CO, USA\\
$^{4}$SLAC National Accelerator Laboratory, 2575 Sand Hill Road, Menlo Park, CA 94025, USA\\
$^{5}$Astronomy Department, University of Michigan, Ann Arbor, MI 48109, USA\\
$^{6}$Dipartimento di Fisica, ``E. Pancini'' Universit\`{a} ``Federico II'' di Napoli, \\ Compl. Univ. Monte S. Angelo Ed. G, Via Cinthia, I-80126 Napoli (Italy)\\
$^{7}$INFN Sez. di Napoli, Compl. Univ. Monte S. Angelo Ed. G, Via Cinthia, I-80126 Napoli (Italy)\\
$^{8}$Physics Department, University of Michigan, Ann Arbor, MI 48109, USA\\
$^{9}$RIKEN Cluster for Pioneering Research, Theoretical Quantum Physics Laboratory, 2-1 Hirosawa, Wako, Saitama, Japan 351-0198\\
$^{10}$Interdisciplinary Theoretical \& Mathematical Science Program, RIKEN (iTHEMS), 2-1 Hirosawa, Wako, Saitama, Japan 351-0198\\
$^{11}$RIKEN Cluster for Pioneering Research, Astrophysical Big Bang Laboratory (ABBL), 2-1 Hirosawa, Wako, Saitama, Japan 351-0198\\
$^{12}$Scuola Superiore Meridionale, Università di Napoli Federico II Largo San Marcellino 10, 80138 Napoli (Italy)\\
}

\date{Accepted 2022 April 19. Received 2022 April 19; in original form 2021 December 1}

\pubyear{2022}
\begin{document}
\maketitle

\begin{abstract}
Gamma-Ray Bursts (GRBs), can be employed as standardized candles, extending the distance ladder beyond Supernovae Type Ia (SNe Ia, $z=2.26$). We standardize GRBs using the 3D fundamental plane relation (the Dainotti relation) among the rest-frame end time of the X-ray plateau emission, its corresponding luminosity, and the peak prompt luminosity. Combining SNe Ia and GRBs, we constrain $\Omega_{\text{M}}= 0.299 \pm 0.009$ assuming a flat $\Lambda$CDM cosmology with and without correcting GRBs for selection biases and redshift evolution. Using a 3D optical Dainotti correlation, we find this sample is as efficacious in the determination of $\Omega_{\text{M}}$ as the X-ray sample. 
We trimmed our GRB samples to achieve tighter planes to simulate additional GRBs. 
We determined how many GRBs are needed as standalone probes to achieve a comparable precision on $\Omega_{\text{M}}$ to the one obtained by SNe Ia only. We reach the same error measurements derived using SNe Ia in 2011 and 2014 with 142 and 284 simulated optical GRBs, respectively, considering the errorbars on the variables halved. These error limits will be reached in 2038 and in 2047, respectively. Using a doubled sample (obtained by future machine learning approaches allowing a lightcurve reconstruction and the estimates of GRB redhifts when z is unknown) compared to the current sample, with errorbars halved we will reach the same precision as SNe Ia in 2011 and 2014, now and in 2026, respectively. If we consider the current SNe precision, this will be reached with 390 optical GRBs by 2054.
\end{abstract}


\section{Introduction} \label{sec:intro}
The tension on the Hubble constant cosmological parameter between different measurements is at the center of the debate in the astronomical and cosmological communities, and asks for a firm theoretical background at fundamental level \citep{2020FoPh...50..893C, 2021ApJ...912..150D,2022Galax..10...24D}. Using the Planck satellite to study the Cosmic Microwave Background (CMB) Radiation, distance measurements have been performed to derive the Hubble constant, $H_{0}$, with a very high precision. The same measurements have been used to derive the matter content of the Universe today, $\Omega_{\text{M}}$. However, discrepancies in the range between $4-6$ $\sigma$ in $H_0$ arise when we compare the measurements of these quantities by the CMB, an early Universe probe, with those obtained by Supernovae Ia (SNe Ia), or other late Universe probes, such as Cepheids. For years, SNe Ia have been studied and used as standard candles, because their nearly uniform luminosities (absolute magnitude M $\simeq$ -19, \cite{2001LRR.....4....1C}) allow us to use them as reliable cosmological tools. Although they reach higher redshifts than other standard candles such as Cepheid Variables \citep{2021jwst.prop.1685R} and the Tip of the Red Giant Branch (TRGB, \cite{2017ApJ...835...86C, 2018MNRAS.480.3879A, 2018AAS...23135105B, 2020A&A...643A.165B, 2020arXiv200710716E, 2020ApJ...891...57F, 2021MNRAS.504..300C, 2021ApJ...919...16F, 2021A&A...647A..72K, 2021JCAP...05..009L}), which are the previous step in the so-called cosmological ladder, they have still only been observed up to $z = 2.26$ \citep{2015AJ....150..156R}. 
Although there are also many groups that find lower $H_0$ values with larger error bars from low-$z$ data that are reasonably consistent with the Planck 2018 value \citep{2017ApJ...835...86C, 2018MNRAS.480.3879A, 2020A&A...643A.165B, 2020arXiv200710716E, 2021MNRAS.504..300C, 2021ApJ...919...16F, 2021A&A...647A..72K, 2021JCAP...05..009L}, to further try to resolve the issue of the Hubble tension, it is crucial to develop new cosmological probes at high-$z$ to test cosmological models even further. If we consider an evolutionary trend existing in the SNe Ia data as it has been shown in \cite{2021ApJ...912..150D, 2022Galax..10...24D}, we may bridge the gap among these probes by employing additional, \emph{standardizable} candles that can be detected at high redshifts: Gamma-Ray Bursts (GRBs) and possibly quasars (QSOs) \citep{Lusso,Bargiacchi}.

For the use of GRBs as cosmological tools, one of the correlations that can be employed to make them a standardizable candle is the fundamental plane relation between the peak prompt luminosity, the rest-frame end time of the plateau, and its corresponding luminosity. In this work we do include a SNe Ia sample, although solely in conjunction with GRBs rather than as a calibrator.
In the case of GRBs, there have been many attempts to use them as standard candles through a series of relationships among the prompt emission, the main explosion in $\gamma$-rays, and in its counterpart in X-rays and optical.
The caveat in the use of the quasars with the Risaliti-Lusso relation is that selection biases in this correlation have only been addressed very recently \citep{ep2022}. These results show that this correlation is indeed intrinsic to the QSO physics and it is not due to selection biases, but it undergoes redshift evolution. Thus, these evolutionary effects need to be taken into account for the application of quasars-cosmology. Further, there is an ongoing discussion on the use of this correlation at high-$z$ (beyond $z=1.5$) and its effectiveness; for details see \cite{2021arXiv210707600K, 2021MNRAS.502.6140K}.

Specifically, the application of GRBs as cosmological tools is worthwhile because they have been detected up to redshift $z = 9.4$ \citep{2011ApJ...736....7C}, and in principle, they can be detected up to $z = 20$ \citep{Lamb_2003}. Other cosmological probes cannot be seen so far away; even the record-setting quasar J0313-1806, discovered very recently, reaches up to only $z = 7.642$ \citep{2021ApJ...907L...1W}. Thus, GRBs have fascinated the astrophysical community ever since their initial detection, and several attempts have been made to use them as cosmological tools or distance estimators \citep{2008MNRAS.391..577A, Izzo2, Izzo, 2008MNRAS.391L...1K, 2011ApJ...730..135D, 2012MNRAS.426.1396D, 2016A&A...585A..68W, 2017A&A...598A.112D, 2017A&A...598A.113D, 2020A&A...641A.174L}. However, this is a complex task as GRBs are not yet standard candles in the sense that their observed luminosities vary greatly from one another. This issue regarding the heterogeneity of GRB luminosities remains an open topic due to the ambiguous nature of their origins. Certain types of GRBs may originate from the core collapse of a very massive star, as described by the “collapsar model" \citep{1993ApJ...405..273W,1998ApJ...494L..45P,1999ApJ...524..262M, 2001ApJ...550..410M}, while other types may arise from the merger of two neutron stars (NS) in a binary system, or a NS–black hole (BH) system merger.

Studies are further complicated given the existence of many classes of GRBs. The scientific community began to categorize these objects by adopting the short and long GRB identifications based on the duration of their prompt emission, $T_{90}$, which indicates the time over which a burst emits $5\%$ to $95\%$ of its total measured counts \citep{1981Ap&SS..80....3M, 1993ApJ...413L.101K, 2013ApJ...764..179B, 2014MNRAS.442.1922L}. Short GRBs (SGRBs) are defined by $T_{90} \le 2$s, whereas long GRBs (LGRBs) have a $T_{90} > 2$s. However, in recent years, the Neil Gehrels Swift Observatory (hereafter Swift) has detected a subsequent phase following the prompt emission referred to as the afterglow phase. Furthermore, Swift has seen that $60\%$ of GRB light curves (LCs) \citet{2020ApJ...904...97D, 2021PASJ...73..970D} present a plateau in the afterglow emission followed by a power-law (PL) decay phase, as pinpointed by \cite{2007Ap&SS.311..167O, 2007AAS...210.1004S, 2007ApJ...662.1093W, 2019ApJS..245....1T, 2019ApJ...883...97Z}. Due to these added details, the two large classes of SGRBs and LGRBs have been scrutinized into many subclasses: X-ray flashes (XRFs) which have greater X-ray fluence ($2-30$ keV) than $\gamma$-ray fluence ($30-400$ keV), X-ray rich (XRR) which have the ratio of the X-ray fluence over the $\gamma$-ray fluence ranging to values peculiar to the regular GRBs and the XRFs, GRB-SNe Ib/c associated (SNe-GRBs), ultra-long GRBs (ULGRBs) with $T_{90}> 2000$ s \citep{2013arXiv1308.1001G, 2014atnf.prop.6334P, 2017hst..prop15349L}, and short GRBs with extended emission (SEEs, \cite{2006ApJ...643..266N, 2007MNRAS.378.1439L, 2010ApJ...717..411N}) which are characterized by mixed features between SGRBs and LGRBs. SEEs are harder in the spectrum than LGRBs, similarly to SGRBs, while intrinsically short (IS) GRBs have $T_{90}/(1+z)<2$ s. The underlying physical differences between sub-classes are still not completely understood, but are hypothesized to come from different GRB progenitors or diverse environments, such as either a constant interstellar medium (ISM) or a wind medium \citep{2021PASJ...73..970D}. A more recent interpretation involves the sorting of these sub-classes into a different classification system based on previously proposed GRB progenitor physics that can be deduced by phenomenological and physical features \citep{2006ApJ...642..354Z}; Type I GRBs occur when two compact objects collide, and Type II emerge from massive star collapse. LGRB, XRF, XRR, SNe-GRB and ULGRB afterglow observations are usually consistent with the Type II-defined origins. SGRBs, SEEs, and IS GRBs are similarly ascribed within the Type I-defined origins. Still, there are some exceptions that do not allow us to match the two broad classifications exactly; for instance, some SGRBs have actually been classified as Type II \citep{2009ApJ...703.1696Z}. In this study, we use a subclass of a well-defined sample of Type II GRBs from Swift (with $z$-range of $0.34-5.91$) as a cosmological tool through a well-established correlation involving properties of the plateau emission.

Regarding the use of GRBs as cosmological tools, before the launch of Swift, \cite{2002MmSAI..73.1178A} analyzed a small sample of 12 GRBs collected by Beppo-SAX \citep{1997A&AS..122..299B}. They observed correlations that eventually led to their use as cosmological tools through the two-dimensional $E_{\text{p}}-E_{\text{iso}}$ relation (the so-called Amati relation, \cite{2002A&A...390...81A, 2008MNRAS.391..577A}), where $E_{\text{p}}$ is the peak of $\nu F(\nu)$ spectrum, and $E_{\text{iso}}$ is the isotropic energy of the prompt emission. Another prompt correlation has been discovered between $E_{\text{p}}-L_{\text{p}}$, where $L_{\text{p}}$ is the isotropic peak luminosity of the prompt emission \citep{Yonetoku_2004}. A similar correlation between the collimated-corrected energy $\nu$ and the peak flux in the spectrum $F_{\nu}$ has been determined by \citet{2004ApJ...616..331G} for a slightly larger sample of 40 pre-Swift GRBs. Yet another correlation seen shortly thereafter is the \cite{2005ApJ...633..611L} relation between $E_{\text{p}}$, $E_{\text{iso}}$, and the break time of the optical afterglow LCs, $t_{\text{b}}$. All of these relations are focused on the prompt emission properties. Thus, they have all also been employed in conjunction with SNe Ia data in attempts to constrain several cosmological parameters, such as $H_{0}$, $\Omega_{\text{M}}$, and the dark energy parameter $w$.

After the launch of Swift and the addition of many higher-quality GRB detections, some of these aforementioned relations suffered increased scatter and became less reliable \citep{2007A&A...472..395C,2007AAS...211.1801K}.
A novelty in the application of GRBs as cosmological tools is the use of correlations involving the plateau emission phase \citep{2009MNRAS.400..775C,2010MNRAS.408.1181C, 2013MNRAS.436...82D,2014ApJ...783..126P}. The advantage of using correlations concerning the afterglow phase is that there exists less variability in its features compared to those of the prompt emission phase. In this paper, we leverage the plateau emission properties and use a tight multi-dimensional relation built using Swift GRBs presenting a plateau with known redshift \citep{2008MNRAS.391L..79D}.
The 2D correlation between the X-ray rest-frame end time of the plateau $T_{\text{a,X}}$ and its corresponding luminosity $L_{\text{a,X}}$ defines the so-called Dainotti relation \citep{2008MNRAS.391L..79D} that has been used to study and standardize GRB luminosities. This relation has been extensively confirmed in the following works: \cite{2011ApJ...730..135D, 2013MNRAS.436...82D,  2015ApJ...800...31D, 2016ApJ...828...36D, 2017A&A...600A..98D}. The discovery of the 2D Dainotti relation in X-rays marked the first time an afterglow correlation had been used as a cosmological tool \citep{2009MNRAS.400..775C, 2010MNRAS.408.1181C, 2013ApJ...774..157D, 2014ApJ...783..126P}.
The 2D correlation in X-rays has been interpreted within several models such as the accretion onto the black hole \citep{2009AAS...21361002C, 2010HEAD...11.1404C}, as powered by a fast spinning neutron star \citep{2013MNRAS.430.1061R,2014MNRAS.443.1779R, 2018ApJ...869..155S, 2021arXiv211014840C, 2022arXiv220105245C,  2021MNRAS.507..730H, 2021arXiv210614155W, 2020arXiv201205627X} or as a modification of the microphysical parameters models \citep{2014MNRAS.437.2448L, 2014MNRAS.442.3495V, 2014MNRAS.445.2414V, 2016A&A...589A..37V} which consider jets viewed slightly off-axis \citep{2020MNRAS.492.2847B}.
Although these works deal strictly with the Dainotti relation in X-ray, recently it has been found that there exists also a two-dimensional relationship in optical wavelengths for a sample of 102 GRBs between the optical rest-frame end time, $T_{\text{OPT}}$, and the optical luminosity at the end of the plateau, $L_{\text{OPT}}$ \citep{2020ApJ...905L..26D}. An extension of the $L_{\text{OPT}}-T_{\text{OPT}}$ relation, obtained by adding the energy in the prompt emission, $E_{\text{iso}}$, has been investigated with a sample of 50 GRBs \citep{2018ApJ...863...50S}. Other correlations in optical have been discussed, which could possibly be related to the optical 2D Dainotti relation \citep{2015MNRAS.453.4121O, 2017Galax...5....4O}. This correlation resembles, in its slope, the X-ray Dainotti luminosity-time relation, and it can be interpreted within the magnetar model as well \citep{2014MNRAS.443.1779R, 2015JHEAp...7...64B, 2015MNRAS.448..629G, 2015ApJ...805...89L, 2015ApJ...813...92R, 2017A&A...607A..84K, 2017MNRAS.472.1152R, 2017ApJ...840...12Y, 2018ApJS..236...26L}.
Other attempts have been made to investigate the prompt-afterglow correlations between the luminosity at the end of the plateau emission, $L_{\text{a,X}}$ and the peak prompt luminosity at 1 s ($L_{\text{peak,X}}$, \cite{2011MNRAS.418.2202D, 2015ApJ...800...31D}).

Since these discoveries, a third correlated GRB parameter has been found in X-ray wavelengths, thus defining the now 3D X-ray Dainotti relation. The addition of the peak luminosity in the prompt emission, $L_{\text{peak,X}}$, yields a significant decrease in the intrinsic scatter with respect to the correspondent 2D correlation.
A very precise plane with small intrinsic scatter in a three-dimensional space of (log $ T_{\text{a,X}}^{*}$, log $L_{\text{a,X}}$, log $L_{\text{peak,X}}$) is thus defined, and is known as the GRB \emph{fundamental plane}.
This more reliable X-ray correlation has also been extensively studied: \cite{2016ApJ...825L..20D, 2017ApJ...848...88D, 2020ApJ...904...97D, 2021PASJ...73..970D, 2021ApJS..255...13D, 2020ApJ...903...18S}.
More precisely, in \citet{2017ApJ...848...88D} it has been discussed that the fundamental plane relation is a tool for discriminating among several classes of GRBs. In \citet{2018ApJ...869..155S}, it has been given a reliable physical grounding by explaining it within the magnetar model. In \citet{2020ApJ...903...18S}, the X-ray fundamental plane has been used as a tool to discriminate among several scenarios of slow or fast cooling in a wind or a constant interstellar medium. It has been shown that the GRBs observed by Fermi-LAT and detailed in the Second Fermi GRB Catalog \citep{2019ApJ...878...52A}, which show the existence of the plateau in $\gamma$-rays, obey this correlation as well \citep{2021ApJS..255...13D}.

It is relevant here to briefly discuss other attempts in the literature to use both the Amati and Dainotti correlations to probe the effectiveness of constraints by GRBs on other cosmological parameters as well. For example, \cite{2020MNRAS.499..391K, 2021JCAP...09..042K} validate the Amati relation among six different cosmological models and show that the results obtained by using only GRBs for the constraint of all of the associated cosmological parameters are consistent with those by baryon acoustic oscillations (BAOs) and SNe Ia.
Further, the Dainotti relation (which this paper employs) has been used by \cite{2021arXiv210614155W} to constrain $\Lambda$CDM model parameters, and they present results consistent with the predictions of the flat $\Lambda$CDM model at high GRB redshifts. \cite{2021MNRAS.507..730H} combine a SGRB sample with the \cite{2021arXiv210614155W} LGRB sample to further constrain both $\Omega_{\text{M}}$ and $\Omega_{\Lambda}$ using the Dainotti correlation. The results are again consistent with a flat $\Lambda$CDM model. More recently, \cite{2021arXiv211014840C, 2022arXiv220105245C} have combined both \cite{2021arXiv210614155W} and \cite{2021MNRAS.507..730H} GRB data sets alongside the Amati-correlated GRBs and the results are in agreement with the Hubble parameter (H(z)) and BAO data-derived constraints.
Very recently, the 2D Dainotti relation has also been investigated in radio emission \citep{2022ApJ...925...15L} and it holds with parameters that are compatible with X-rays and optical when we correct for selection biases and redshift evolution \citep{2021Galax...9...95D}. 

Given this series of papers (9 since 2016 dealing with the X-ray fundamental plane relation, the 2D relation in optical, and selection biases) we believe we are now ready to use the fundamental plane in X-rays and optical wavelengths as a cosmological tool. In this paper we test, for the first time in this research field, the novel 3D optical correlation as the extension of the 3D X-ray fundamental plane as a cosmological tool, and check its applicability compared to that of the confirmed X-ray relation. The goal is to employ the fundamental plane as a mean to use GRBs in X-rays and optical as standard candles to constrain cosmological quantities, in analogy to what has been done with SNe Ia. One important question to answer is to what extent the precision on cosmological parameters can be increased by these new probes, alone as well as with SNe Ia.

We show in this paper how the 3D Dainotti correlation has achieved a comparable or smaller intrinsic scatter ($\sigmaint = 0.20 \pm 0.06$) than the aforementioned attempts. In particular, we show the most updated scatter for the discussed alternative correlations: $\sigma=0.41 \pm 0.03$ is achieved with the $E_{\text{p}}-E_{\text{iso}}$ correlation \citep{Amati_2019, 2021arXiv211014840C}. This scatter comes from the highest data quality compilation, which is the A118 compilation of \cite{2020MNRAS.499..391K, 2021JCAP...09..042K}, based on \cite{2016A&A...585A..68W} and \cite{2019ApJ...887...13F}. When the GRB sample is calibrated using H(z) data, $\sigma=0.20 \pm 0.01$ is achieved using the same correlation. Furthermore, after correcting for the jet opening angle, the scatter is reported to be $0.09$ \citep{Ghirlanda_2007}.
Even more recently, \cite{Wang_2018} found a wider scatter for this correlation; a dispersion value is not explicitly given in this report, but the normalization parameter is shown to hold a very high error. Lastly, for the $E_{\text{p}}$-$E_{\text{iso}}$-$t_{\text{b}}$ correlation, a dispersion equal to $0.15$ has been found \citep{Wang_2018}. The main advantage of our method over these is that the 3D correlation here proposed has already been corrected for both selection biases and redshift evolution \citep{2013MNRAS.436...82D, 2015MNRAS.451.3898D, 2020ApJ...904...97D, 2018PASP..130e1001D} with the reliable statistical \citet{1992ApJ...399..345E} method. This is contrary to other relationships which have not been corrected for such biases; see \cite{2012ApJ...747...39C} on this topic. For reviews on the topic of GRB correlation both in the prompt and afterglow emission see \cite{2013ApJS..209...20G, 2017NewAR..77...23D, 2018AdAst2018E...1D, 2019gbcc.book.....D}. Because our data is corrected in this manner, we can be sure that the correlation is intrinsic to the GRBs' physics and not due to selection biases.

The main goal of this work is to investigate the possibility of using GRBs as cosmological tools, together with SNe Ia as well as alone. To achieve this goal, we need to have the smallest possible scatter in the GRB fundamental plane correlation, given the difficulty in their standardization. To this end, we consider the data currently available, as well as data expected to be gathered in the next years by present and future deep-space observational missions and campaigns. In Sec.\S \ref{sec:sample_selection}, we describe the GRB and SNe Ia data samples. In Sec.\S \ref{sec:PLAT+SNe} we detail the calculations regarding $\Omega_{\text{M}}$ and its error measurements. In Sec.\S \ref{opt sect} we present the results using the same methodology shown in Sec.\S \ref{sec:PLAT+SNe}, but considering a GRB sample in optical wavelengths. We perform redshift evolution and bias correction on all GRB samples in Sec.\S \ref{sec:EP_method}. We then describe the techniques used to simulate additional GRBs in X-ray and in optical to constrain $\Omega_{\text{M}}$  with the same precision reached by the SNe Ia (Sec.\S \ref{sec:simulations}), and define the minimum number of GRBs needed to do so at the end of Sec.\S \ref{sec:simulations_trimmed}. In Sec.\S \ref{sec:future_surveys} we use this  number to define a timeline in which we will reach it through current and proposed deep-space satellite surveys. Finally, in  Sec.\S \ref{conclusions}, we abridge our findings and conclusions. In the Appendix Sec.\S \ref{MCMC error}, methods concerning the quantification of numerical Monte Carlo Markov Chain (MCMC) sampling error are discussed to assert the validity of our results. In the Appendix Sec.\S \ref{retro choices}, we present more specific details on the sample selection process concerning the Sec.\S \ref{sec:simulations} simulations.

\section{Sample Selection for GRBs and SNe Ia} \label{sec:sample_selection}

We select our X-ray GRB sample from an initial set of all 372 GRBs observed by Swift from 2005 January to 2019 August for which a redshift has been observed taken from Swift+BAT+XRT repository \citep{Evans2009}. From these, only those that are successfully fit by the \cite{2007ApJ...662.1093W} model and showing a reliable plateau are chosen. This reduces the starting sample to one of 222 GRBs. Furthermore, it is imperative to define phenomenological GRB categories to reach a reduction in the intrinsic scatter within this 3D relation. Thus, we focus only on the use of LGRBs from which XRFs, XRR, GRB-SNe Ib/c, and ULGRBs are removed. This is because  \cite{2016ApJ...825L..20D, 2017ApJ...848...88D, 2020ApJ...904...97D} have shown that the segregation in classes is essential for a) pinpointing a more homogeneous physical mechanism, and b) reducing the scatter of the correlation at the minimal point. Alongside these initial efforts to find a suitable sample, within the chosen LGRB class we also apply morphological criteria to the GRB LCs. This defines our final, “platinum" subsample, as also defined in \cite{2020ApJ...904...97D}, hereafter called the PLAT sample. To build this set, we have excluded the LCs of all the GRBs with at least one of the following features:
\begin{enumerate}
    \item \ an ill-defined onset point of the plateau phase, starting from its beginning;
    \item \ an observed time at the end of the plateau, $T_{\text{a}}$, that falls within a large observational gap;
    \item \ a short-duration plateau ($<500$ s);
    \item \ flares at anytime during the plateau phase;
    \item \ less than 5 observational points before the plateau phase;
    \item \ an inclination of the plateau larger than 41°, similarly to what has been done in previous papers \citep{2016ApJ...825L..20D,2017ApJ...848...88D, 2020ApJ...904...97D}.
\end{enumerate}
After these exclusions, a sample of 50 GRBs remains and defines our PLAT sample, with redshift range 0.055 \textless \hspace{0.9 mm} $z$ \textless \hspace{0.9 mm} 5. As a consequence of this choice, the PLAT fundamental plane will produce increasingly accurate estimates on cosmological parameters, and in this specific case, a better constraint on $\Omega_{\text{M}}$. Regarding the optical data, we select all GRBs taken from \cite{2020ApJ...905L..26D}, the GCN, and from private communication from Liang and Kann presenting both a plateau and a peak in the optical prompt emission (Dainotti et al. 2021 [in prep]). In total, these sources provide a full optical sample of 45 GRBs. Details of the data gathering and the selection and fitting criteria are presented in \cite{2020ApJ...905L..26D}. The fitting procedure for determining the presence of the optical plateaus is again determined based on the phenomenological \cite{2007ApJ...662.1093W} function, as it was for the X-ray sample. In regards to a similar “platinum" trim for these optical LCs, we find that the optical sample so far is still too small to allow such scrutiny. Future analysis will allow us to increase the sample size and to uniformly use the definition of the platinum sample in optical too. Besides the morphological investigation, we account for biases in our selection process and redshift evolutionary effects using the \citet{ 1992ApJ...399..345E} methodology, as it has been done in previous works \citep{2000AAS...197.2506L, 2013MNRAS.436...82D, 2015MNRAS.451.3898D, 2015ApJ...806...44P, 2020ApJ...904...97D}.

Regarding the use of other cosmological probes in this study, our SNe Ia sample is the “Pantheon Sample" (PS) built by \cite{2018ApJ...859..101S}; it is an aggregation of 1048 SNe Ia which ranges from 0.01 \textless \hspace{0.9 mm} $z$ \textless \hspace{0.9 mm} 2.3. It is worth noting that our 50 PLAT GRBs have been selected from a total number of 1305 GRBs observed by Swift from January 2005 up to 2019 August. In comparison, the PS has been slimmed down from a total of 3473 SNe Ia events from the full samples of each survey used in the catalogue. The total number of SNe Ia events is almost three times as large as the total number of GRBs. Our selection of the PLAT sample trims drastically the full data set to the $30\%$ of the total sample, while the SNe Ia trims the sample of $4\%$ \citep{2018ApJ...859..101S}.

\subsection{Comparison of the X-ray and optical chosen samples vs the full X-ray and optical sample} \label{sec:subample_test}

We here show how much the properties of the X-ray and optical chosen samples are representative of population of GRBs if we select a sample composed of bright GRBs. Bright GRBs overcome the problem of selection biases because GRBs with a high luminosity will be less affected by the Malmquist bias effect \citep{1922MeLuF.100....1M} induced by the missing population of faint events. Thus, we perform a cut in luminosity within the full X-ray GRB sample so that log $L_{\text{a,X}} >46.5\ \text{erg}\  \text{s}^{-1}$ and log $ L_{\text{peak,X}} >49.5 \ \text{erg}\  \text{s}^{-1}$. These values do not change their respective luminosity functions significantly even if they are not corrected for selection biases (for details on the computation of the luminosity functions corrected for selection biases see \cite{2013MNRAS.436...82D, 2015ApJ...806...44P}). To ensure that the PLAT sample is representative of the entire population, we perform the two-sample Kolmogorov–Smirnov (KS) test distribution of the $L_{\text{a,X}}$, $T_{\text{a,X}}^{*}$, and $L_{\text{peak,X}}$ of the PLAT sample. The resulting KS statistic highlights whether or not the two samples come from the same parent distribution.
The cuts of log $T_{\text{a,X}}^*/\text{s}=1.76 $  and $z=0.54$ are performed so that we choose the sample to have plateaus which are not too short, so that there is no ambiguity on the existence of the plateau itself. Indeed, LCs with very small plateaus could be in principle compatible with a simple power law fit. We choose the minimum redshift of the PLAT sample. We perform the same comparison between the $L_{\text{a,OPT}}$, $T_{\text{a,OPT}}^*$ of the chosen sample versus the full sample of 181 optical LCs with plateaus. The cut for the full optical sample is log $ L_{\text{a,OPT}}=43.5 \ \text{erg}\  \text{s}^{-1}$, log $ T_{\text{a,OPT}}^*\text{s}=2.41$, and $z=0.34$.

With these cuts in time, luminosity, and redshift, we then perform the KS test on each data set to compare the chosen samples used in this paper with respect to the full samples from which we have chosen them from.  This test was performed on each aforementioned trimmed variable set, and the resulting p-values are shown in Table \ref{tab:pvals}.

\begin{table*}
\caption{This table showcases the p-values for each variable achieved by the KS test when comparing the full population of GRBs to the chosen sampling distribution. The first row shows the p-values for the X-ray GRB parameters, and the second shows the same for the optical GRB parameters. \label{tab:pvals}}
\begin{tabular}{|c|c|c|c|c|}
\hline
GRB Sample & KS($L_{\text{a}}$) & KS($L_{\text{peak}}$)  & KS($T_{\text{a}}^*$) & KS($z$)  \\\hline

X-ray & 0.460 & 0.068 & 0.240 & 0.340 \\
Optical & 0.670 & N/A & 0.004 & 0.960 \\ \hline
\end{tabular}
\end{table*}

The KS test was performed with a null hypothesis stating that the underlying continuous distributions are identical. Before testing, we defined the p-value for which we will either reject or fail to reject the null hypothesis: $p=0.05$. Table \ref{tab:pvals} shows that for all cases excluding the optical $T_{\text{a,OPT}}^*$ values, we fail to reject the null hypothesis and thus we can conclude that these chosen samples are indeed well representative of their populations. It is important to note first that there was no testing performed on the $L_{\text{peak,OPT}}$ values for the optical set because our chosen optical sample are only those GRBs whose $L_{\text{peak,OPT}}$ have been measured, so there are no additional values of $L_{\text{peak,OPT}}$ to be compared with. Our current chosen optical sample is the largest in the literature to date with $L_{\text{peak,OPT}}$ measurements. Therefore, in noting the fact that we must reject the null hypothesis for the optical $T_{\text{a,OPT}}^*$ values, it must be recognized that the optical sample is nevertheless the largest analyzed so far with current GRB data. In the near future, when additional analysis of new data is available we will be able to increase the chosen optical sample for the 3D correlation and thus overcome the differences in the rest-frame end-time population. We conclude that, with the current sample, all X-ray variables and optical luminosities are compatible with their respective parent populations.

\section{Deriving \texorpdfstring{$\Omega_{\text{M}}$}{TEXT}  with the Full X-ray GRB Sample + SNe Ia Data} \label{sec:PLAT+SNe}

We begin this analysis using GRB emission data in X-ray in conjunction with SNe Ia data, utilizing the samples defined in the previous section. We here describe the methodology regarding the analysis of GRBs and SNe Ia to derive $\Omega_{\text{M}}$ keeping $H_{0}$ fixed at $H_0=70$ km $\text{s}^{-1} \text{Mpc}^{-1}$ and $w=-1$. In particular, we present the computations performed to derive this cosmological parameter by using the fundamental plane correlation built with the PLAT sample, both with and without the correction of selection biases and redshift evolution. First, we describe the GRB fundamental plane used in combination with the SNe Ia data. In all subsections below, the driving methodology is the same: we aim to find the best fitting three-dimensional coefficients of the fundamental plane together with the best fit value for $\Omega_{\text{M}}$ using a Bayesian approach.
We now start by describing the equation of the fundamental plane for a given fixed cosmological model. Later in our calculations, we vary $\Omega_{\text{M}}$ together with the other fundamental plane variables. In order to build the likelihood pertaining to the GRBs, we start from the 3D Dainotti correlation which is described by the following equation:
\begin{equation} \label{Fundamental_plane}
\log_{10}L_{\text{a}} = a \cdot \log_{10}T_{\text{a}}^{*} +b \cdot \log_{10}L_{\text{peak}} + c
\end{equation}
where $c$ is the normalization parameter, and $a$ and $b$ are the best fit scaling parameters for the PLAT sample,  related to $\log_{10} T_{\text{a}}^{*}$ and $\log_{10} L_{\text{peak}}$, respectively. 
The equation above is general and is used for the other planes described in the paper. We here use different notations for the parameters according to the sample used, such as in X-rays or in optical or the full sample or trimmed.
Here, $T_{\text{a,X}}^*$ refers to the X-ray rest-frame end time of the plateau, and $L_\text{a,X}$ refers to its corresponding luminosity. We retrieve these parameters $L_{\text{peak,X}}$ and $L_\text{a,X}$ by selecting directly from our sample LCs once they have been fit to the phenomenological \cite{2007ApJ...662.1093W} model. This model's functional form for a LC contains the parameters of the fluxes and times at the end of the plateau emission which do not depend on any assumption of a cosmological model nor SNe Ia calibration.

Bayesian, rather than frequentist, methods are utilized to derive precise estimates of the resulting fundamental plane parametric values. Specifically, we fit and compute all following parameters using the D’Agostini \citep{2005physics..11182D} method because it takes into account all variable uncertainties, both statistical and systematic, and it directly infers the intrinsic scatter. 
We here write a generic function for the likelihood that can be applied to any plane (X-rays, optical for the full sample and for a trimmed one).
The fitting algorithm is defined by the following function of log-likelihood:
\begin{multline}
\mathcal{L}(a,b,c,\sigmaint) = - \frac{1}{2} \cdot (ln(\sigmaint^{2}+b^{2} \cdot \log_{10}^{2}(L_{\text{peak\ err}})+a^{2} \cdot \log_{10}^{2}(T^*_{\text{a\ err}})+\log_{10}^{2}(L_{\text{a\ err}})))- \\
\frac{1}{2} \cdot \frac{(\log_{10}(L_{\text{a,th}})-\log_{10}(L_{\text{a}}))^{2}}{\sigmaint^{2}+b^{2} \cdot \log_{10}^{2}(L_{\text{peak\ err}})+a^{2} \cdot \log_{10}^{2}(T^*_{\text{a\ err}})+\log_{10}^{2}(L_{\text{a\ err}})},
\label{likelihood no ev}
\end{multline}
where $T^*_{\text{a\ err}}$, $L_{\text{peak\ err}}$, and $L_{\text{a\ err}}$ are the errors on the rest-frame time corresponding to the end of the plateau, the peak luminosity of the emission, and the luminosity at the end of the plateau, respectively, and $L_{\text{a,th}}$ is the theoretically-computed luminosity at the end of the plateau as shown in Equation (\ref{Fundamental_plane}); $\sigmaint$ is the intrinsic scatter on the plane which depends on an unknown source of scatter. To calculate $a$, $b$, $c$, and $\sigmaint$, we make use of MCMC sampling. We use the python MCMC sampler \emph{cobaya}~\citep{2019ascl.soft10019T, 2021JCAP...05..057T} to allow the plane parameters and the scatter of the plane to vary together. This is a crucial point in our methodology, since it avoids the so-called circularity problem. We also define very reasonable priors, so that we do not risk to incur in parameters which are not physically possible and not pertinent to the physical parameter space of the fundamental plane relation. The results we have obtained by using this method on the PLAT sample (denoted by `X, PLAT' subscripts) are as follows: $\axfull = -0.88 \pm 0.12$, $\bxfull = 0.55 \pm 0.12$, $\cxfull = 22.56 \pm 6.37$. The only assumption we make to derive these values is that the fundamental plane exists and is reliable along with its parameters. As computed for this plane, $\sigmaintxfull = 0.36 \pm 0.04$. These results are visualized in Fig. \ref{fig:PLAT_GRB+SNE} (upper left). 

To use these relations for computing the best fit value for $\Omega_{\text{M}}$, we begin by comparing the observed distance modulus, $\mu_{\text{obs,GRB}}$, to the theoretical $\mu_{\text{theory}}$ value. We first define the theoretical distance modulus:
\begin{equation}
\mu_{\text{theory}} = 5 \cdot \log_{10}d_{\text{L}}(z,H_{0},\Omega_{\text{M}})+25,
\label{mu_theory}
\end{equation}

where $d_{\text{L}}$ is the luminosity distance, and $z$ is redshift. Using a flat $\Lambda$CDM model, we define

\begin{equation}
\frac{H_{z}}{H_{0}} = \sqrt{\Omega_{\text{r}}(1+z)^{4}+\Omega_{\text{M}}(1+z)^{3}+\Omega_{\text{k}}(1+z)^{2}+\Omega_{\Lambda}}
\label{dimensionless hubble}
\end{equation}

where $\Omega_{\text{r}}$ is the radiation energy density, the curvature of the Universe $\Omega_{k}$ is considered flat, and $H_0=70$ km $\text{s}^{-1} \text{Mpc}^{-1}$. We choose to neglect $\Omega_{\text{r}}$ in our computations because the Universe seems to be closely represented by $\Omega_{\text{r}}$ = 0. Finally we define the theoretical luminosity distance as

\begin{equation}
d_{\text{L}}(z,H_{0},\Omega_{\text{M}}) = (1+z)\frac{c}{H_{0}}\int_0^z\frac{dz'}{\sqrt{\Omega_{\text{M}}(1+z')^{3}+\Omega_{\Lambda}}}
\label{lum dist}
\end{equation}

where the dark energy density $\Omega_{\Lambda} = 1 - \Omega_{\text{M}}$ given this flat cosmological model in the Friedmann Lemaitre Robertson Walker metric. Using this definition of luminosity distance, we hereby define the likelihood for the full GRB sample. We compare Equation (\ref{mu_theory}) with the observed distance modulus that we can derive from the fundamental plane correlation in Equation (\ref{Fundamental_plane}), by isolating the luminosity distance in the following way:

\begin{equation}
    \log_{10}(d_\text{L})=
    \frac{a \log_{10} T^{*}_\text{a}}{2 (1-b)}
    +
    \frac{b \cdot (\log_{10} F_{\text{peak}}+\log_{10}K_{\text{peak}})}{2 (1-b)}
    +
    \frac{(b-1)\log_{10}(4\pi)+c}{2 (1-b)}
    -
    \frac{\log_{10}F_{\text{a}} + \log_{10}K_{\text{a}}}{2 (1-b)}
\end{equation}

where $K_{\text{peak}}$ and $K_{\text{a}}$ are the $K$-corrections for cosmic expansion \citep{2001AJ....121.2879B} computed for the prompt and the afterglow, respectively. This relation was achieved by defining $L_\text{a}$ as $4 \pi d_L^2  F_{\text{a}}$, with $F_{\text{a}}$ as the flux at $T_\text{a}^*$, and defining $L_{\text{peak}}$ as $4 \pi d_{\text{L}}^2 F_{\text{peak}}$, with $F_{text{peak}}$ as the peak flux in the prompt emission. Using now the definition of the distance modulus and the new variables definitions
$a_1=a/(2(1-b))$; $b_1=b/(2(1-b))$; $c_1=((b-1)\log_{10}(4\pi)+C)/(2(1-b))$; $d_1=-1/(2(1-b))$; $F_{\text{peak,cor}}= F_{\text{peak}} \cdot K_{\text{peak}}$; and $F_{\text{a,cor}}= F_{\text{a}} \cdot K_{\text{a}}$,
we obtain:

\begin{equation}
\mu_{\text{obs, GRB}} = 5 \cdot (a_{1}\ \log_{10}(T_{\text{a}}^{*})+b_{1}\ \log_{10}(F_{\text{peak,cor}})+c_{1}+d_{1}\ \log_{10}(F_{\text{a,cor}}))+25.
\label{mu_obs, GRB}
\end{equation}
This allows us to define the following likelihood that has the advantage to use the distance modulus directly, as the one related to SNe Ia does:

\begin{equation}
\mathcal{L}_{\text{GRB}}=\sum_i\bigg( ln\bigg(\frac{1}{\sqrt{2\pi}\sigma_{\mu,i}}\bigg)-\frac{1}{2}\bigg(\frac{\mu_{\text{th,GRB,}i}-\mu_{\text{obs,GRB,}i}}{\sigma_{\mu,i}}\bigg)^{2}\bigg)
\label{GRB likelihood}
\end{equation}

where $\sigma_{\mu,i}$ is the error on the observed distance moduli.
The observed distance moduli obtained by the GRBs through the variables pertinent to the fundamental plane is compared with the theoretical GRB distance moduli. Specifically, the observed quantities of the GRBs do not depend on cosmology, since they are the observed flux, $F_{\text{a}}$, its rest-frame time at the end of the plateau emission, $T^{*}_{\text{a}}$ and the peak prompt flux, $F_{\text{peak}}$. The parameters $a_1$, $b_1$ and $c_1$ are defined from $a$, $b$, and $c$ which are left free to vary and converge to the theoretical distance of the GRBs. They are completely independent from the distance of the SNe Ia. We calculate the theoretical distance within large priors of 0 < $\Omega_{\text{M}}$ < 1. The parameters that minimize the difference between the theoretical and observational moduli then provide the most probable values of $\Omega_{\text{M}}$. We fix reasonable priors of the $a$, $b$ and $c$ and implicitly in $a_1$, $b_1$ and $c_1$ so that the parameters of the fundamental plane remain roughly constant in  ranges compatible with the expected underlying physics of the correlation. By keeping these reasonable ranges the reliability of the plane is thus preserved. In this regard, GRBs through the fundamental plane relation can be considered standardizable candles because the parameters of the correlations are left to vary within physical allowed ranges. For example, we cannot allow the $a$ parameter to be greater than zero, because it would then imply a different physics and as a consequence the energy reservoir of the plateau would not be constant (see the magnetar model for the evaluation of this parameter being closer to $-1$, \citep{2018ApJ...869..155S}). Similarly, we cannot allow the $b$ parameters to be less than zero. This would not respect the physical observations that the more kinetic energy is transferred in the prompt, the more to the afterglow as it is demonstrated in \cite{2011AIPC.1358..113D, 2015ApJ...800...31D}. This assumption is supported by theoretical modeling \citep{2014MNRAS.442.3495V, 2014MNRAS.445.2414V}.

We then also compare $\mu_{\text{theory}}$ to the observed distance modulus of the SNe Ia sample, $\mu_{\text{obs, SNe}}$, and finally compute the likelihood for the full sample. We add SNe Ia by allowing the total likelihood to encompass all samples:
\begin{equation} \label{updated likelihood}
\mathcal{L}_{\text{Tot}}= \mathcal{L}_{\text{GRB}} + \mathcal{L}_{\text{SNe}}
\end{equation}
where $\mathcal{L}_{\text{SNe}}$ is the minimizing functions defined canonically for the SNe Ia sample:
\begin{equation}
\mathcal{L} =- \frac{1}{2}\sum_{i} \frac{(\mu_{\text{obs}}^{i}-\mu_{\text{theory}}^{i})^{2}}{(\epsilon^{i}_{\mu_{\text{obs}}})^{2}}.
\end{equation}
We generalize the above equation into
\begin{equation}
\mathcal{L} = -\frac{1}{2}\bmath{\Delta\mu}^{T}\mathbfss{{C}}^{-1}\bmath{\Delta\mu}
\end{equation}
where $\mathcal{C}$ is the covariance matrix, including both statistical uncertainties diagonally and systematic contributions in the opposed diagonal.

We now allow $\Omega_{\text{M}}$ to vary together with the fundamental plane parameters with a uniform prior in the interval $0 \le \Omega_{\text{M}} \le 1$. When we use this GRB sample in conjunction with the SNe Ia, we observe an $\Omega_{\text{M}}= 0.299 \pm 0.009$. Conversely, previous results for this cosmological parameter, probing only with SNe Ia data, yield an $\Omega_{\text{M}}= 0.298 \pm 0.008$. Slightly increased error bars on $\Omega_{\text{M}}$ exist due to the addition of the errors carried out  from the GRB sample. However, we show in the following sections how the sample can be resized so that we can reduce the GRB scatter to produce the smallest error bars on $\Omega_{\text{M}}$ yet when we consider evolution (see Sec.\S \ref{sec:EP_method}).

\subsection{Deriving \texorpdfstring{$\Omega_{\text{M}}$}{TEXT}  with the Trimmed X-ray GRB Sample + SNe Ia} \label{sec:PLATtrim+SNe}

To make the errors on the parameters computed in this analysis as small as possible, we now look for the tightest fundamental plane correlation by considering a subset of the X-ray PLAT sample composed of only the GRBs whose plane has an intrinsic scatter near zero. The goal is to reduce the error bars on $\Omega_{\text{M}}$ from what was computed before with SNe Ia + PLAT. This is done by calculating the closest GRBs to the X-ray fundamental plane. We choose this number to be 10 GRBs from the full PLAT sample, and hereby refer to this subsample as the trimmed platinum sample (PLATtrim). These 10 GRBs constitute a large enough sample to define a plane, yet still give a $\sigmaintxtrim$ near zero. By increasing the sample, $\sigmaintxtrim$ also increases. In these calculations, a near-zero intrinsic scatter is one on the order of $10^{-2}$ or smaller. In this section, we consider this new subsample of the X-ray GRBs using GRBs + SNe Ia to again derive $\Omega_{\text{M}}$.

\begin{figure*}
\centering
    \begin{tabular}{cc}
        \includegraphics[width=0.50\textwidth]{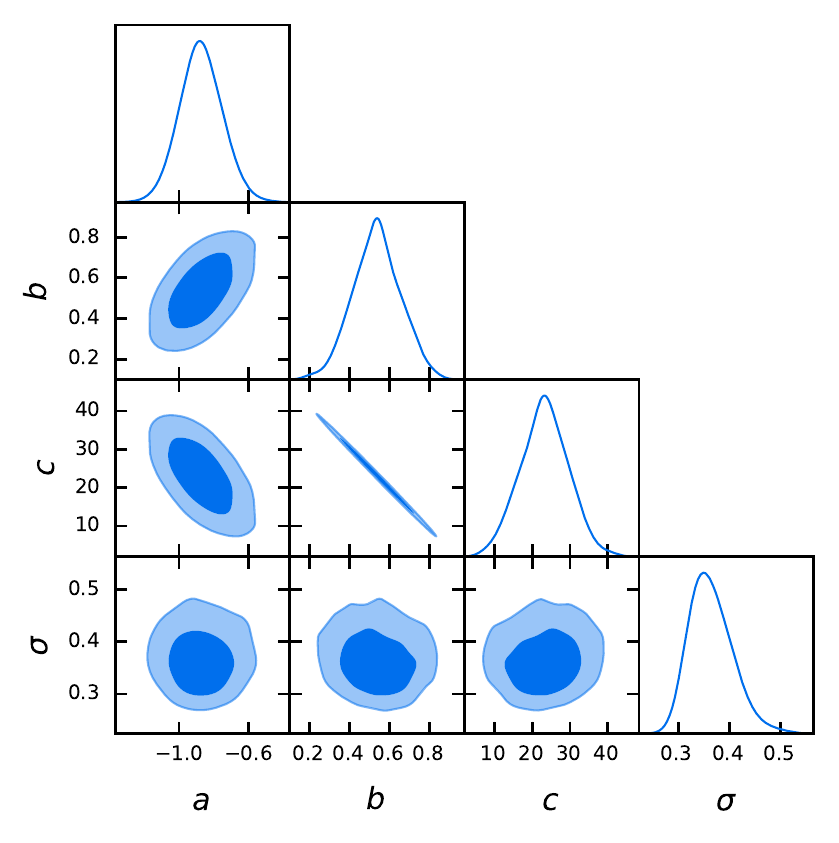} &
        \includegraphics[width=0.50\textwidth]{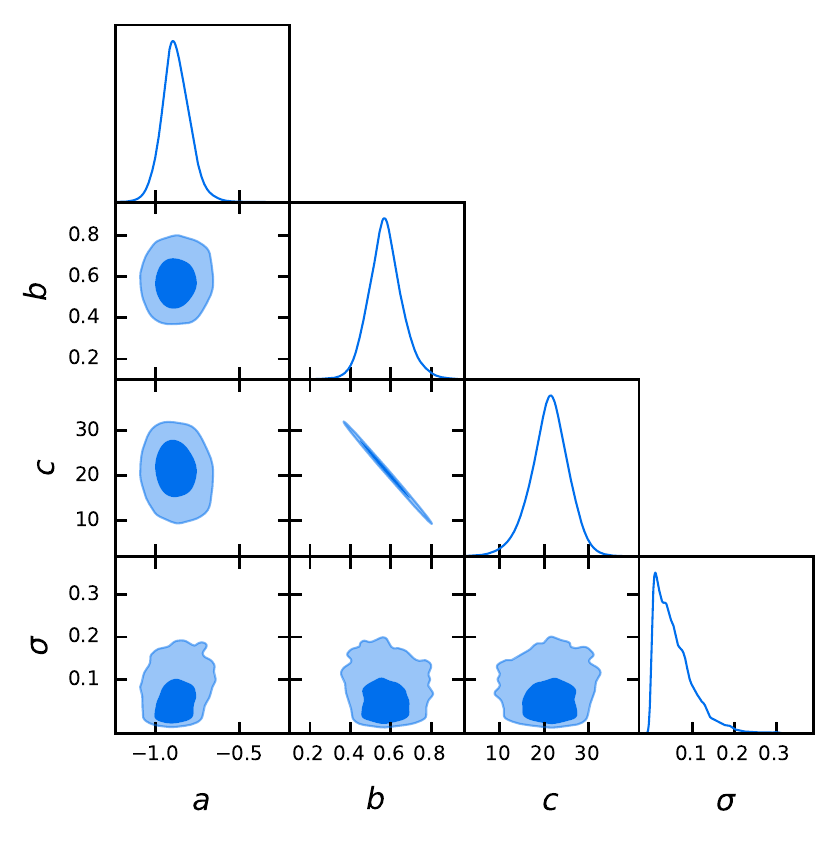} \\
        \includegraphics[width=0.50\textwidth]{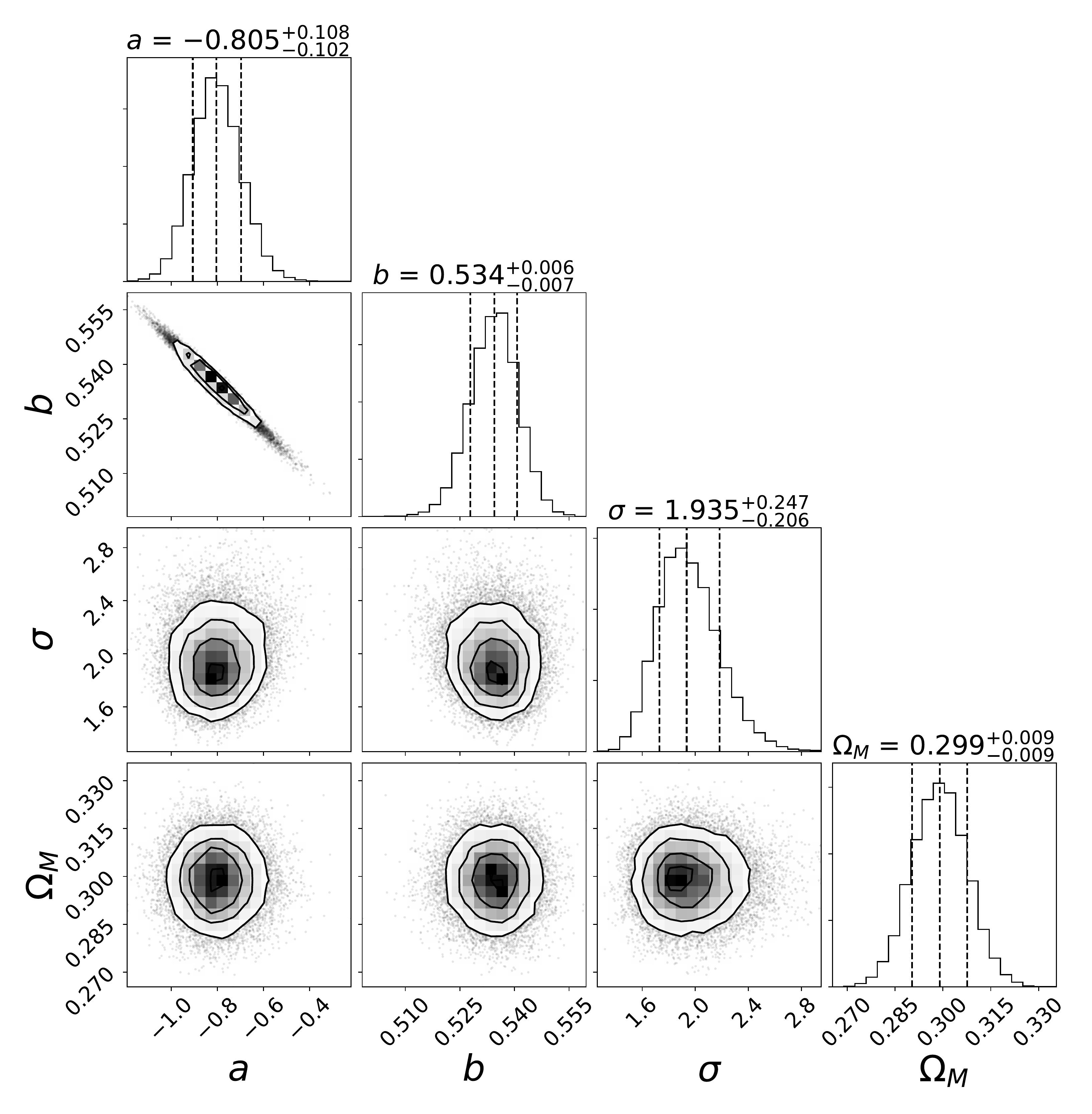} &
        \includegraphics[width=0.50\textwidth]{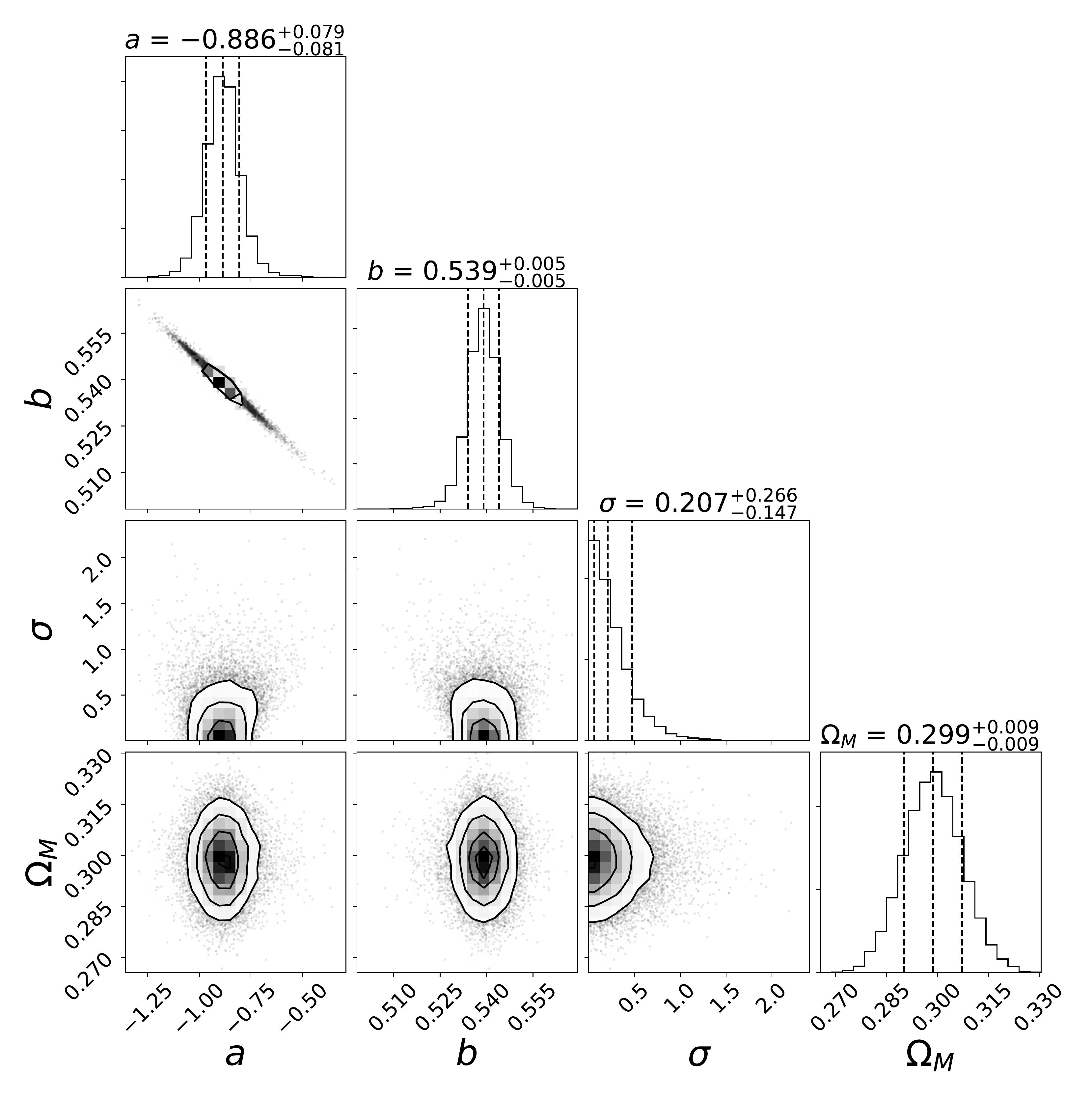} \\
    \end{tabular}
\caption{This figure compares the PLATtrim over PLAT when both are paired with SNe Ia data. The upper panels show the fundamental plane fitting for the full GRB PLAT sample (upper left) and the PLATtrim  (upper right)  calculated with \emph{cobaya}. The bottom panels show the derivation of the $\Omega_{\text{M}}$ correspondent to the upper panels by the SNe Ia combined with both the full PLAT and PLATtrim samples. Each plot shows the 2D posterior contours.}
\label{fig:PLAT_GRB+SNE}
\end{figure*}

The results for this new plane fitting of the PLATtrim sample are as follows:  $\axtrim = -0.89 \pm 0.08$, $\bxtrim = 0.54 \pm 0.005$, $\cxtrim = 20.14 \pm 4.05$, and $\sigmaintxtrim = 0.05 \pm 0.05$. These results are seen in Fig. \ref{fig:PLAT_GRB+SNE} (upper right). By the trimming of the PLAT sample, we use this newly-defined fundamental plane for which a smaller intrinsic scatter exists in comparison to the full PLAT sample. Again, we perform cosmological computation together with SNe Ia data, and we obtain $\Omega_{\text{M}}= 0.299 \pm 0.009$. It should be noted that the errors on the uncertainties on $\Omega_{\text{M}}$ determined by the MCMC calculations are one order of magnitude less than the uncertainties on $\Omega_{\text{M}}$ itself; for details, see Appendix Sec. \S \ref{MCMC error}.

The full compilation of results are compared in Table \ref{tab:Effects of using PLATtrim}. With the inclusion of the SNe Ia, we do not yet see an improvement in the results by the combination of probes by trimming the PLAT sample down to the 10 GRBs closest to the plane. The table makes clear that the PLATtrim sample has yet to be more efficacious in the reduction of the overall scatter, and consequently, in the error on $\Omega_{\text{M}}$.
We reach the same precision on $\Omega_{\text{M}}$ as the one obtained by the SNe Ia when the evolutionary parameters shown in Table \ref{tab:kcorrs} are considered (see Table \ref{tab:ev}). 
In Sec.\S \ref{sec:simulations}, the true effects of the PLATtrim sample becomes visible and efficacious when we run simulations; the precise plane that the PLATtrim sample define is used successfully as a base for simulating additional GRBs.

As a final note, we must also stress that there is no calibration of the PLAT sample related to the SNe Ia, but we do fix the flat $\Lambda$CDM model to perform a comparison with the uncertainties derived with SNe Ia. However, it is important to note that the goal of the paper is to explore the reliability of the fundamental plane as a cosmological tool in comparison with SNe Ia, and not comparing different cosmological models. We additionally point out that this procedure of trimming the sample is meant to show how many GRBs should be used in the future once more data is available. This is the reason why the trimmed sample is the basis of our MCMC simulations and will inform us on how many of these GRBs close to the fundamental plane need to be chosen in order to have similar precision on the $\Omega_{\text{M}}$ parameter compared to the SNe Ia.



\section{Exploring the Efficacy of Deriving \texorpdfstring{$\Omega_{\text{M}}$}{TEXT}  with Optical GRB Samples + SNe Ia Data} \label{opt sect}

\begin{table*}
\caption{The first column refers to the sample used, while the second column refers to the results of $\Omega_{\text{M}}$. Results are obtained without the correction for evolution. The errors reported in this table are the corresponding to the $68\%$ confidence limit.} \label{tab:Effects of using PLATtrim}
\begin{tabular}{|c|c|c|c|}
\hline
Sample & $\Omega_{\text{M}}$  \\\hline
SNe & $0.299 \pm 0.008$  \\
PLAT+SNe Ia & $0.299 \pm 0.009$ \\
PLATtrim+SNe Ia & $0.299 \pm 0.009$\\
OPT+SNe  & $0.299 \pm 0.009$ \\
OPTtrim+SNe Ia & $0.299 \pm 0.009$\\
\hline
\end{tabular}
\end{table*}

In this section, we shift from the use of the X-ray GRB emission data to test the reliability of optical GRB data. Similarly to the methodology used in the case of the X-ray fundamental plane, we calculate the number of GRBs closest to the optical plane that hold the intrinsic scatter to near-zero values. We investigate the 3D Dainotti relation at optical wavelengths to see how tight the plane is for a sample of 45 GRBs. The ability of this subsample to infer the parameters is then compared to that of the entire optical sample.

\begin{figure*}
\centering
    \begin{tabular}{cc}
        \includegraphics[width=0.50\textwidth]{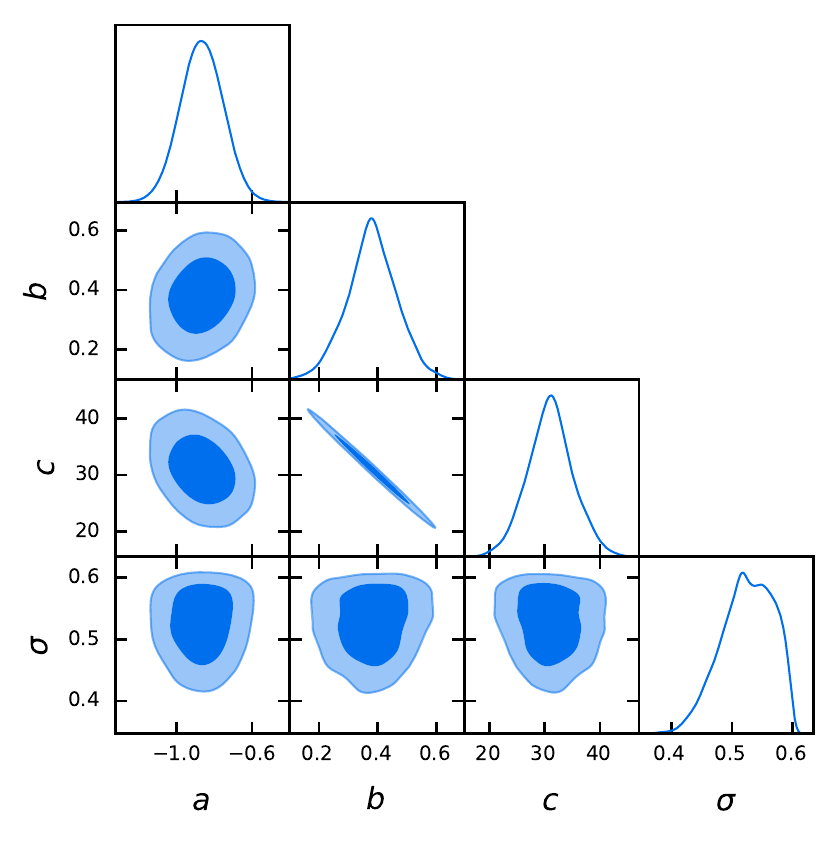} &
        \includegraphics[width=0.50\textwidth]{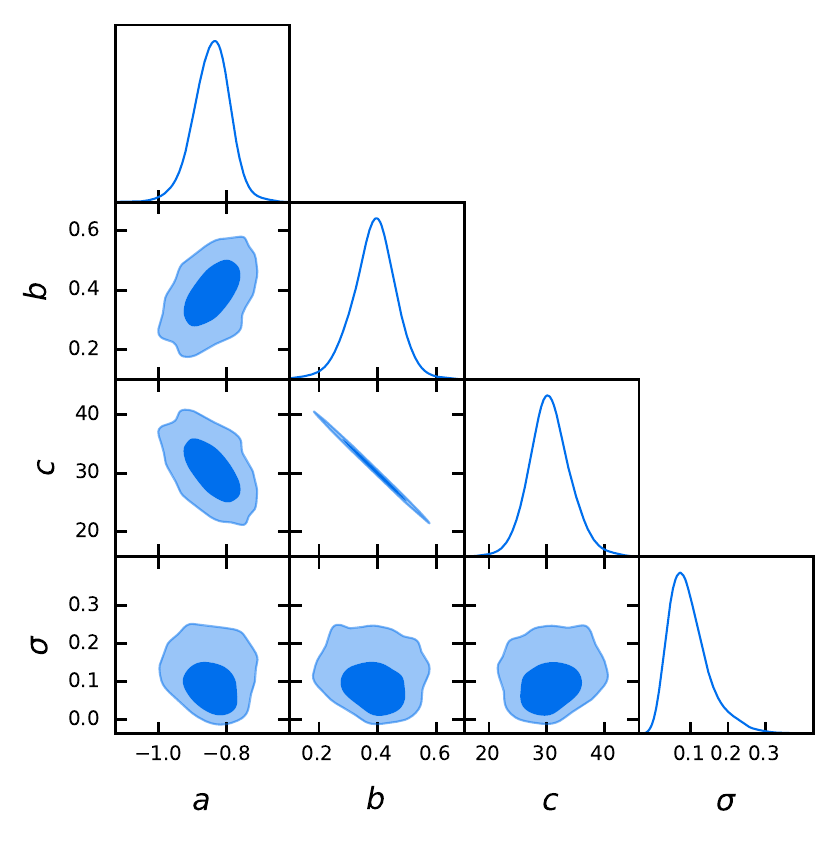} \\
        \includegraphics[width=0.50\textwidth]{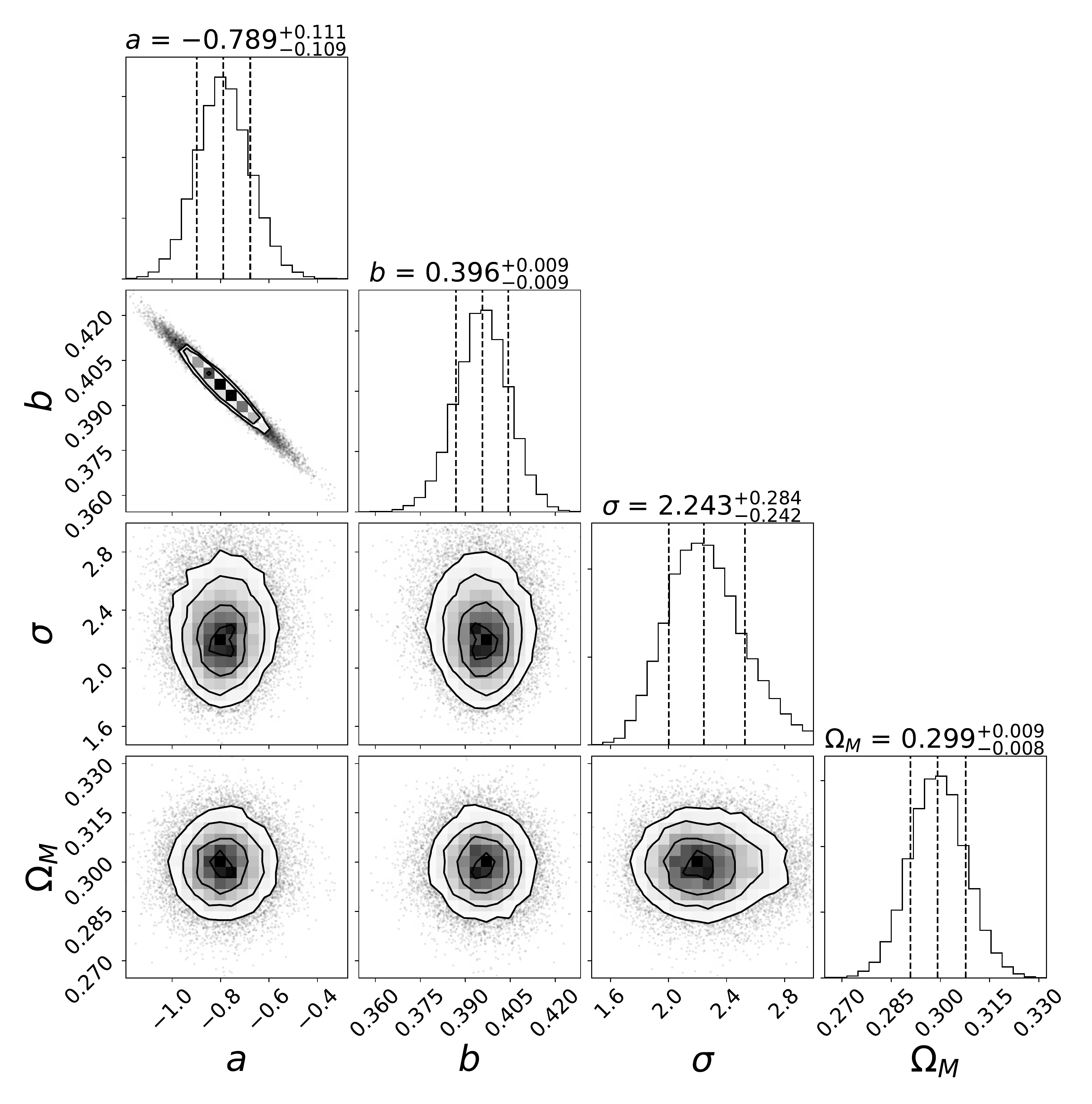} &
        \includegraphics[width=0.50\textwidth]{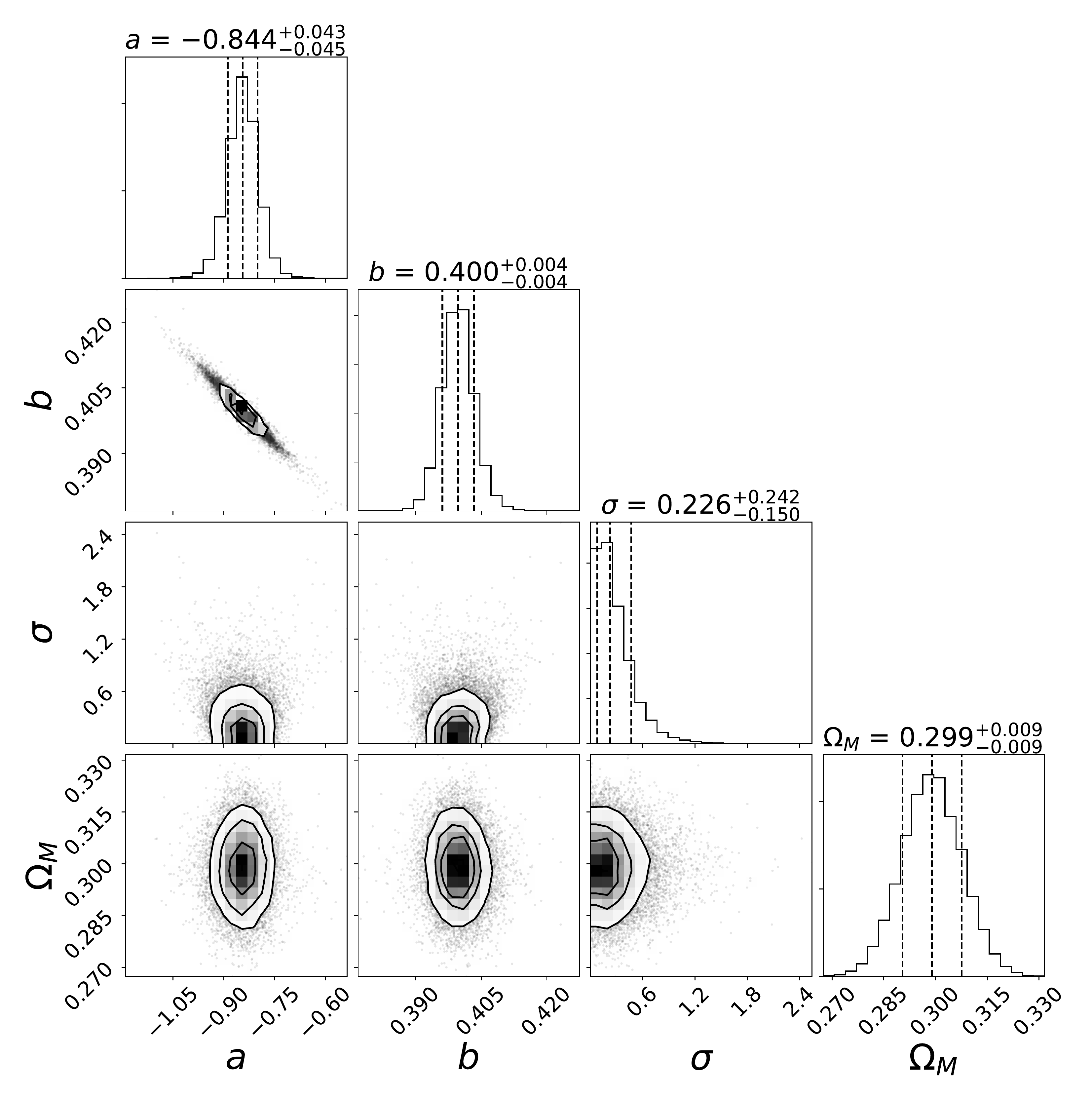} \\
    \end{tabular}
  \caption{The upper panels show the fundamental plane fitting for the full GRB OPT sample (upper left) and the OPTtrim  (upper right)  calculated with \emph{cobaya}. The bottom panels show the derivation of the $\Omega_{\text{M}}$ correspondent to the upper panels by the SNe Ia combined with both the full optical and the trimmed optical sample. Each plot shows the 2D posterior contours.}
  \label{fig2 optvopttrim}
\end{figure*}

Once again using the D'Agostini methodology, we compute the 3D fundamental plane parameters and the correspondent intrinsic scatter of the full optical GRB sample. The results are the following:
$\aoptfull = -0.87 \pm 0.11$, $\boptfull = 0.37 \pm 0.08$, $\coptfull = 31.46 \pm 4.07$, and $\sigmaintoptfull = 0.53 \pm 0.04$, and are shown in Fig. \ref{fig2 optvopttrim} (upper left). Combining the full optical sample (OPT) with SNe Ia data, we obtain $\Omega_{\text{M}} = 0.299 \pm 0.009$ (Fig. \ref{fig2 optvopttrim}, lower left). This analysis has been performed fixing the value of the parameter $\coptfull=30$ following the same strategy of \citet{Amati_2019}.
This result is novel and beneficial; the error is on par with that obtained by X-ray GRB samples. This means that the use of optical GRB samples may prove just as or perhaps more efficacious in constraining cosmological parameters with a future larger sample. This leads us to perform a similar trim on this optical data as to the one performed on the X-ray sample in an attempt to better understand the behavior of the optical data.

We find and consider only the optical GRBs closest to the plane with the aim to produce a sample with near-zero intrinsic scatter, hereby referred to as the optical trimmed sample (OPTtrim). We determine through testing that the maximum number of optical GRBs that maintain this level of accuracy is 10, as it was also in X-ray, and the best-fit results are as follows: $\aopttrim = -0.84 \pm 0.07$, $\bopttrim = 0.40 \pm 0.09$, $\copttrim = 29.87 \pm 4.12$, and $\sigmaintopttrim = 0.11 \pm 0.08$ (Fig. \ref{fig2 optvopttrim}, upper right).

Again, $\sigmaintopttrim$ peaks around zero. We again include SNe Ia, resulting in an $\Omega_{\text{M}} = 0.299 \pm 0.009$ (Fig. \ref{fig2 optvopttrim}, lower right). We here note in Table \ref{tab:Effects of using PLATtrim} that the error bars derived by the optical emission data is comparable with that of the X-ray sample, asserting that optical GRB data can be just as serviceable as X-ray ones in constraining cosmological parameters. Therefore, we continue the computations considering both wavelengths independently.

\section{Considering Redshift Evolution Correction} \label{sec:EP_method}

\begin{figure*}
\centering
    \begin{tabular}{cc}
        \includegraphics[width=0.45\textwidth]{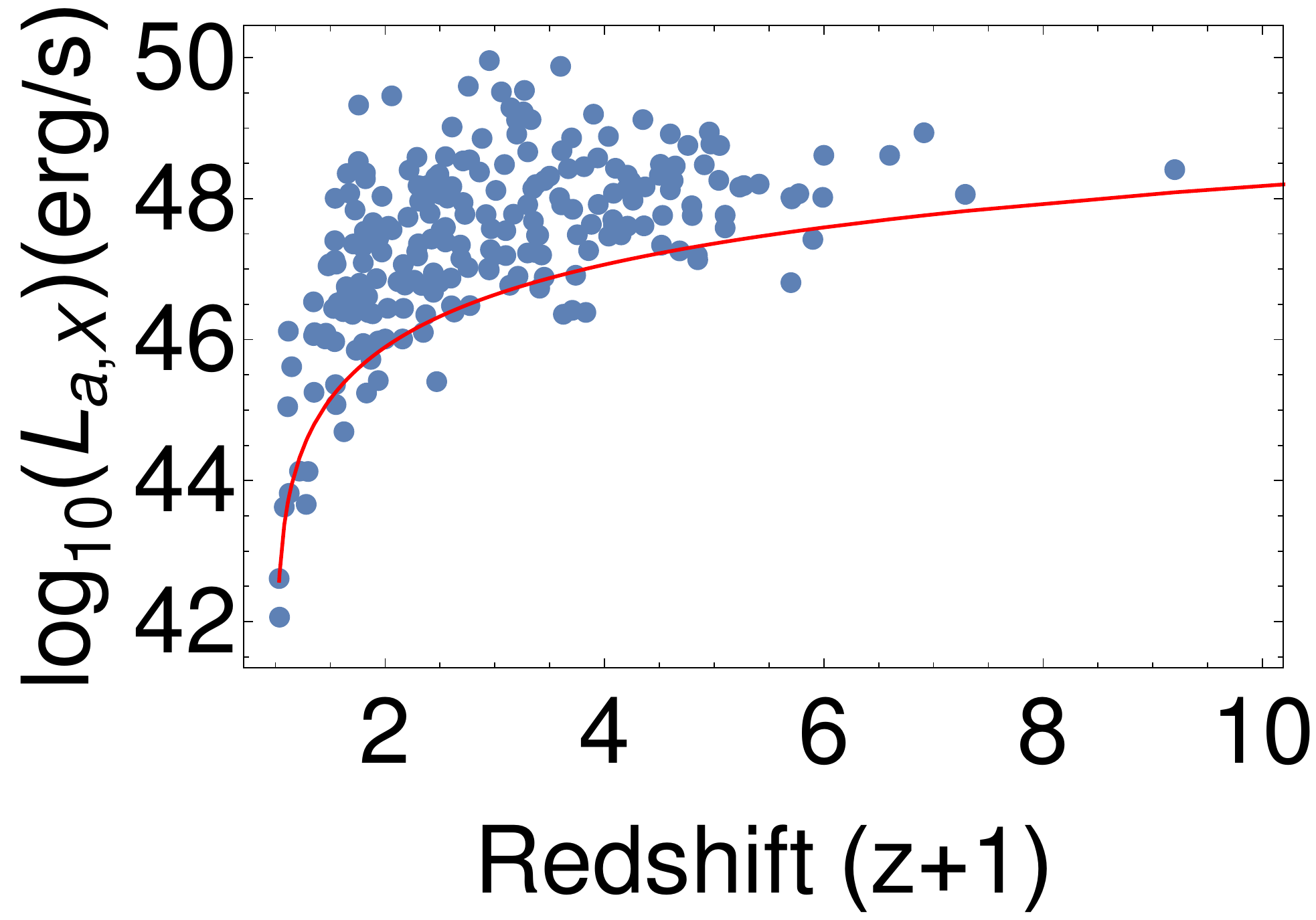}\label{fig:EP X-ray luma} &
        \includegraphics[width=0.45\textwidth]{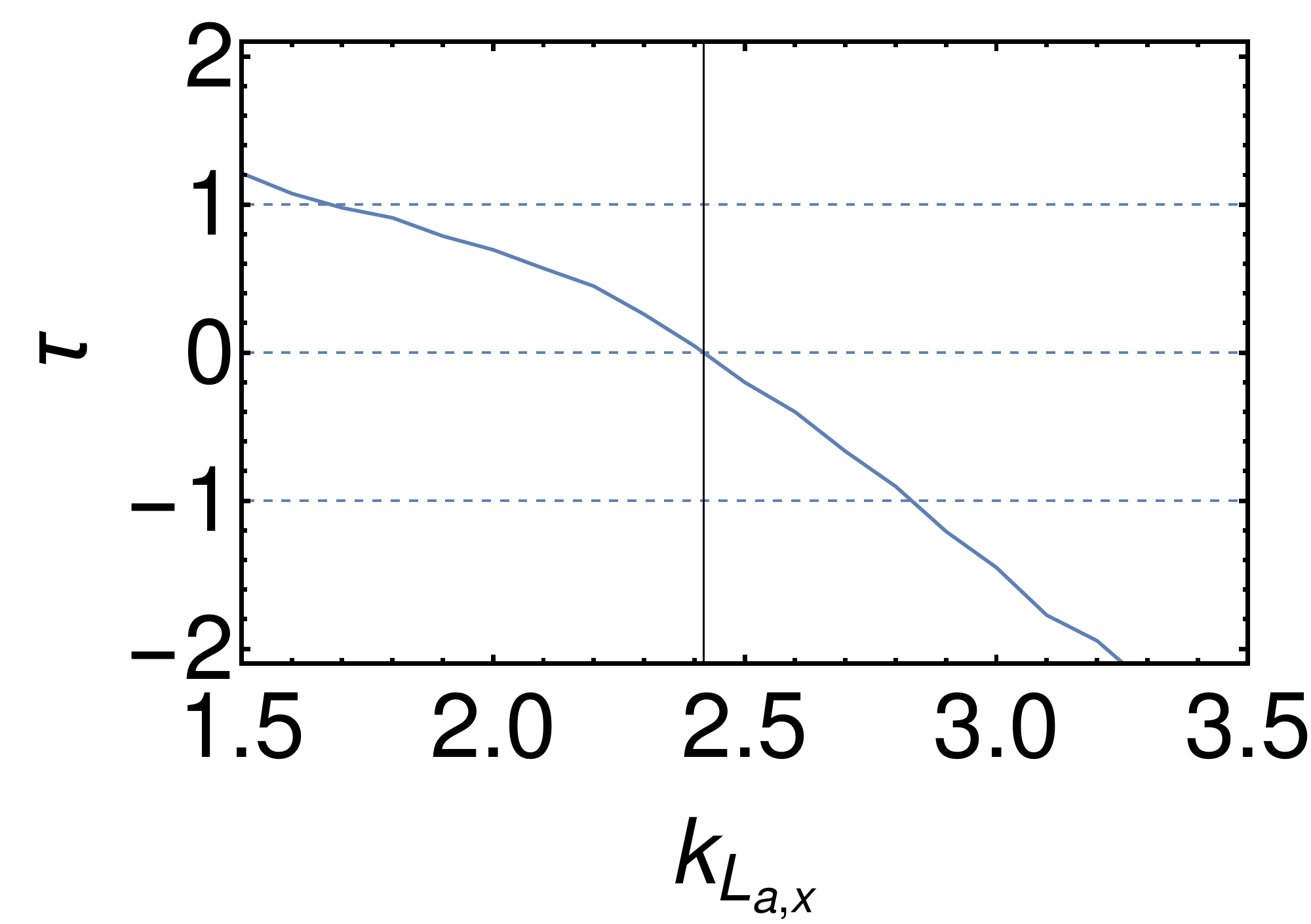}\label{fig:EP X-ray luma tau} \\
        \includegraphics[width=0.45\textwidth]{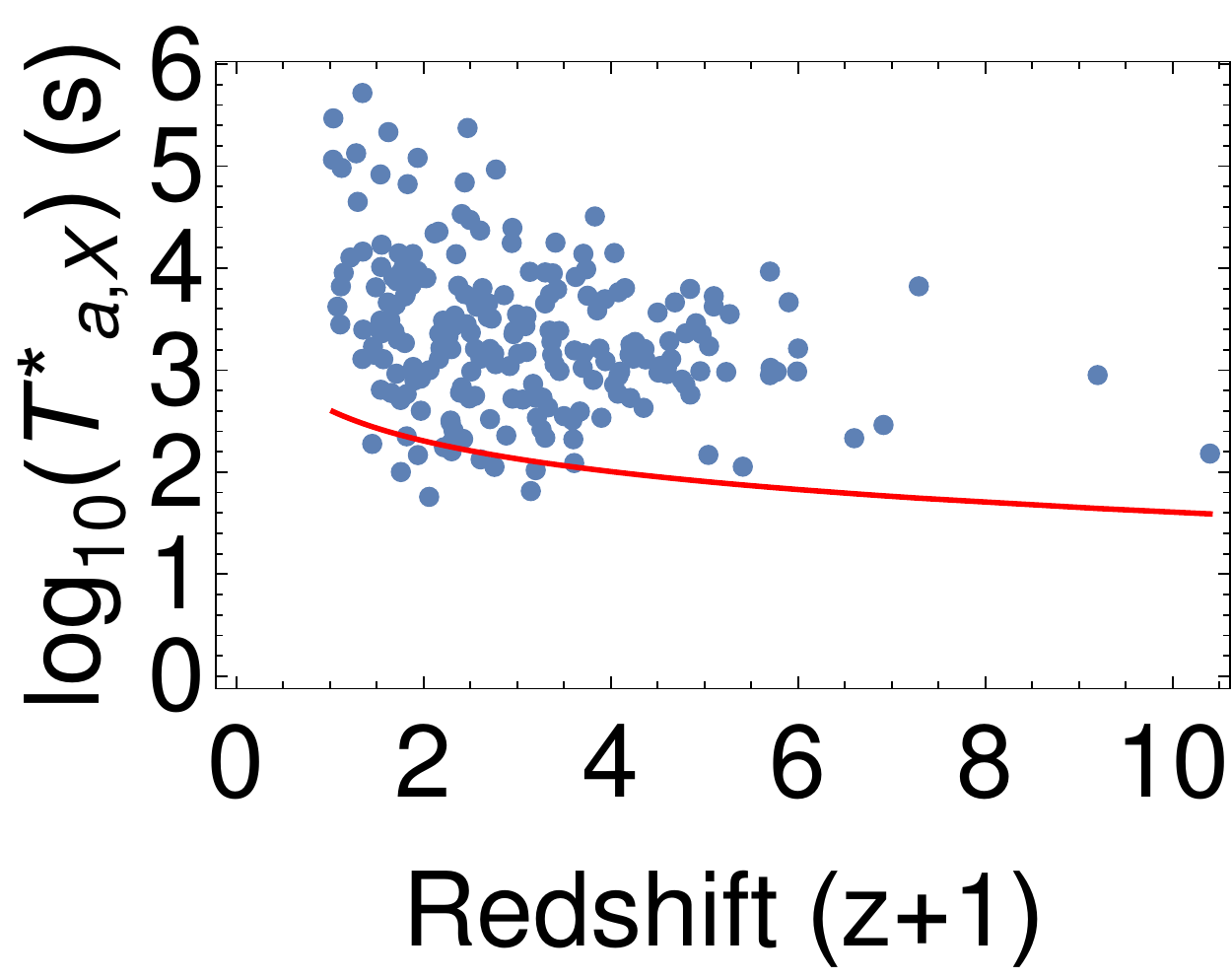}\label{fig:EP xray ta} &
        \includegraphics[width=0.45\textwidth]{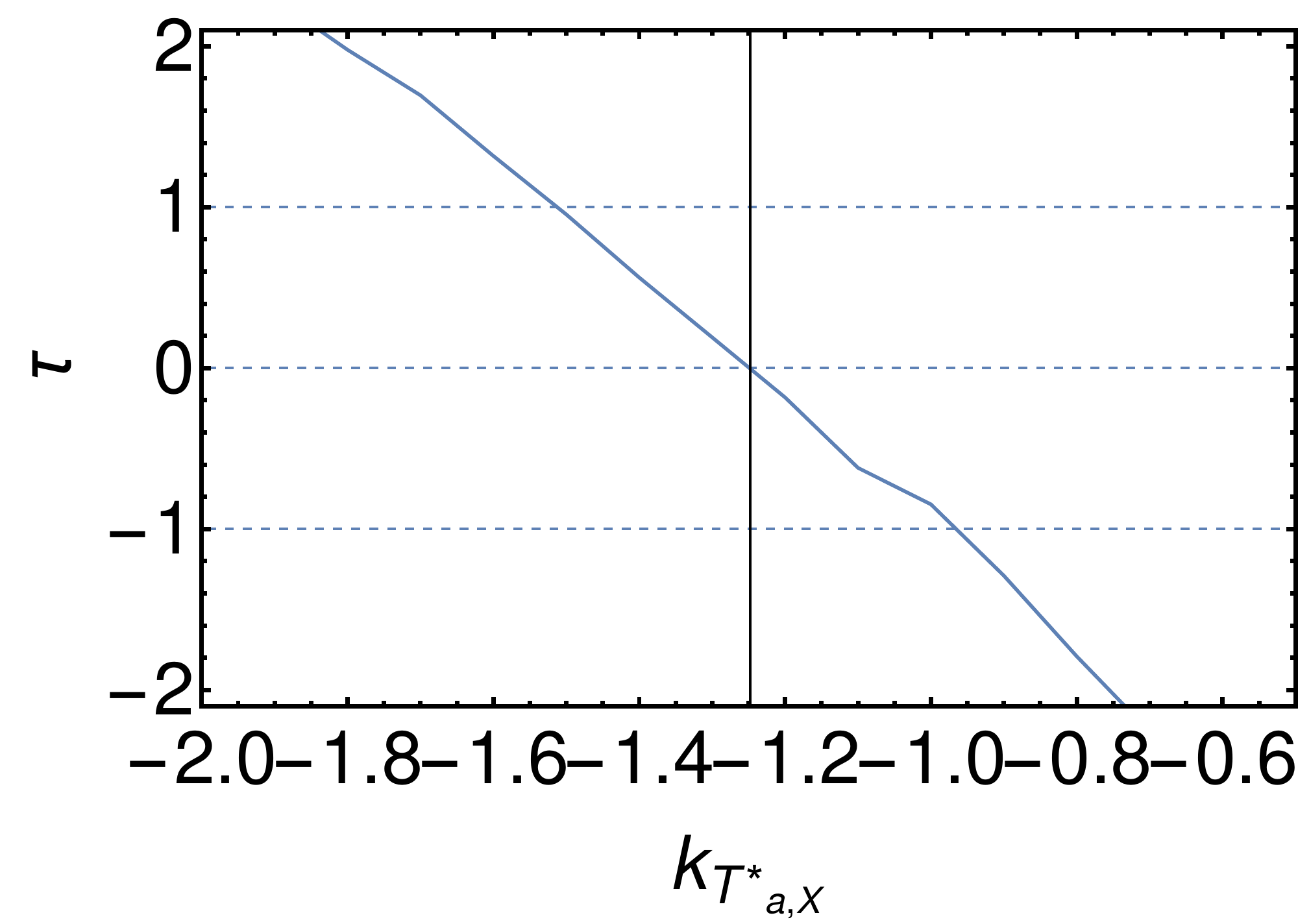}\label{fig:EP xray ta tau} \\
        \includegraphics[width=0.45\textwidth]{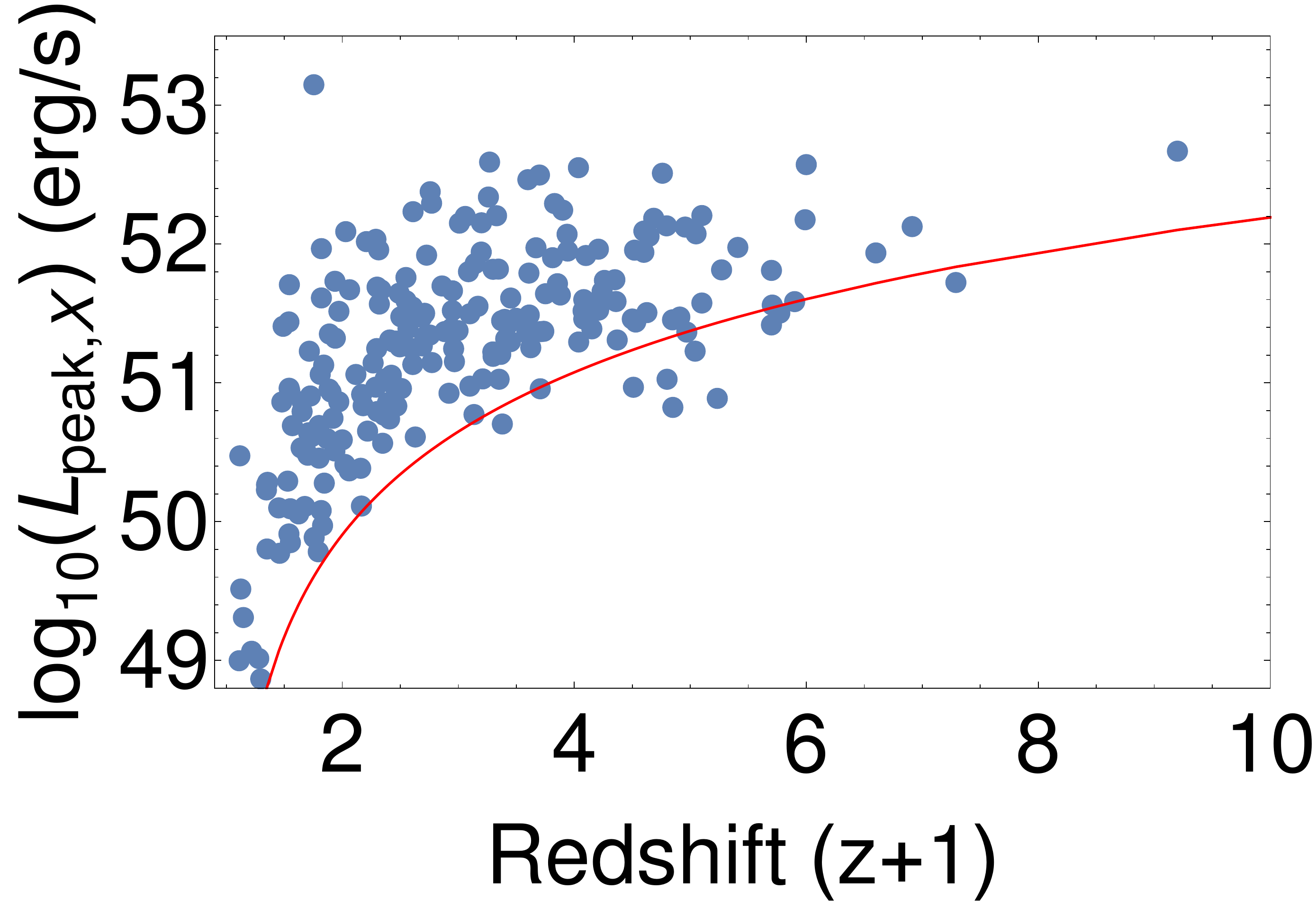}\label{fig:EP xray lpeak} &
        \includegraphics[width=0.45\textwidth]{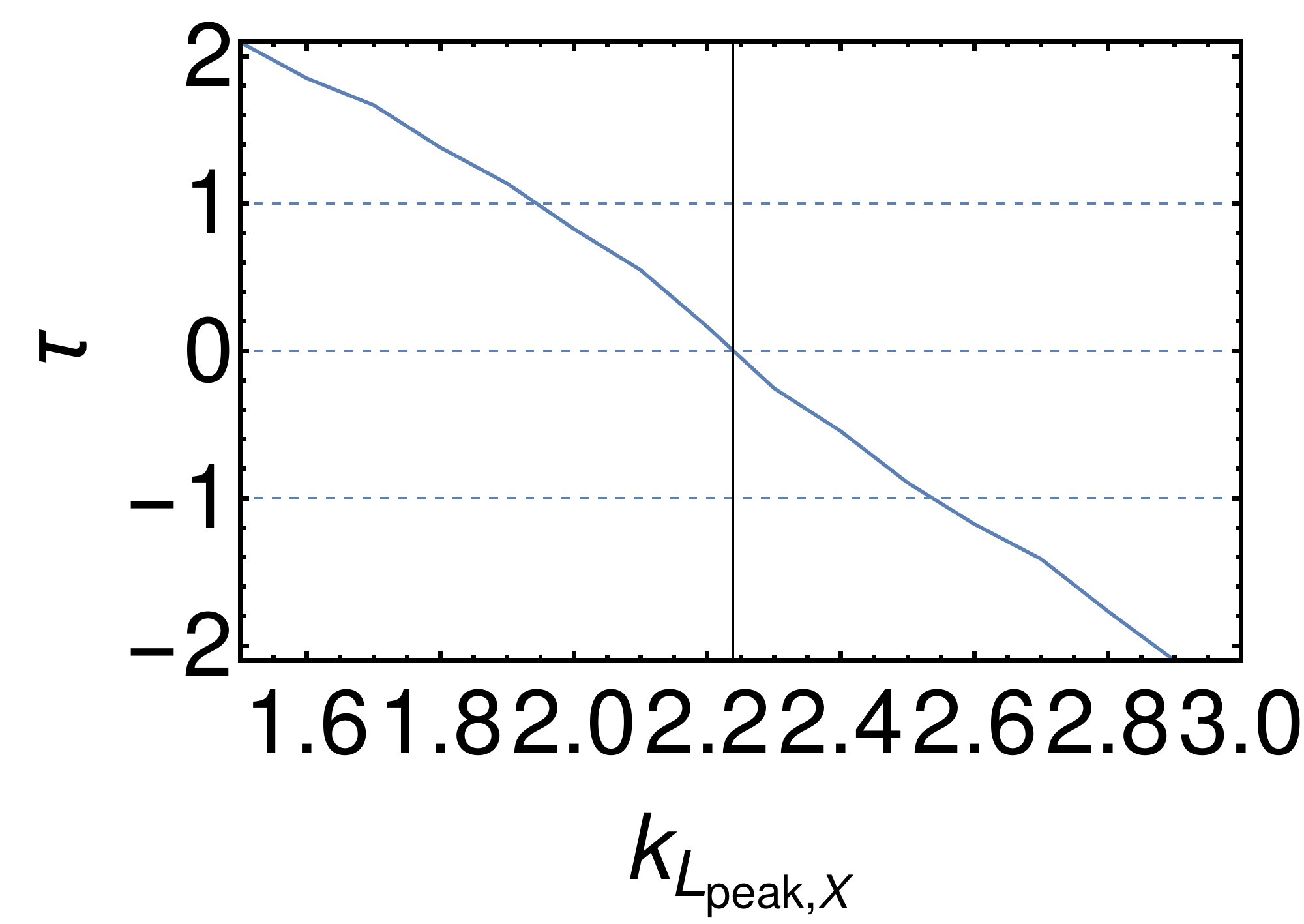}\label{fig:EP xray lpeak tau} \\
    \end{tabular}
\caption{This figure shows how the limiting luminosities and the Kendall $\tau$ vs the slope of the evolutionary functions for the full X-ray GRB sample. The panels show the evolution of $L_{\text{a,X}}$, $T_{\text{a,X}}^{*}$, and $L_{\text{peak,X}}$ vs. redshift. The limiting line is plotted in red. The right panels show the evolution of $\tau$. The middle dashed line is $\tau=0$ and the dashed lines are the defined bounds of $+1\sigma$ and $-1\sigma$, while the red line corresponds to the best-fit value of $\tau$.}
\label{fig:EP xray}
\end{figure*}

\begin{figure*}
\centering
    \begin{tabular}{cc}
        \includegraphics[width=0.50\textwidth]{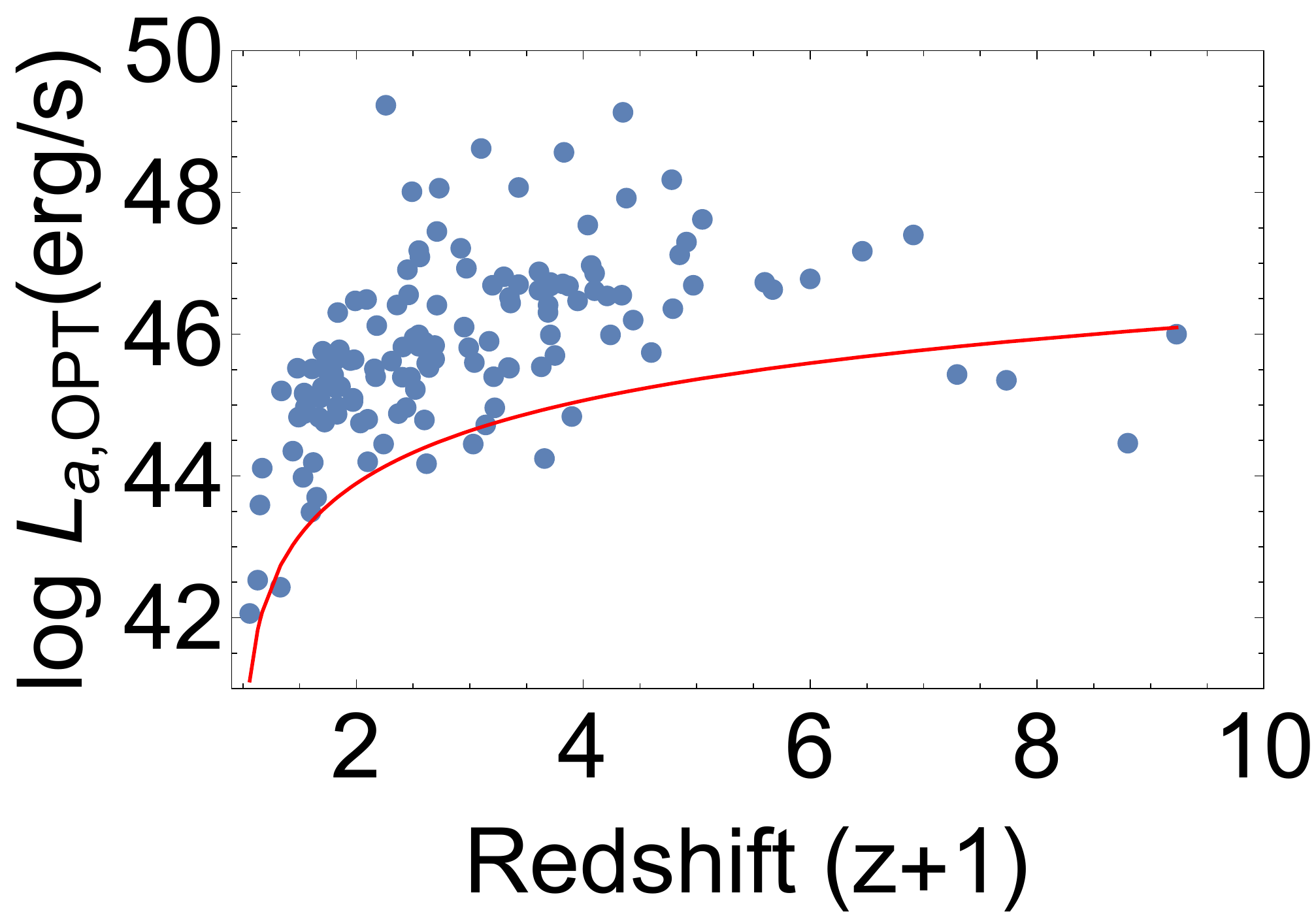}\label{fig:EP opt luma} &
        \includegraphics[width=0.50\textwidth]{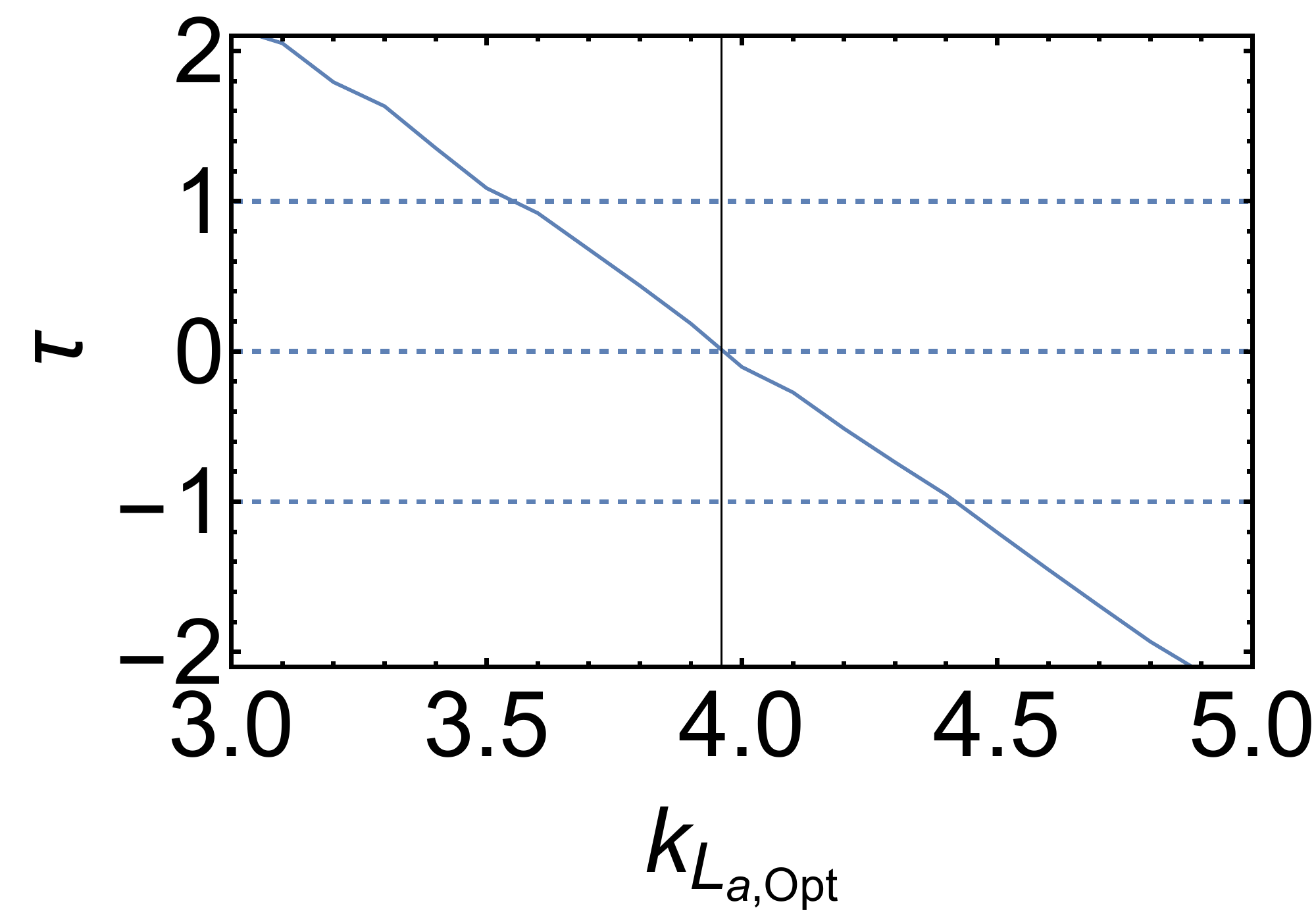}\label{fig:EP opt luma tau} \\
        \includegraphics[width=0.50\textwidth]{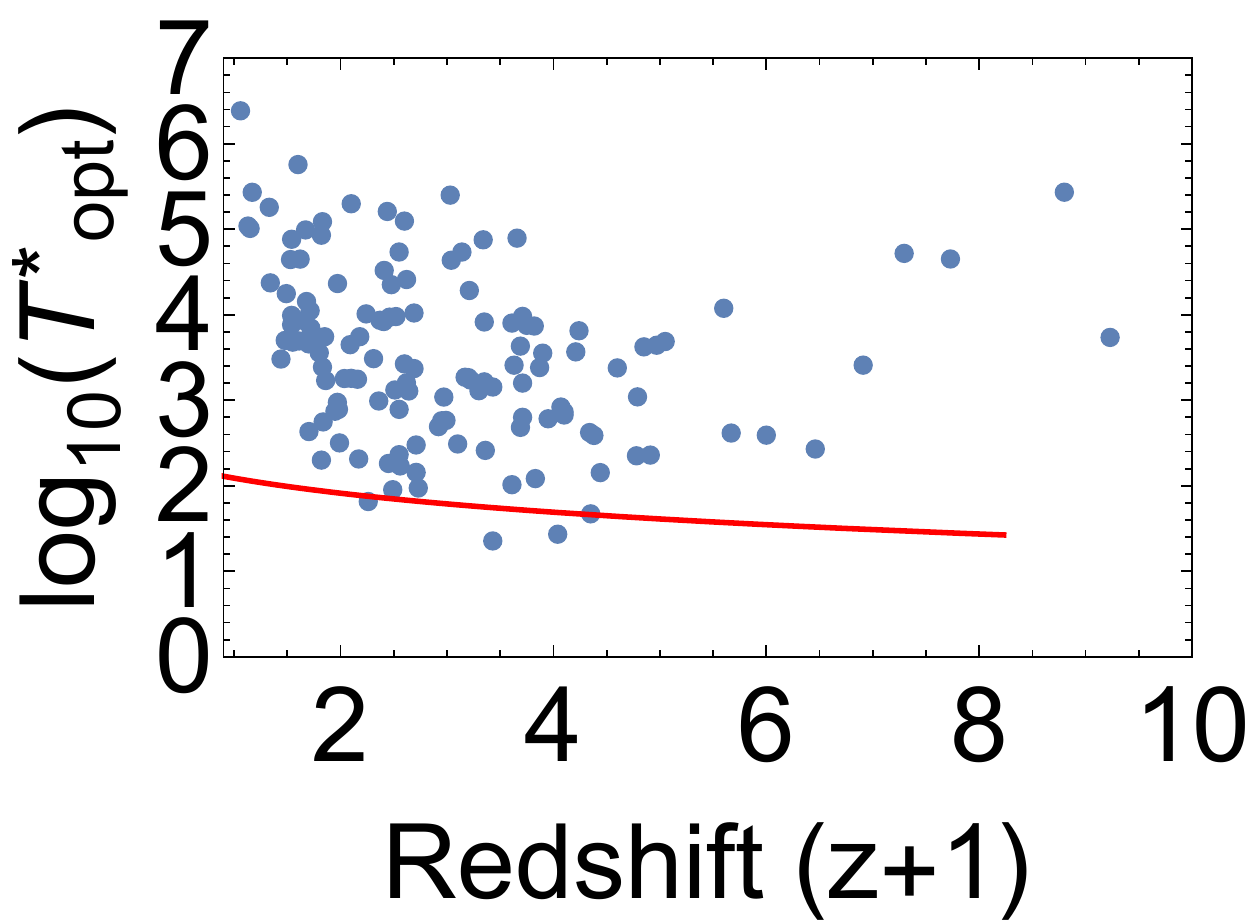}\label{fig:EP opt ta} &
        \includegraphics[width=0.50\textwidth]{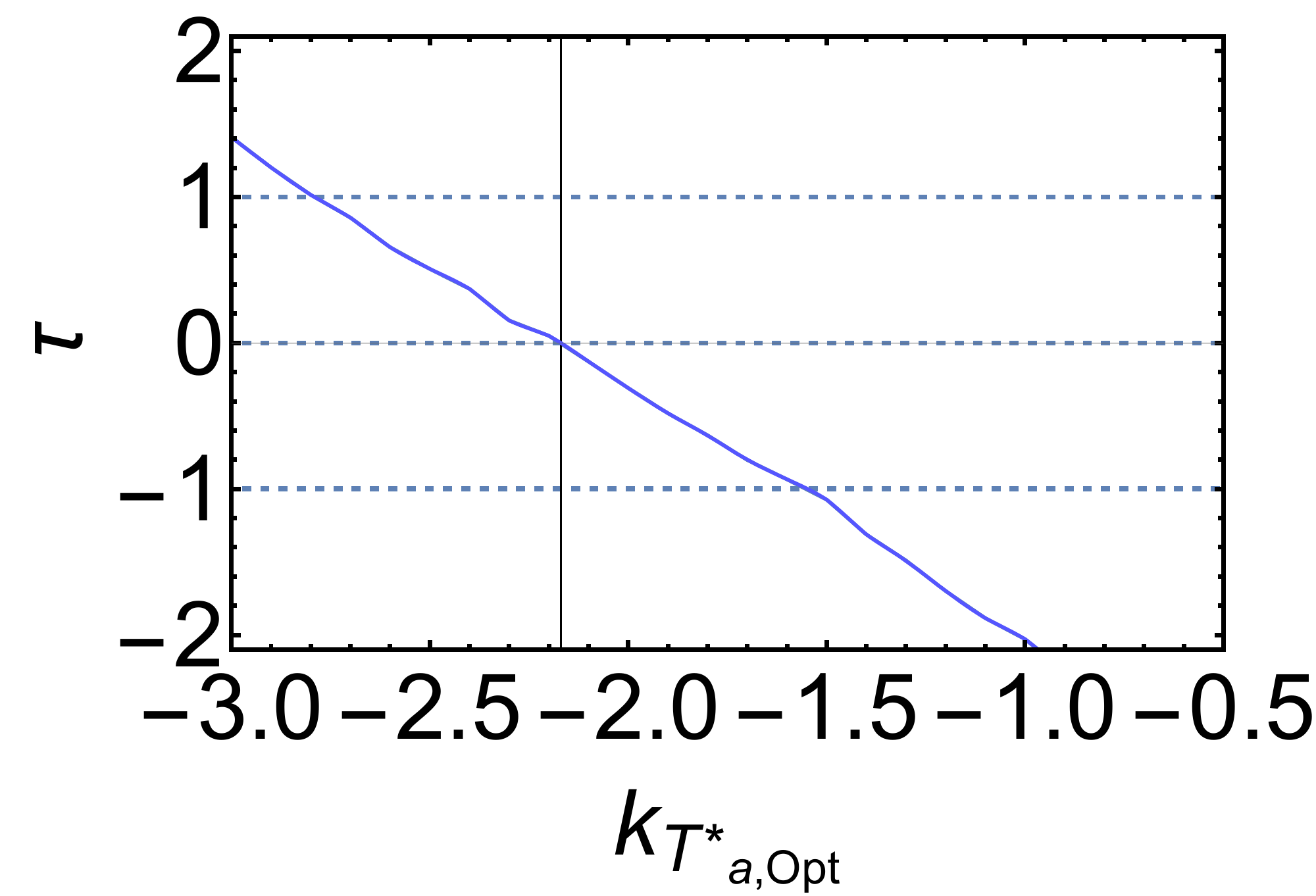}\label{fig:EP opt ta tau} \\
        \includegraphics[width=0.50\textwidth]{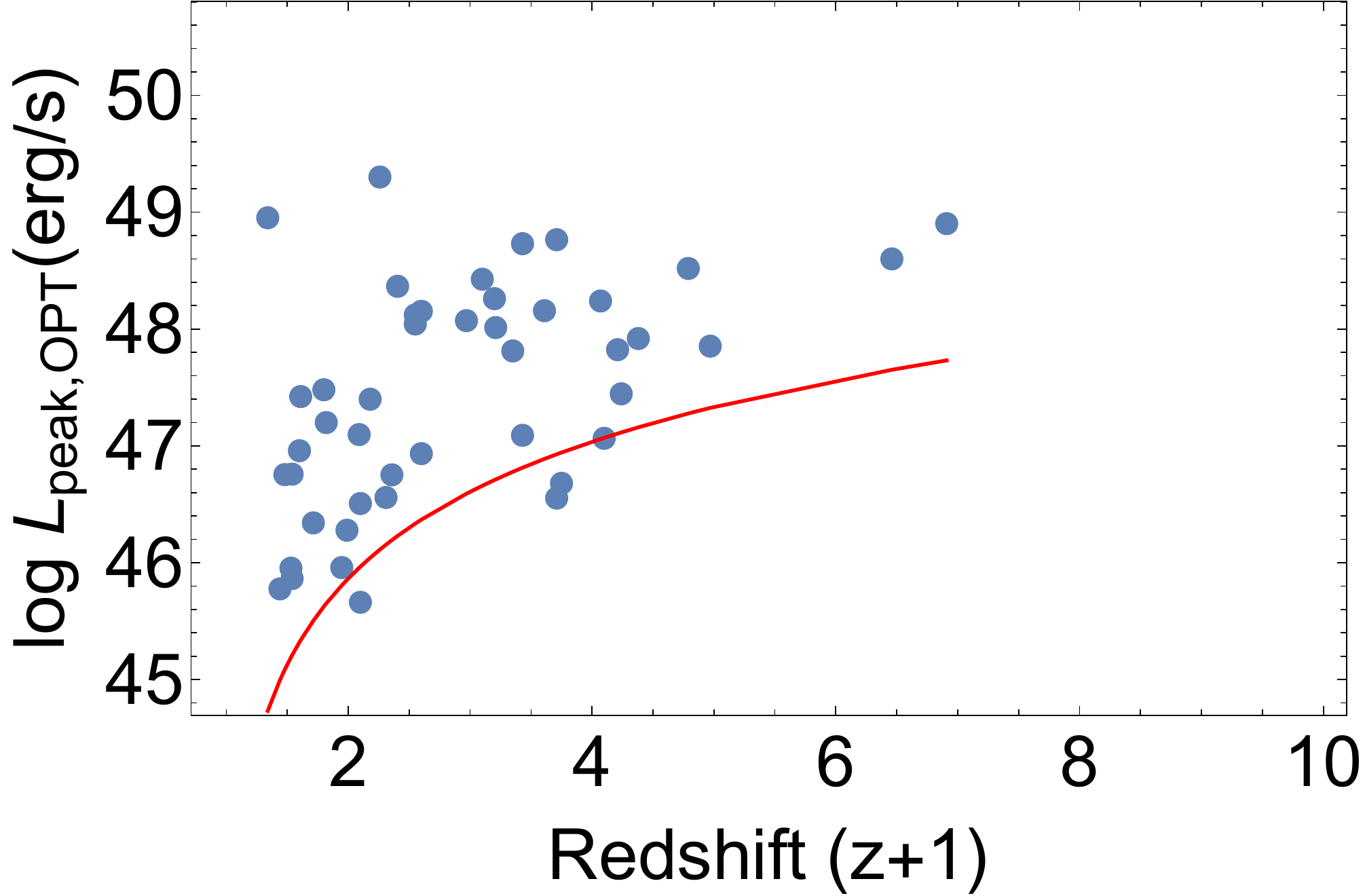}\label{fig:EP opt lpeak} &
        \includegraphics[width=0.50\textwidth]{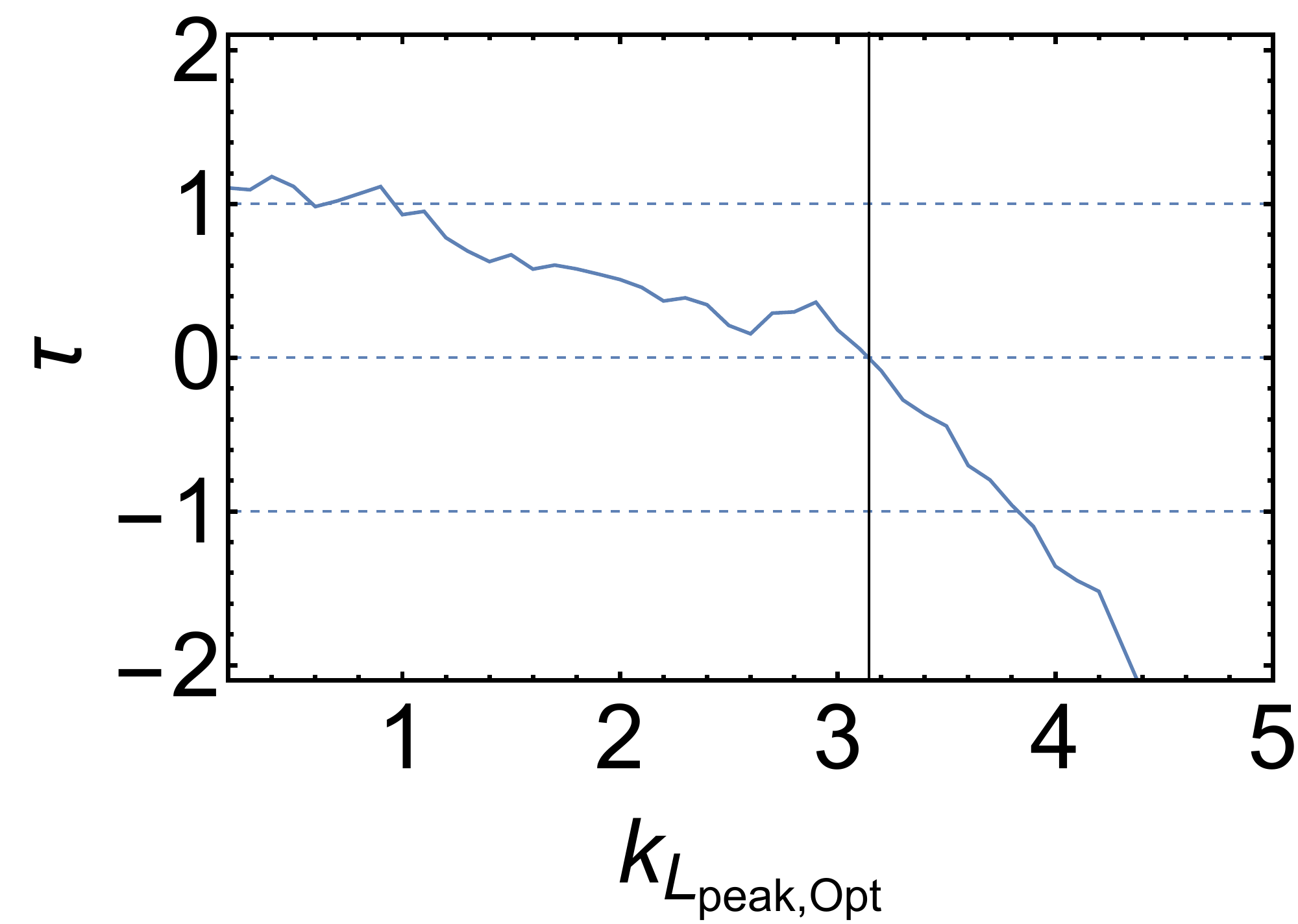}\label{fig: EP OPT lpeak k} \\
    \end{tabular}
\caption{This Figure shows how the limiting luminosities and the Kendall $\tau$ vs the slope of the evolutionary functions for the full optical GRB sample. The panels show the evolution of $L_{\text{a,OPT}}$, $T_{\text{a,OPT}}^{*}$, and $L_{\text{peak,OPT}}$ vs. redshift. The limiting line is plotted in red. The right panels show the evolution of $\tau$. The middle dashed line is $\tau=0$ and the dashed lines are the defined bounds of $+1\sigma$ and $-1\sigma$, while the red line corresponds to the best-fit value of $\tau$.}
\label{fig:EP opt}
\end{figure*}

In this section, we perform the same analysis on all the combinations of samples studied in the previous sections, although now we account for redshift evolutionary effects. Because we deal with astronomical objects observed at large distances ($z \geq 0.033$), and because there also exists a dependence between GRB luminosity and redshift, we are aware of the data truncation due to the Malmquist bias and the Eddington effect \citep{1913MNRAS..73..359E, 1922MeLuF.100....1M}. To correct for these, we employ techniques as described by the \cite{1992ApJ...399..345E} methodology on the full sample of 222 GRBs presenting X-ray plateaus. The EP method allows us to overcome the problems of redshift evolution and selection biases by introducing a modification of the Kendall rank correlation coefficient. This coefficient can be written in the following way:
\begin{equation}
    \tilde{\tau} = \frac{n_{\text{c}}-n_{\text{d}}}{\frac{1}{2}n(n-1)}
\end{equation}
where $n_{\text{c}}$ is the number concordant and $n_{\text{d}}$ is the number discordant. The EP test statistic is a similar non-parametric test, but the main improvement it makes upon Kendall's $\tilde{\tau}$ \citep{10.1093/biomet/30.1-2.81} is that it can work with both one-sided or doubly truncated data. It is defined as follows:
\begin{equation}
    \tau = \frac{\sum_{i}(R_{i}-E_{i})}{\sqrt{\sum_{i}V_{i}}}
\end{equation}
where $E$ is the expectation value, $V$ is the variance, and $R$ is rank. In order to use the EP method, we need to define the associated sets for which we will calculate the $\tau$ values. On the other hand, the associated sets are defined given a limiting value of a particular distribution. In our case we have a 3-variate distribution among $L_{\text{peak}}$, $T^{*}_\text{a}$ and $L_\text{a}$. Thus, we define limiting fluxes for $L_{\text{peak,X}}$ and $L_{\text{a,X}}$ in X-ray as $1.54\cdot10^{-8}$ and $1.5\cdot10^{-12}$ ergs s$^{-1}$, respectively, and for $L_{\text{peak, OPT}}$ and $L_{\text{a,OPT}}$ in the optical as $1.40\cdot10^{-12}$ and $1.50\cdot10^{-14}$ ergs s$^{-1}$, respectively. These limits have been chosen in a conservative way so that no more than the $10\%$ of the total sample for each physical parameter is cut via the EP method. This allows us to compute the uncertainty of the evolutionary effects without affecting the statistical significance of the samples. Simulations in \cite{2013ApJ...774..157D} have shown the reliability of this method. The distribution of the limiting fluxes in X-ray and optical are shown in the two uppermost plots of Figs. \ref{fig:EP xray} and \ref{fig:EP opt}, respectively; the limiting times in X-rays and optical are shown in the lower plots of the same figures.
Once we correct for redshift evolution, we define new variables for GRB luminosities and times. These are the de-evolved variables, indicated with $'$. We define $L^{'}=L_X/(1+z)^{g(z)}$, where $g(z)=(1+z)^{k}$ is the function that mimics the redshift evolution for the X-ray luminosity at the end of the plateau emission. The same procedure has been applied for the peak luminosity in the prompt emission and for the time at the end of the prompt emission. The value of the exponent $k$ is determined by the EP method (right panels of Figs. \ref{fig:EP xray} for X-rays and \ref{fig:EP opt} for optical).
With the new evolutionary functions computed, we can then write the corrected by redshift evolution and selection bias fundamental plane relation:
\begin{equation}
    \log_{10}(L_{\text{a,theory}}) = c+a \times (\log_{10}(T_{\text{a}}^{*})-\log_{10}((1+z)^{k_{T_{\text{a}}^{*}}}))+b \times (\log_{10}(L_{\text{peak}})-\log_{10}((1+z)^{k_{L_{\text{peak}}}}))+\log_{10}((1+z)^{k_{L_{\text{a}}}})
\end{equation}
so that $L_{\text{a}}^{'}$, $L_{\text{peak}}^{'}$, and $T_{\text{a}}^{*'}$ become independent of redshift. Here, the $k$-corrections derived by the EP methodology for both X-ray and optical samples are defined in Table \ref{tab:kcorrs}.


The fundamental plane fitting for the full OPT GRB sample, including evolution correction, produces the following parameters:
 $\aoptfullev = -0.74 \pm 0.11$, $\boptfullev = 0.22 \pm 0.08$, $\coptfullev = 37.52 \pm 3.78$ and $\sigmaintoptfullev = 0.41 \pm 0.06$.
When we consider the optical trimmed sample with evolution the parameters are the following:
$\aopttrimev=-0.77\pm 0.15$, $\bopttrimev=0.22 \pm 0.15$, $\copttrimev= 37.61 \pm  2.97$, $\sigmaintopttrimev=0.13 \pm 0.12$. 
Furthermore, we compute the plane fitting for the full X-ray GRB sample with evolution corrections and find the following: 
$\axfullev = -0.85 \pm 0.12$, $\bxfullev = 0.48 \pm 0.12$, $\cxfullev=25.64 \pm 6.55$ and $\sigmaintxfullev = 0.20 \pm 0.06$.
The X-rays trimmed and corrected with evolution has the following parameters:
$\axtrimev = -0.79 \pm 0.17$, $\bxtrimev = 0.49 \pm 0.19$, $\cxtrimev =25.29 \pm 9.87$ and $\sigmaintxtrimev = 0.13 \pm 0.09$.
The percentage decrease regarding $\sigmaintxtrimev$ for the OPT sample considering the evolutionary effects is $68\%$, while for the X-ray one is $35\%$.
It is interesting to note that the coefficient of the X-ray planes and the scatter results shown here, for which the most updated sample has been used, are within the $\sigmaint= 0.22 \pm 0.10$ presented in \cite{2020ApJ...904...97D}.

\begin{figure*}
\centering
    \begin{tabular}{cc}
        \includegraphics[width=0.45\textwidth]{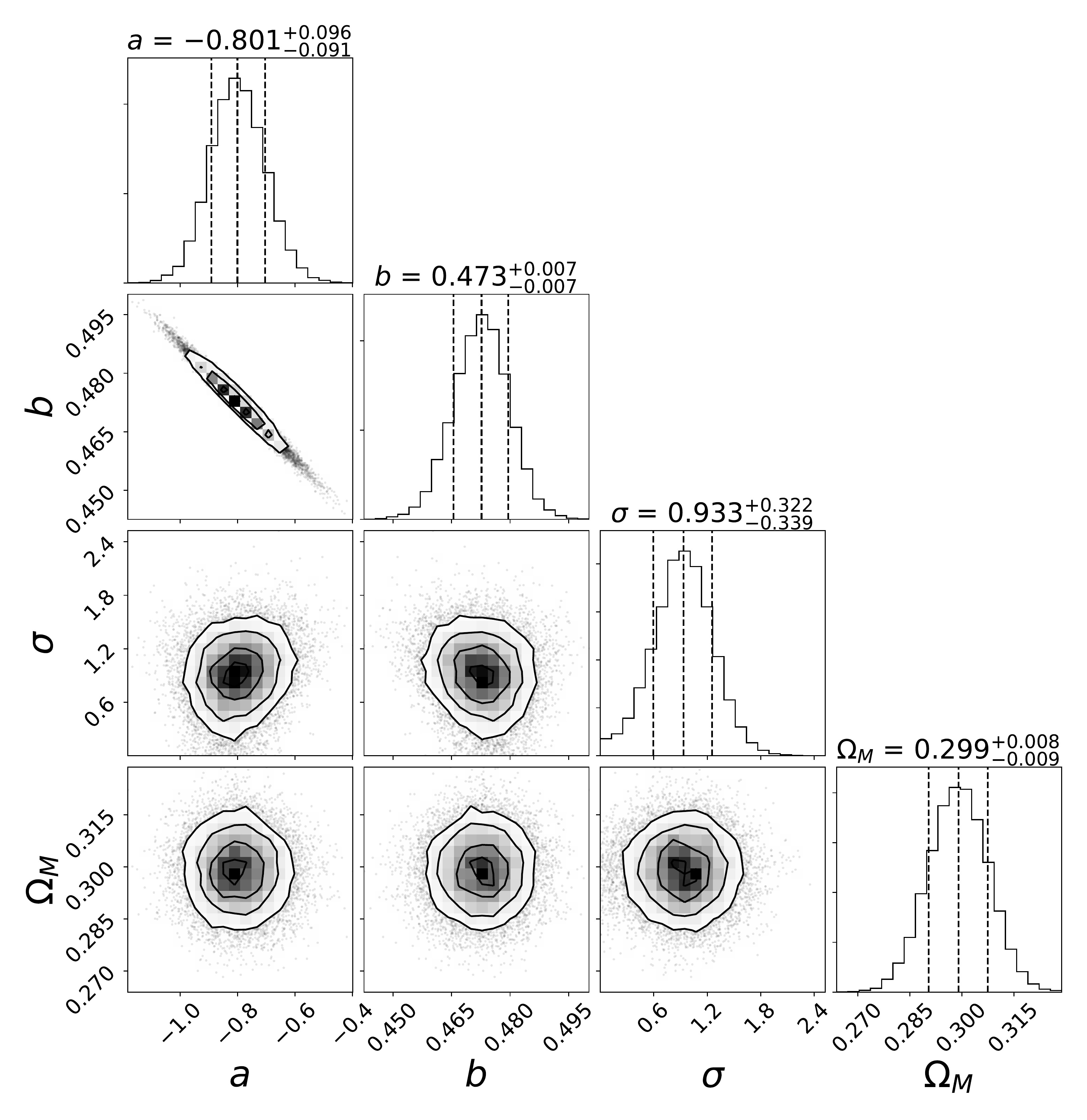} &
        \includegraphics[width=0.45\textwidth]{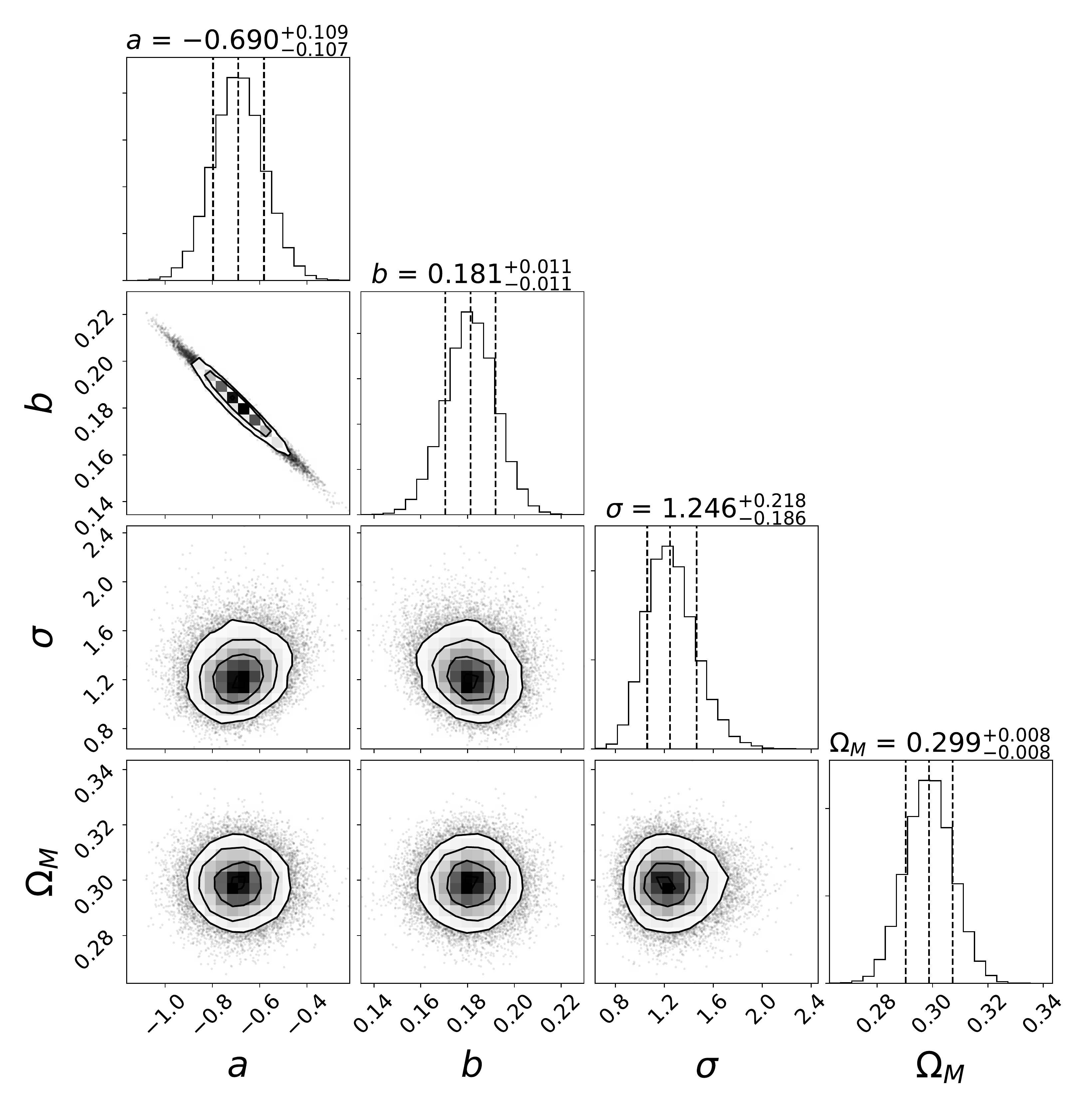} \\
        \includegraphics[width=0.45\textwidth]{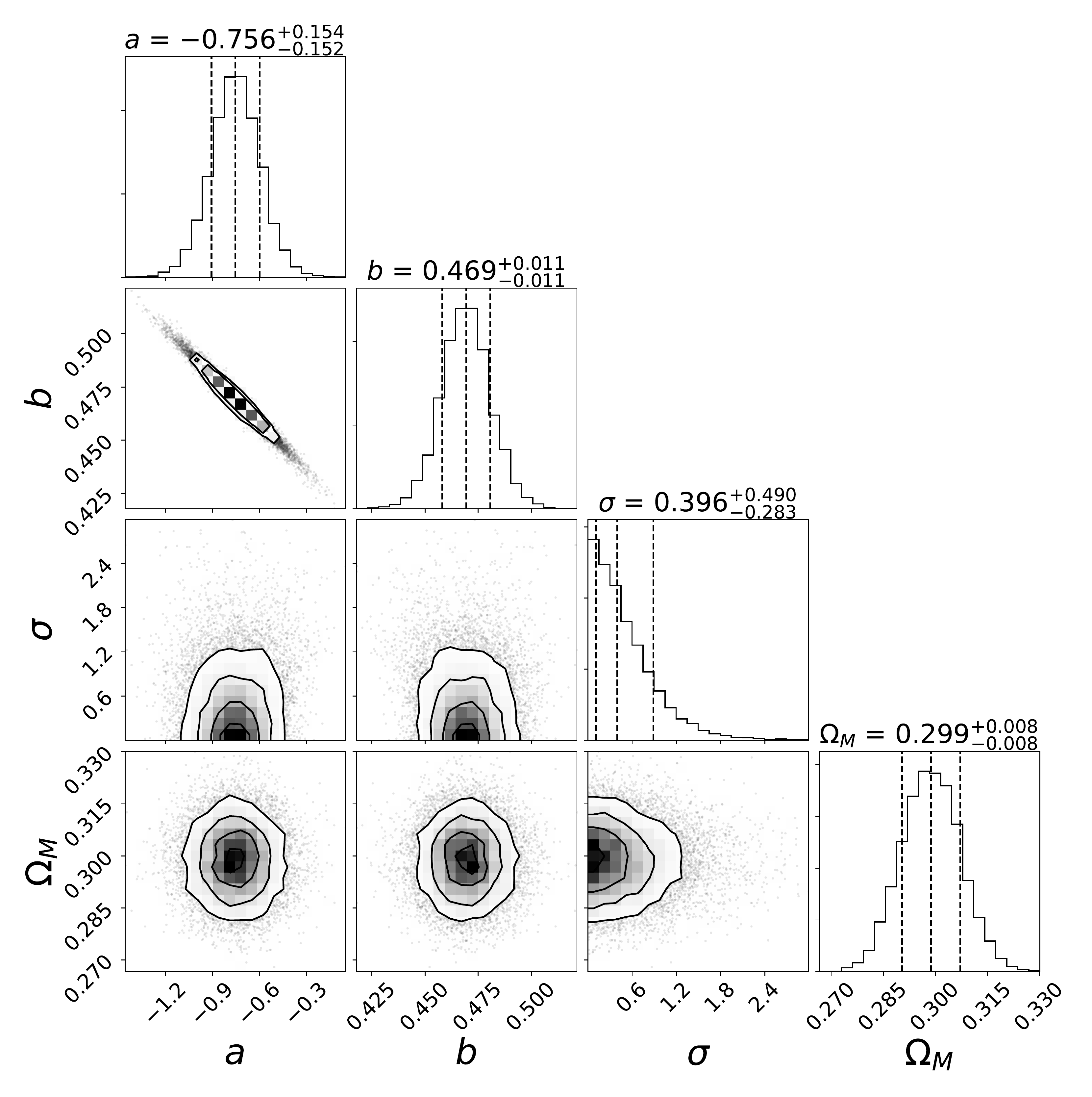} &
        \includegraphics[width=0.45\textwidth]{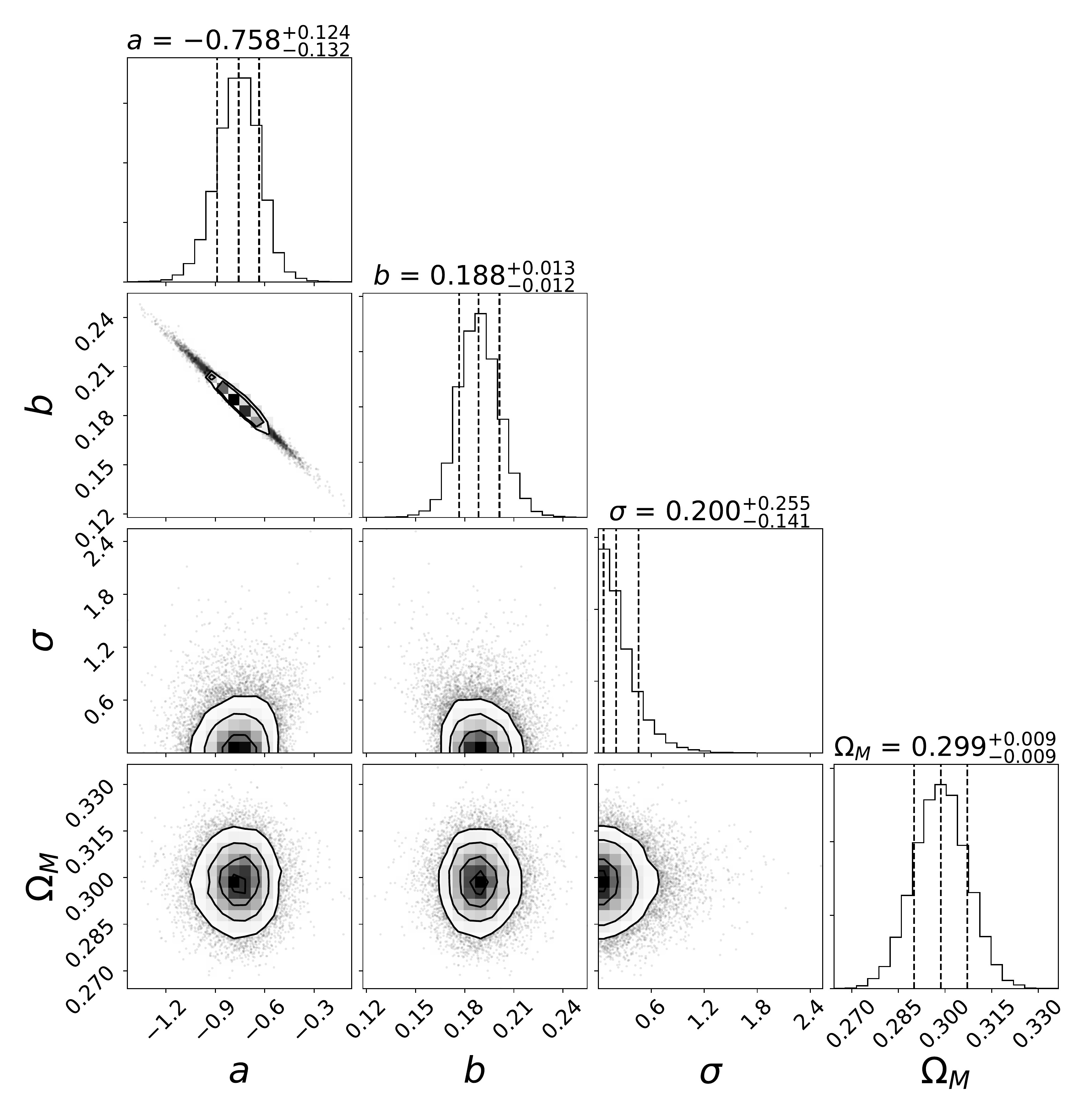} \\
    \end{tabular}
\caption{The upper panels show the fundamental plane fitting for the full GRB PLAT sample (upper left) and the OPT sample  (upper right) with evolution performed contemporaneously with SNe Ia to  derive the $\Omega_{\text{M}}$ values. The bottom panels show the same derivation of $\Omega_{\text{M}}$ with both the PLAT and optical trimmed samples (left and right panel, respectively). Each plot shows the 2D posterior contours as well as the 1D histograms for each parameter.}
\label{fig5 optvopttrim-with-evo}
\end{figure*}

All evolution-corrected results are displayed in Table \ref{tab:ev} for comparison and some of the results for the full platinum and the optical samples and for their respective trimmed samples are shown in Fig. \ref{fig5 optvopttrim-with-evo} . We note that they are very similar to the ones without correction; this could be due to the fact that the results with evolution carry a larger uncertainty on the variables, and this may lead to comparable results. More GRB data must be gathered so that the corrections for selection biases and redshift evolution carry less uncertainty. Thus, in the following sections when we consider the simulated data, we limit ourselves to the non-evolutionary cases.

\begin{table*}
\caption{The table shows the coefficient of evolutionary functions for the luminosity at the end of the plateau emission, $L_{\text{a}}$, its correspondent rest frame, $T^{*}_\text{a}$ and the peak prompt luminosity, $L_{\text{peak}}$ for both X-ray (first row) and optical (second row). \label{tab:kcorrs}}
\begin{tabular}{|c|c|c|c|}
\hline
Sample & $kL_{\text{a}}$ & $kT_{\text{a}}^{*}$  & $kL_{\text{peak}}$  \\\hline

X-ray & $2.42 \pm 0.58$ & $-1.25 \pm 0.28$ & $2.24 \pm 0.30$\\
Optical & $3.96 \pm 0.43$ & $-2.11 \pm 0.49$ & $3.10 \pm 1.60$\\\hline
\end{tabular}
\end{table*}


\begin{table*}
\centering
\caption{The first column shows the evolution-corrected sample, while the second shows the correspondent values of $\Omega_{\text{M}}$. The errors reported here are variances, corresponding to the $68\%$ confidence limit. \label{tab:ev}}
\begin{tabular}{|c|c}
\hline
Sample & $\Omega_{\text{M}}$  \\\hline
PLAT+SNe Ia (EV)& $0.299 \pm 0.009$ \\
PLATtrim+SNe Ia (EV)& $0.299\pm 0.008$\\
OPT+SNe Ia (EV) & $0.299 \pm 0.008$\\
OPTtrim+SNe Ia (EV) & $0.299 \pm 0.009$\\
\hline
\end{tabular}
\end{table*}

\section{Simulating GRBs from the Full Sample Fundamental Planes} \label{sec:simulations}

\begin{figure*}
\centering
    \begin{tabular}{cc}
        \includegraphics[width=0.45\textwidth]{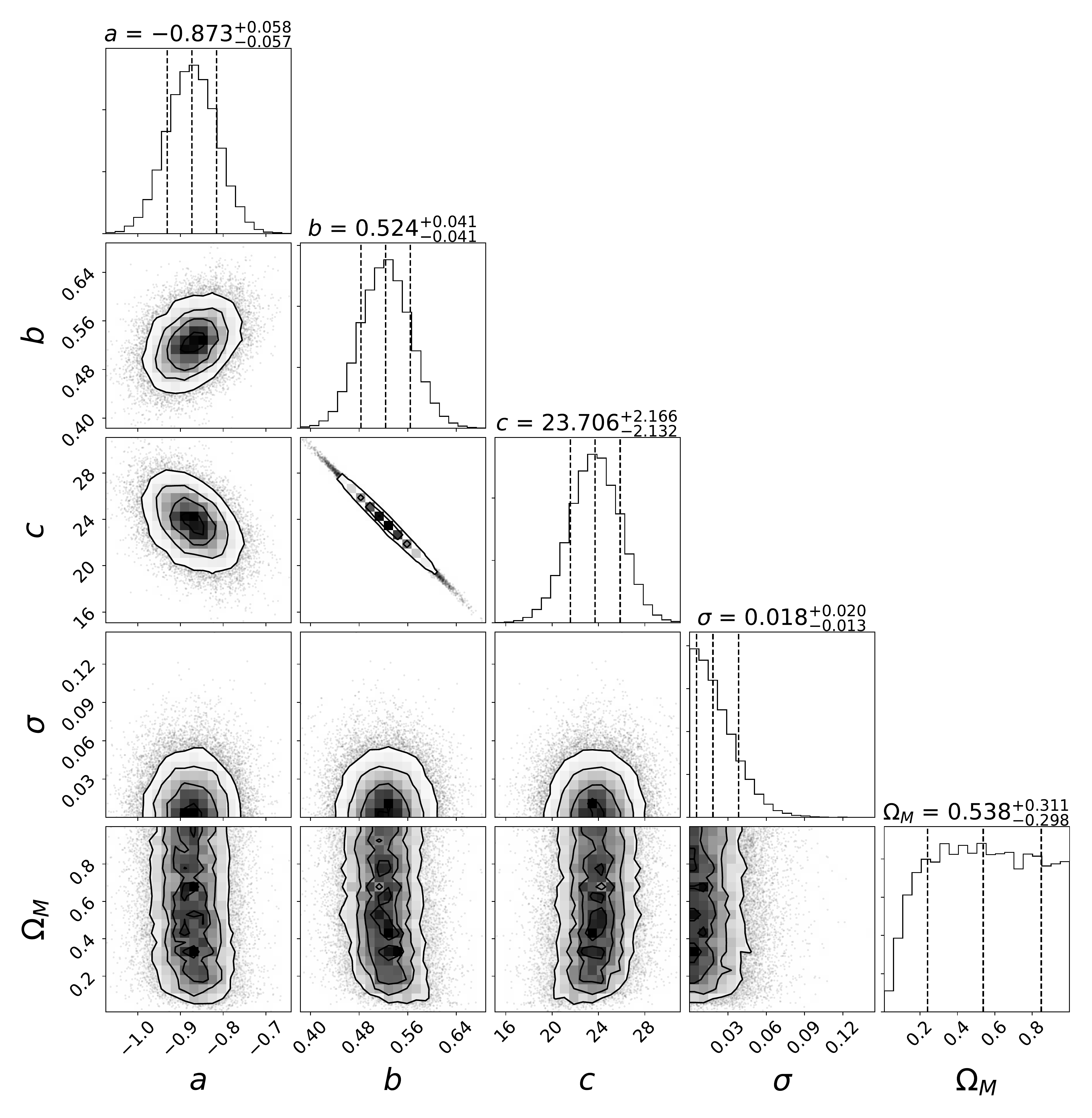}\label{fig:PLATxN=50} &
        \includegraphics[width=0.45\textwidth]{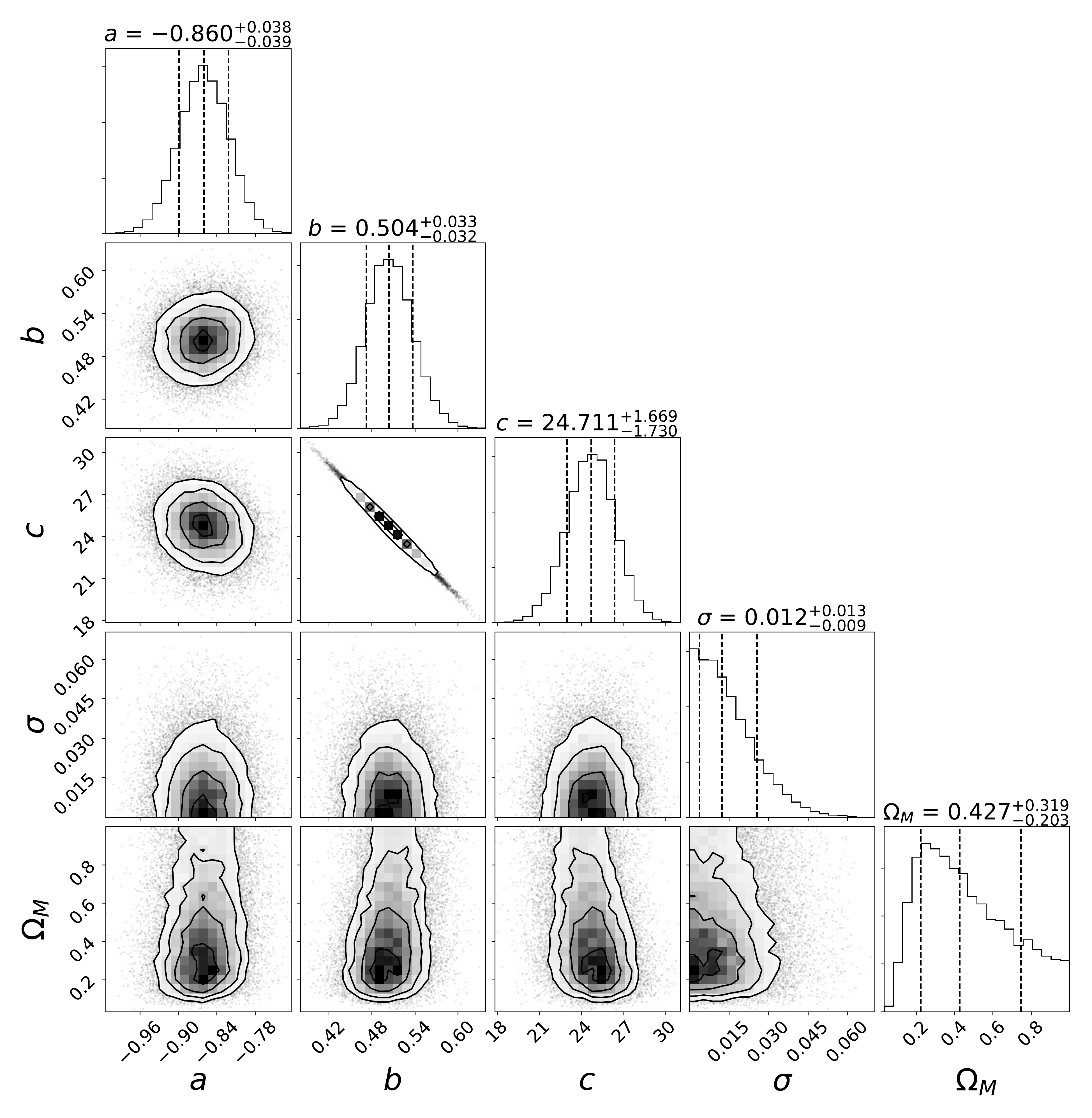}\label{fig:PLATxN=100} \\
        \includegraphics[width=0.45\textwidth]{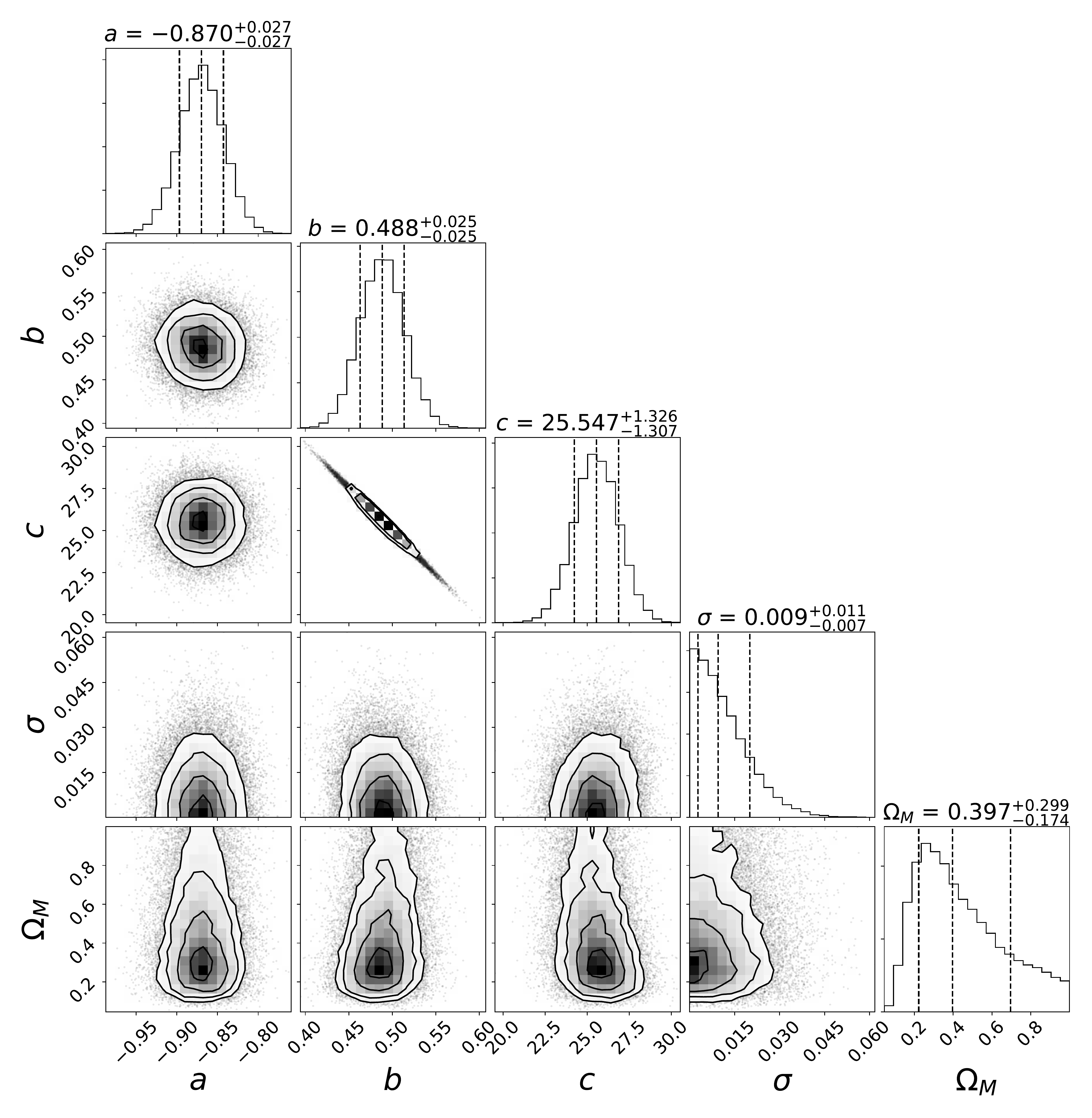}\label{fig:PLATxN=150} &
        \includegraphics[width=0.45\textwidth]{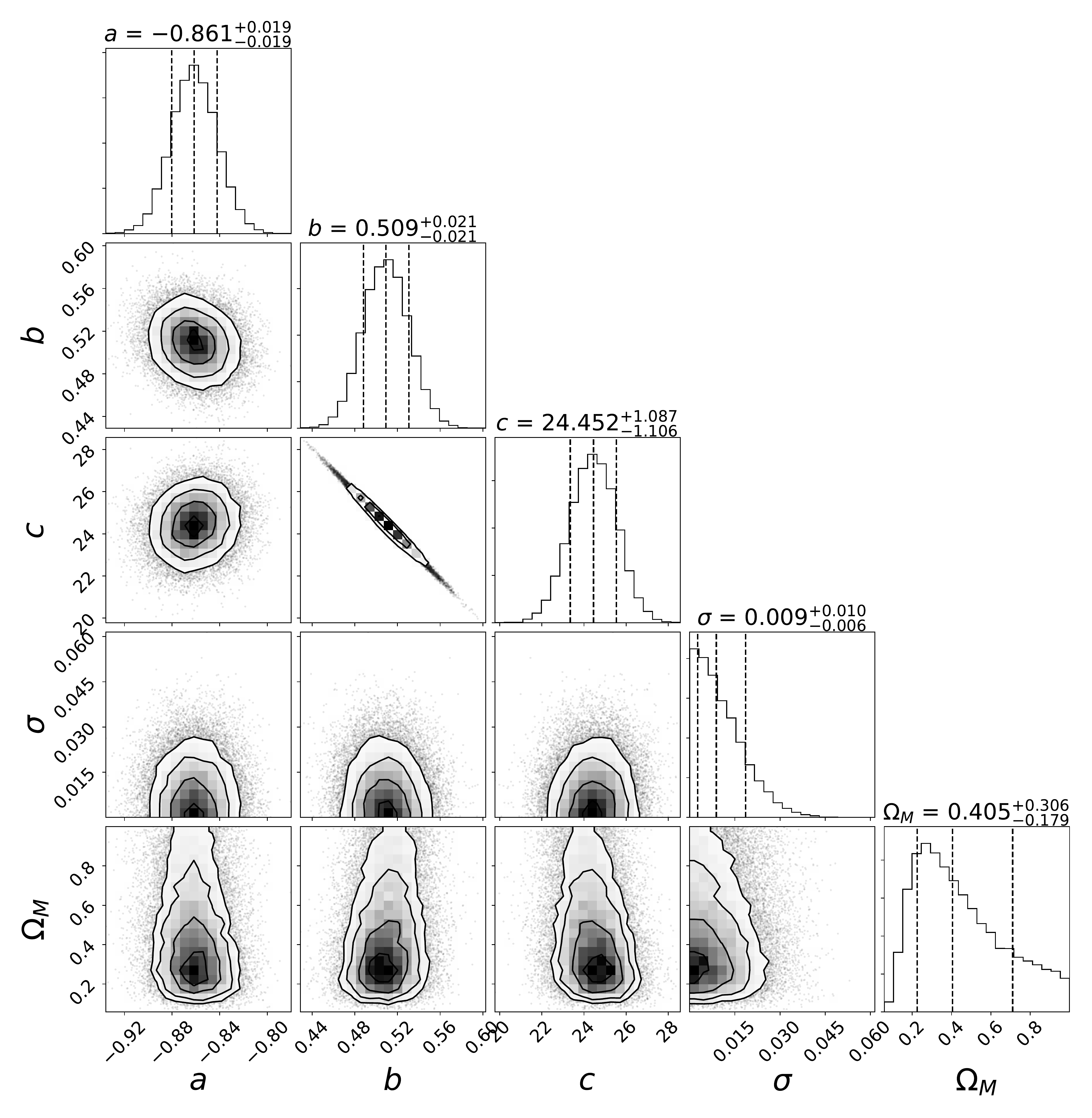}\label{fig:PLATxN=200} \\
    \end{tabular}
\caption{This figure shows the progression of closing contours around an $\Omega_{\text{M}}$ value for 50 (top left), 100 (top right), 150 (bottom left), and 200 (bottom right) GRBs simulated off the PLAT fundamental plane.}
\label{fig:PLATxN Runs}
\end{figure*}

We now use the 3D fundamental planes in X-ray as defined by the 50 GRBs of the PLAT sample, and in optical (all 45 GRBs) as a base for simulating GRB events. We first perform these simulations to compute the number of GRBs in X-rays needed to achieve closed-contours around the mean value computed for $\Omega_{\text{M}}$, without requiring any upper limit on the error. We begin by simulating Gaussian distributions resembling the observed $T_{\text{a,X}}^{*}$, $L_{\text{a,X}}$, and $L_{\text{peak,X}}$, and the $K$-correction distributions from all GRBs belonging to the PLAT X-ray fundamental plane. The errors on all of these variables have also been simulated in this way. We run this data through python's MCMC sampler \emph{emcee}~\citep{2013PASP..125..306F} to randomly simulate different numbers of GRBs. The \emph{emcee} sampler has been chosen for simulation production due to its ease of parallelization over \emph{cobaya}. We find that only 150 GRBs are needed to provide a reasonable $\Omega_{\text{M}}$ value of $0.387 \pm 0.473$, as shown in bottom left of Fig. \ref{fig:PLATxN Runs}. However, the error bars on this value are undesirable; much greater accuracy has been achieved by using SNe Ia as sole probes. We define three desired error limits as those determined by SNe Ia data as a standalone standard candle; \cite{2011ApJS..192....1C} determined a symmetrized error of $\sigma = 0.10$ from 472 SNe Ia, \cite{2014A&A...568A..22B} obtained a standard deviation of $\sigma = 0.042$ from 740 SNe Ia, and \cite{2018ApJ...859..101S} obtained a standard deviation of $\sigma = 0.022$ from 1048 SNe Ia. The goal with simulating differently trimmed data is to reach a value for the error that is less than or equal to these error limits using GRBs as a standalone probe. We then use the number of GRBs needed to achieve this to infer the number of years needed, given present and future deep-space survey missions, to make this constraint possible. We start by choosing the errors found by \cite{2011ApJS..192....1C}, because it has a sample size of SNe Ia more comparable to our sample size of 222 GRBs than other, more recent studies. Thus, we explore methods of reaching this error limit first by increasing the number of simulated GRBs off of the full PLAT sample, and then in the following subsections, trimming our PLAT sample to reach SNe Ia accuracies and define the smallest error on $\Omega_{\text{M}}$ yet.

\begin{figure*}
  \centering
  \subfloat[X-RAY | Simulation Results for the Full PLAT Base with Undivided Errors ]
  {\includegraphics[width=8.5cm]{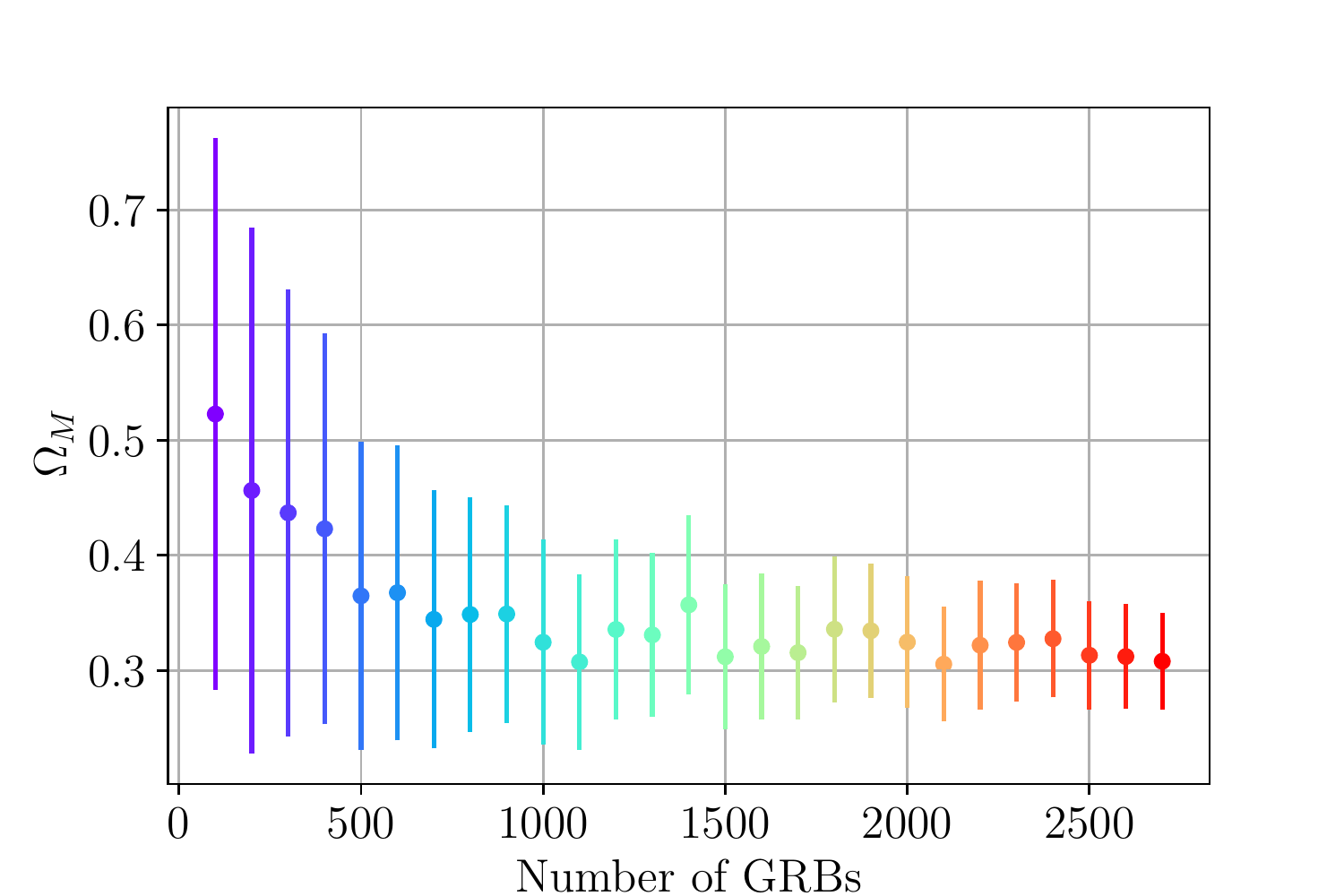}\label{fig:FULL PLAT noev n=1}}
  \hspace{0.4cm}
  \subfloat[X-RAY | Simulation Results for the Full PLAT Base with Halved Errors]
  {\includegraphics[width=8.5cm]{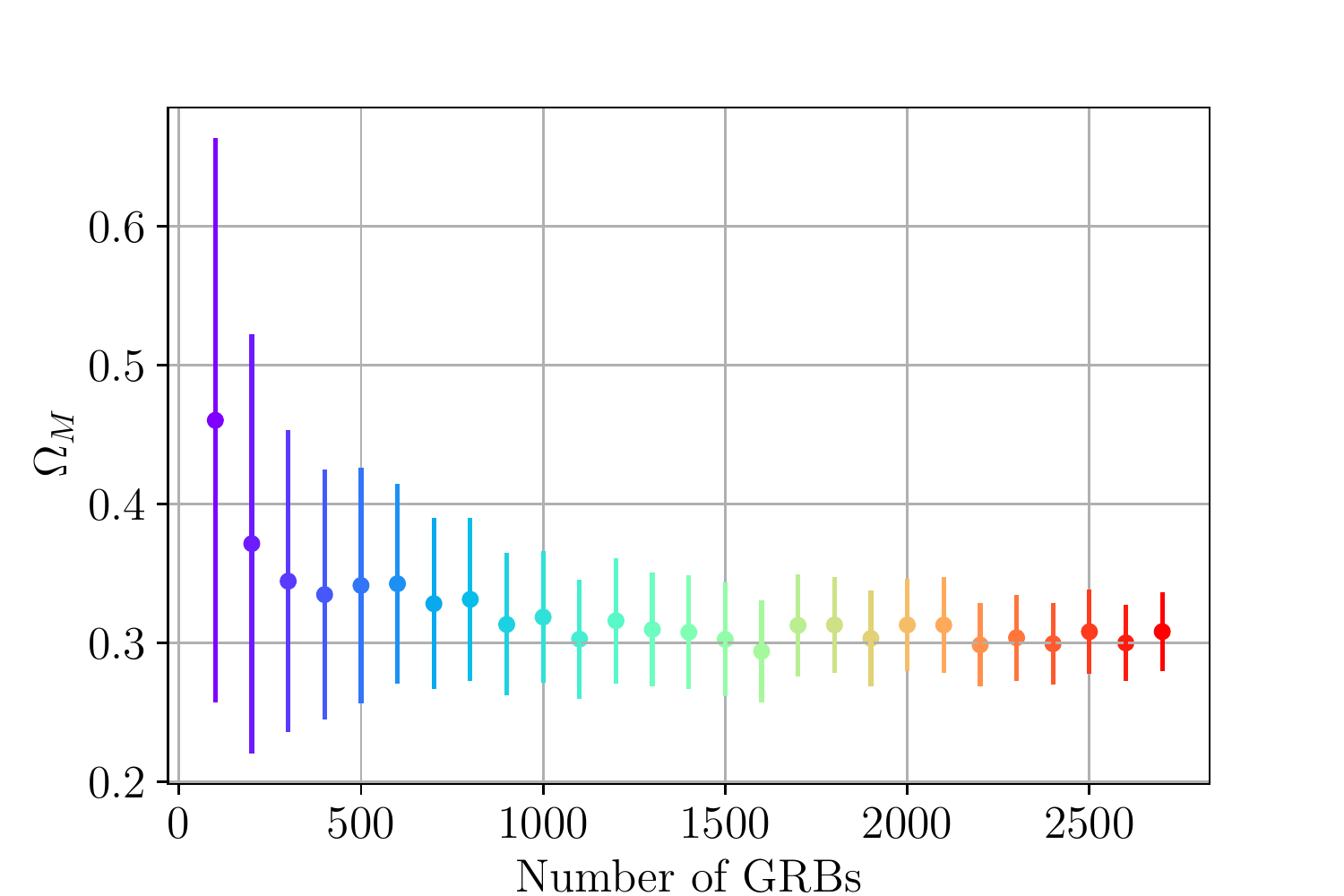}\label{fig:FULL PLAT noev n=2}}
   \hspace{0.4cm}
  \subfloat[X-RAY | Probability Map for Undivided Errors]
  {\includegraphics[width=8.5cm]{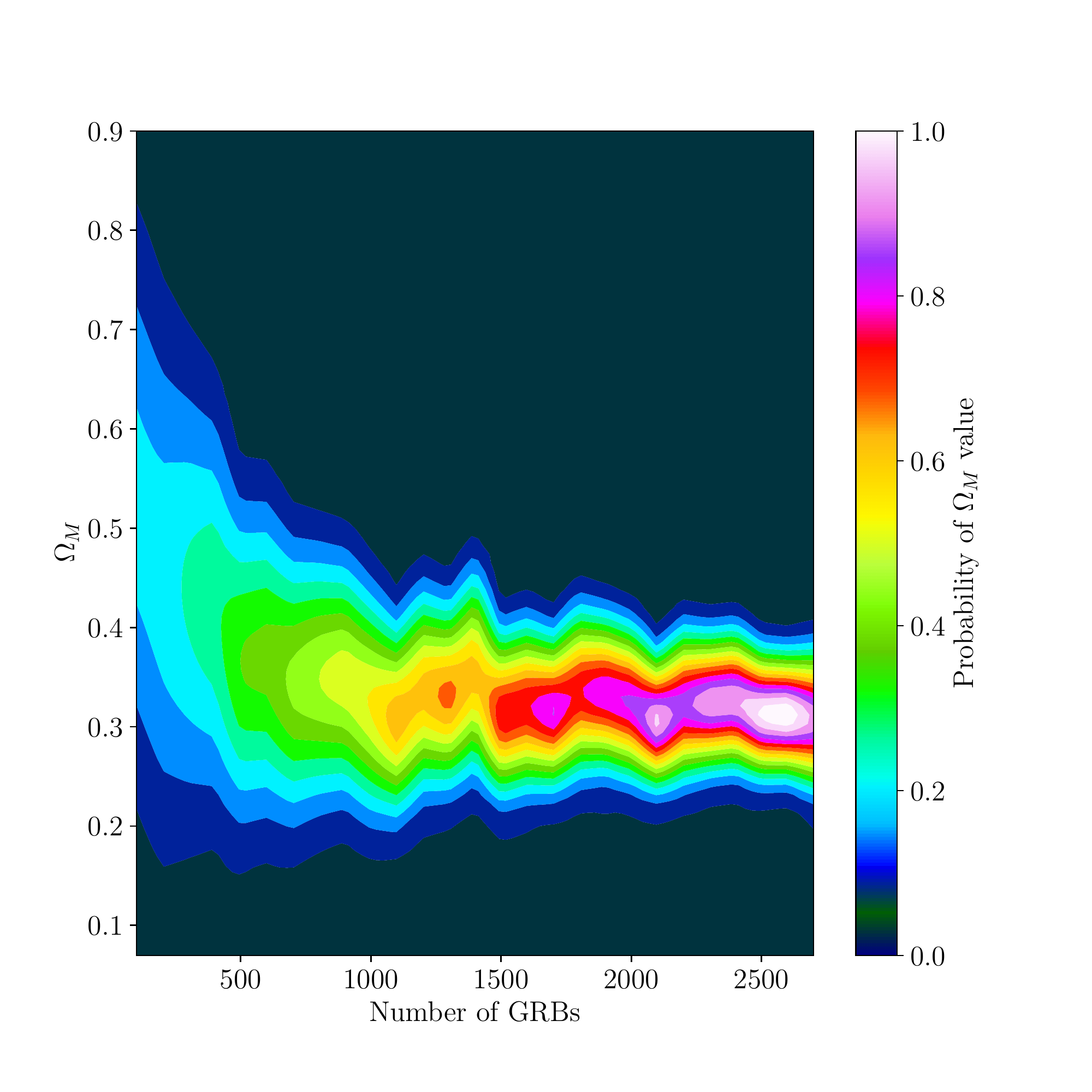}\label{fig:FULL PLAT probmap noev n=1}}
  \hspace{0.4cm}
  \subfloat[X-RAY | Probability Map for Halved Errors]
  {\includegraphics[width=8.5cm]{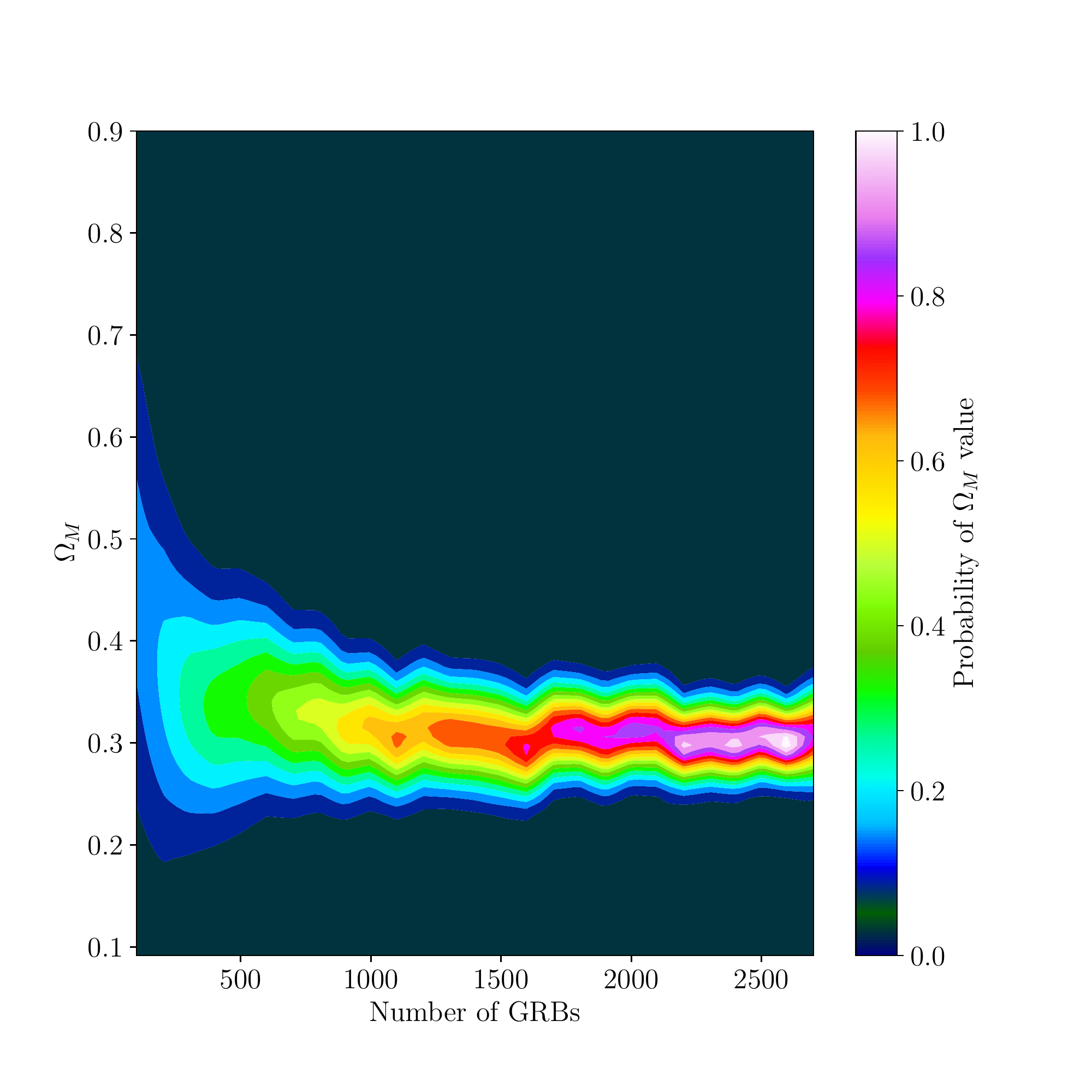}\label{fig:FULL PLAT probmap noev n=2}}
  \caption{Upper left panel: the mean values of $\Omega_{\text{M}}$ vs. the numbers of GRBs obeying the X-ray fundamental plane to converge upon a value of $\Omega_{\text{M}}$ using GRBs as the standalone probe by considering the observed error bars. Upper right panel: the same as the left panel, but considering the error bars divided by 2. The bottom left and right panels show the corresponding probability distributions of the upper left and right panels, respectively.}
  \label{fig:FULL PLAT conv and probs}
\end{figure*}

Using all 50 PLAT GRBs in X-ray as the simulation base first, we test for a large range of simulated GRBs. We ran multiple simulations to see what it is the optimal number of GRBs simulated to increase the precision on our cosmological computation until the desired error limit is reached. We note the convergence of the $\Omega_{\text{M}}$ parameter to a value about $0.3$ (see Fig. \ref{fig:FULL PLAT noev n=1}), as expected. However, we also change the errors on the simulated $L_{\text{a,X}}$, $T_{\text{a,X}}^{*}$, and $L_{\text{peak,X}}$ to test simulated samples of varying quality. The first simulations were run by considering the original errors, and then halving those, resulting in increasingly better quality samples. As predicted, the samples simulated from the GRBs with halved error bars show a convergence to an $\Omega_{\text{M}}$ value with smaller standard deviations, see Fig. \ref{fig:FULL PLAT noev n=2}, than those with undivided errors, as seen in Fig. \ref{fig:FULL PLAT noev n=1}. In comparison to the previously stated $\Omega_{\text{M}}$ value for a sample of 150 simulated n = 1 GRBs, for the same-sized sample of 150 simulated n = 2 GRBs, we now achieve $\Omega_{\text{M}} = 0.416 \pm 0.177$. This represents a near 63\% decrease on the error in $\Omega_{\text{M}}$.

\begin{figure*}
  {\includegraphics[width=8.6cm]{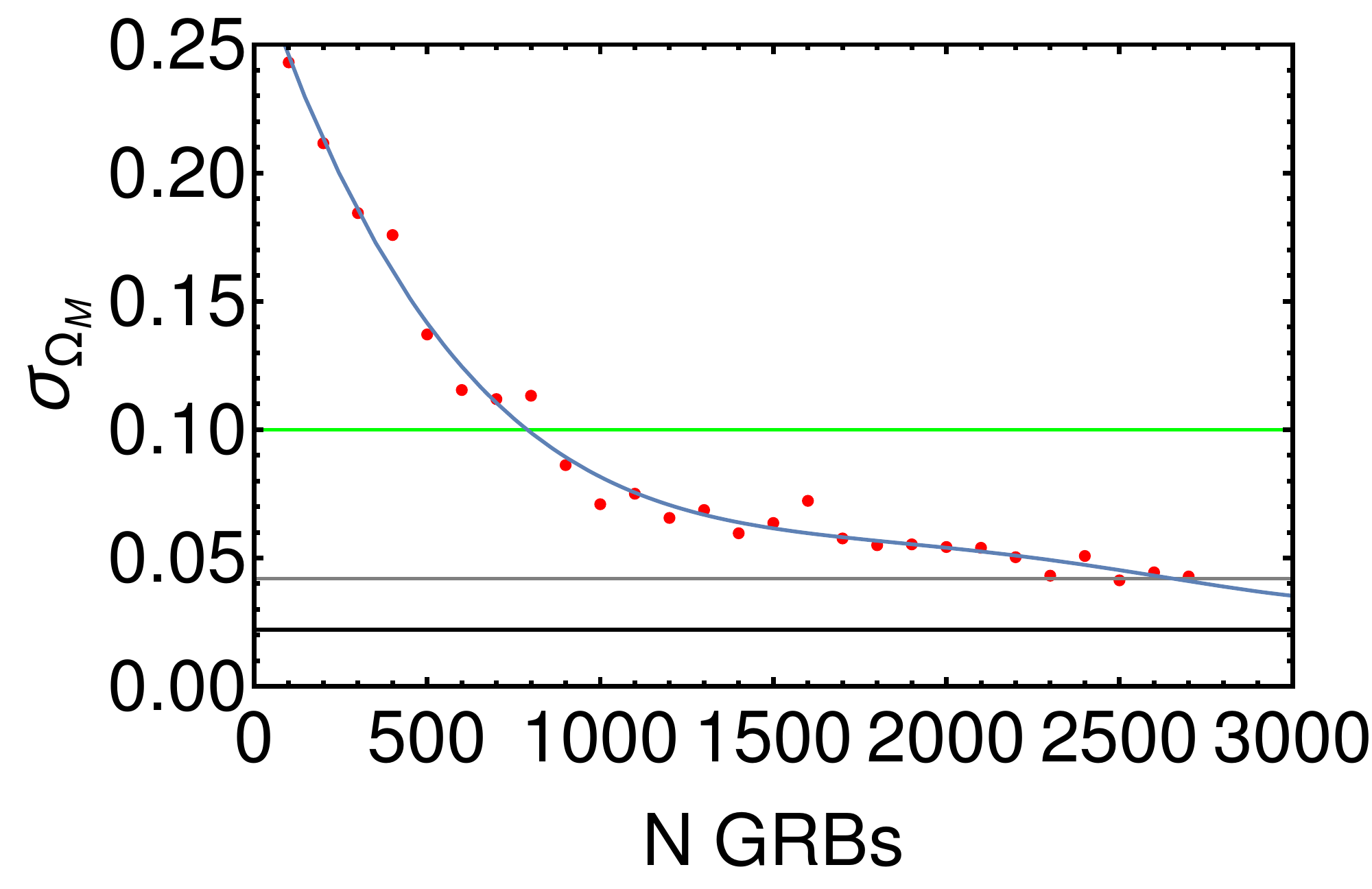}
  }
  \hspace{0.4cm}
    {\includegraphics[width=8.6cm]{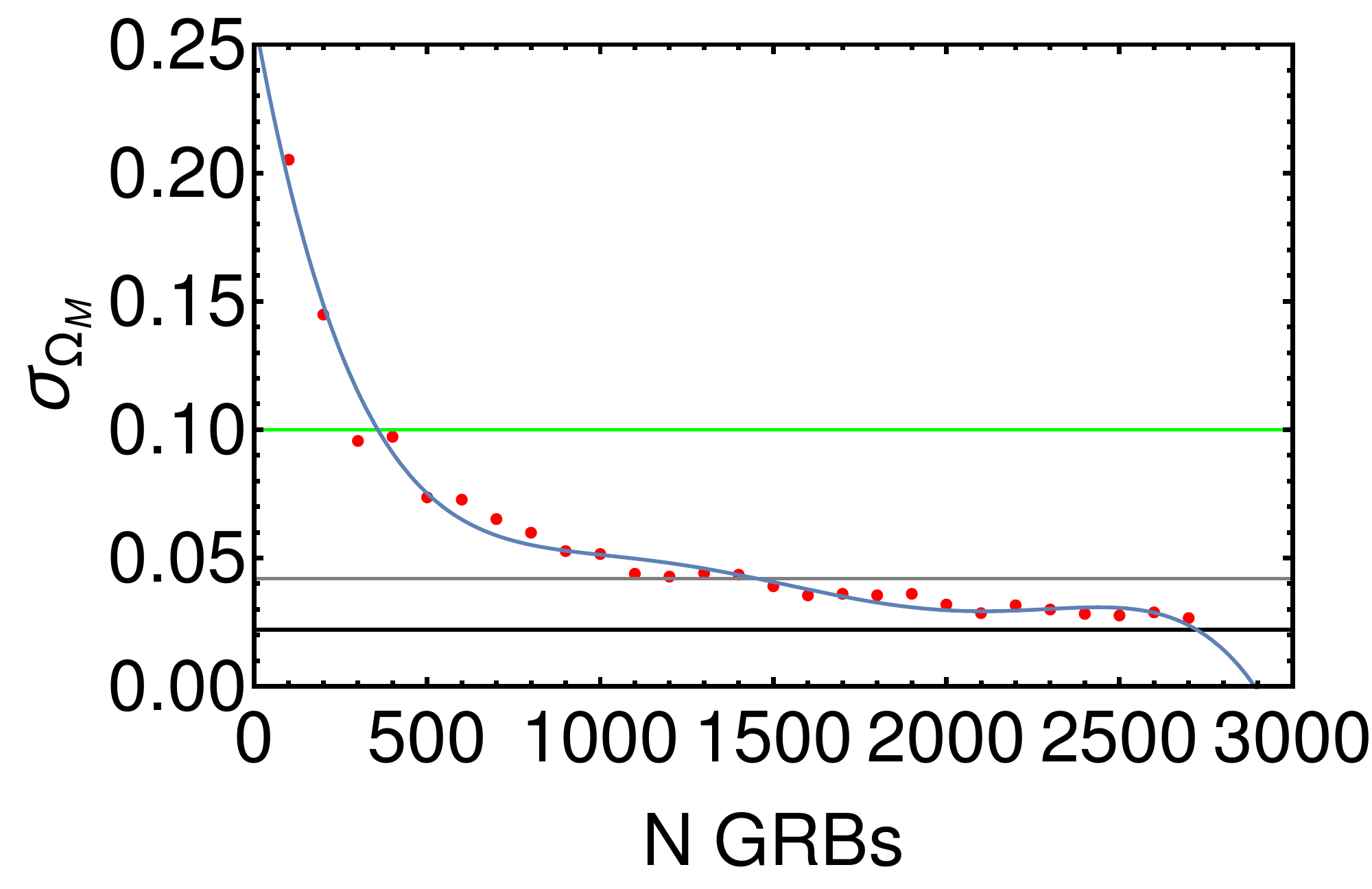}
    }
\caption{Left and right panels show the error plots for undivided and halved error bars, respectively, for the PLAT sample. The green, grey and black lines identify the \citet{2011ApJS..192....1C,2014A&A...568A..22B} and \citet{2018ApJ...859..101S} errors on $\Omega_{\text{M}}$, respectively. The blue line in the right panel denotes a polynomial fitting function used for the extrapolation.    \label{fig:FULL_PLAT_error_limits}}
\end{figure*}

We constructed a probability map on the value of $\Omega_{\text{M}}$ as computed by the simulations. This was created by taking the Monte Carlo chains and computing the probability density function (PDF) on each simulation, and then converting this density to a probability. The PDFs were then linearly interpolated in the three-dimensional space of the number of GRBs, $\Omega_{\text{M}}$, and the probability density from each simulation run to create a probability map. As is evident from Fig. \ref{fig:FULL PLAT probmap noev n=1}, we see no highly probabilistic closed contours for a number of GRBs less than 2100. However, if we are to focus our attention only on achieving the desired precision of \cite{2011ApJS..192....1C} of $\sigma = 0.10$, we reach this goal considering $789$ GRBs. This error limit is represented by the bright green line in the left panel of Fig. \ref{fig:FULL_PLAT_error_limits}.
Thus, we can focus on the plots that show the error on $\Omega_{\text{M}}$ versus the correspondent number of simulated GRBs rather than $\Omega_{\text{M}}$ itself.
$789$ GRBs are needed to be able to use them as standalone standard candles when the errors that enter the likelihood remain undivided. Further, the grey line in the left panel of Fig. \ref{fig:FULL_PLAT_error_limits} shows that the \cite{2014A&A...568A..22B} limit is reached for a minimum of 2653 simulated GRBs. The \cite{2018ApJ...859..101S} limit shown by the black line in Fig. \ref{fig:FULL_PLAT_error_limits} is not reached if we limit to a maximum number of 3000 GRBs.

Considering now the predictions if the errors are divided by two (Fig. \ref{fig:FULL PLAT noev n=2}), we build a new probability map and observe the minimum width in the distribution, corresponding to the probability peak, beginning at around 2100 GRBs (Fig. \ref{fig:FULL PLAT probmap noev n=2}).
We here stress that the above mentioned width of the distribution is correlated with the normalized standard deviation we see in the upper panels of Fig. \ref{fig:FULL PLAT conv and probs}. It should be noted that PDFs for both maps in the lower two panels of Fig. \ref{fig:FULL PLAT conv and probs} are normalized with respect only to the simulations used in each respective map.
What is important in the comparison between the two maps is the spread of the PDF for the number of GRBs that give the most probabilistic value for $\Omega_{\text{M}}$. For the map of Fig. \ref{fig:FULL PLAT probmap noev n=1} (n = 1), the standard deviation on the normalized PDF of $\Omega_{\text{M}}$ is calculated as $\sigma_{\text{pdf}} = 0.037$, whereas the map of Fig. \ref{fig:FULL PLAT probmap noev n=2} (n = 2), it is correspondent to $\sigma_{\text{pdf}} = 0.022$.
When the errors are halved, the $\Omega_{\text{M}}$ values present a smaller uncertainty, for a smaller number of simulated GRBs. Although this result is expected, we nevertheless perform simulations to investigate to which extent the number of GRBs needed to achieve the desired uncertainty on $\Omega_{\text{M}}$ is reduced as much as possible without the need of a relatively great increase in the initial number of GRBs observed. Moreover, it is also important to note from Fig. \ref{fig:FULL PLAT conv and probs} that the probability increases and, thus, the standard deviation decreases consistently for $\Omega_{\text{M}}$ values approaching $0.30$. Specifically, for undivided errors, the most probable value is for $\Omega_{\text{M}} = 0.308 \pm 0.042$ at 2700 GRBs, and for halved errors, $\Omega_{\text{M}} = 0.300 \pm 0.027$ at 2600 GRBs. Furthermore and as expected, as the errors that enter the likelihood decrease, the number of GRBs needed to reach an $\Omega_{\text{M}}$ with the required accuracy also decreases. 
We again recover a lower number of GRBs needed if we look only to the number that falls under the $\sigma = 0.10$ error cutoff. All of the simulations we ran for halved errors had symmetrized errors below this error limit (see Fig. \ref{fig:FULL_PLAT_error_limits}). As for the \cite{2014A&A...568A..22B} limit, Fig. \ref{fig:FULL_PLAT_error_limits} shows that we reach a $\sigma = 0.042$ for 1452 GRBs. 
We here stress that in Fig. \ref{fig:FULL_PLAT_error_limits}, as in the following ones with the exception of the figures in the Appendix, we obtain the limiting numbers of GRBs by extrapolation the polynomial fitting functions which are of various order from order 4th to order 7th. 
Moreover, the \cite{2018ApJ...859..101S} limit is reached with 2724 simulated GRBs in X-ray. These studies have been completed for a number of GRB samples of varying quality as defined by the division of the errors that enter the likelihood equation; we present only the two most likely scenarios of unchanged errors and halved ones both for conciseness and probability of utility. It is reasonable to assume a sample this large of GRBs with halved error bars can be built in a relatively small amount of time due to the recent and rapid progression of efforts in the statistical reconstruction of GRB LCs, proving significant error bar reductions \citep{2022arXiv220312908D}. \cite{2022arXiv220312908D} tested that with the current sample a mean error reduction of $47.5\%$ is viable with LC reconstruction (from now on called LCR) when we consider the error bars on the time at the end of the plateau, $T_\text{a}$ and its correspondent flux. This is why we consider the scenario with confidence in this paper.

\begin{figure*}
  \centering
  \subfloat[OPTICAL | Simulation Results for the Full OPT Base with Undivided Errors ]
  {\includegraphics[width=8.6cm]{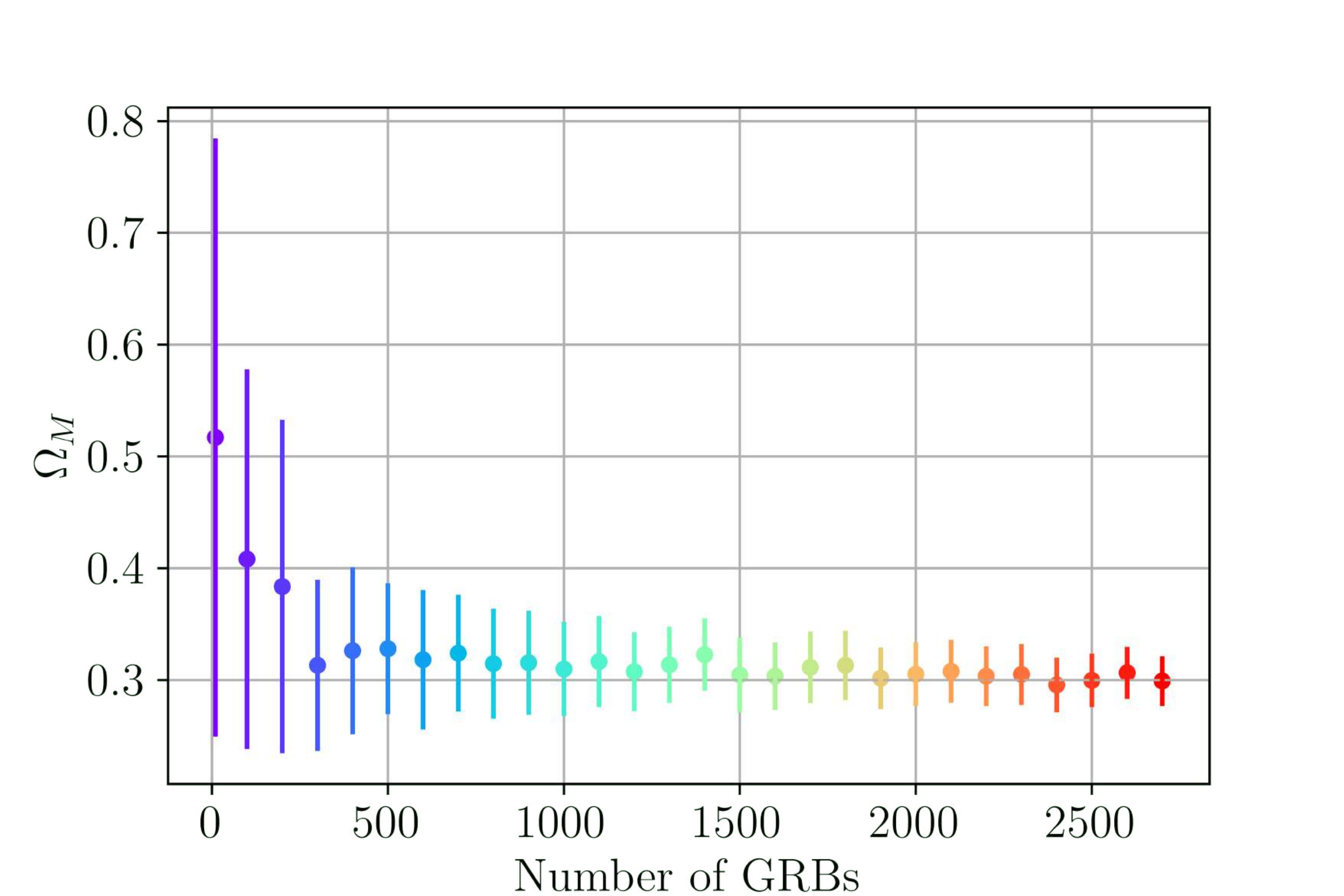}\label{fig:FULL OPT noev n=1}}
  \hspace{0.4cm}
  \subfloat[OPTICAL | Simulation Results for the Full OPT Base with Halved Errors]
  {\includegraphics[width=8.6cm]{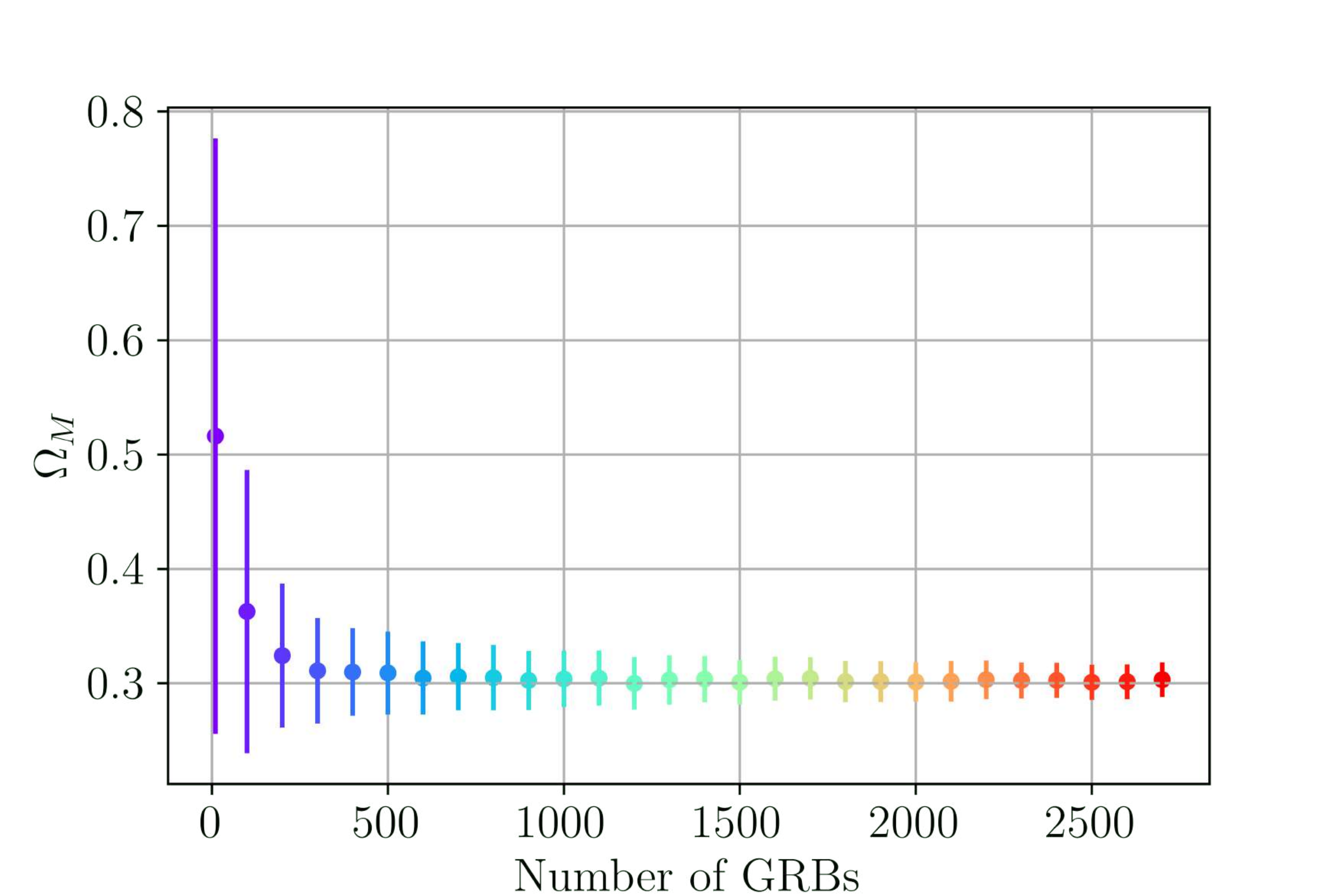}\label{fig:FULL OPT noev n=2}}
  \vspace{0.1cm}
  \subfloat[OPTICAL | Probability Map for Undivided Errors ]
  {\includegraphics[width=8.65cm]{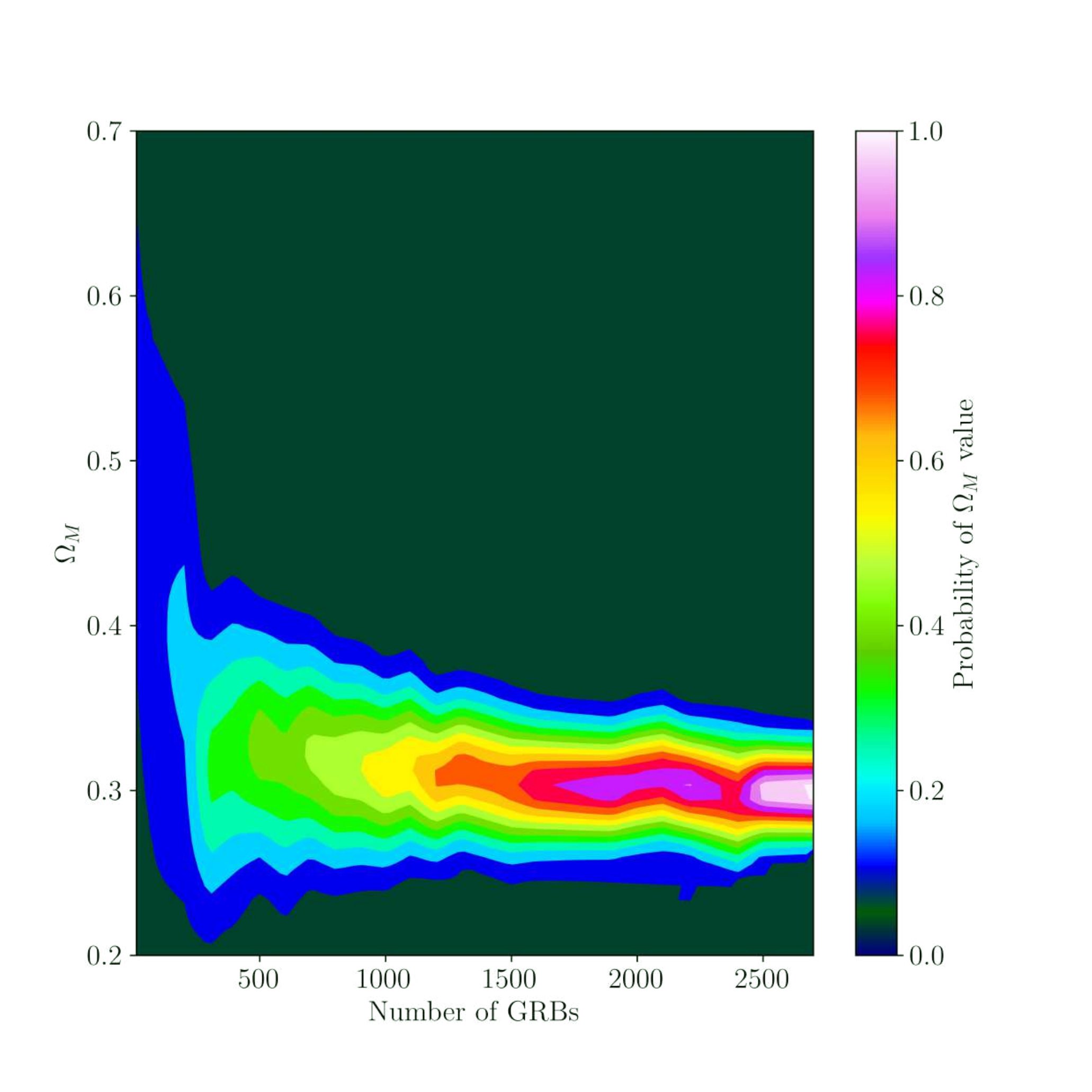}\label{fig:FULL OPT probmap noev n=1}}
  \hspace{0.4cm}
  \subfloat[OPTICAL | Probability Map for Halved Errors ]
  {\includegraphics[width=8.65cm]{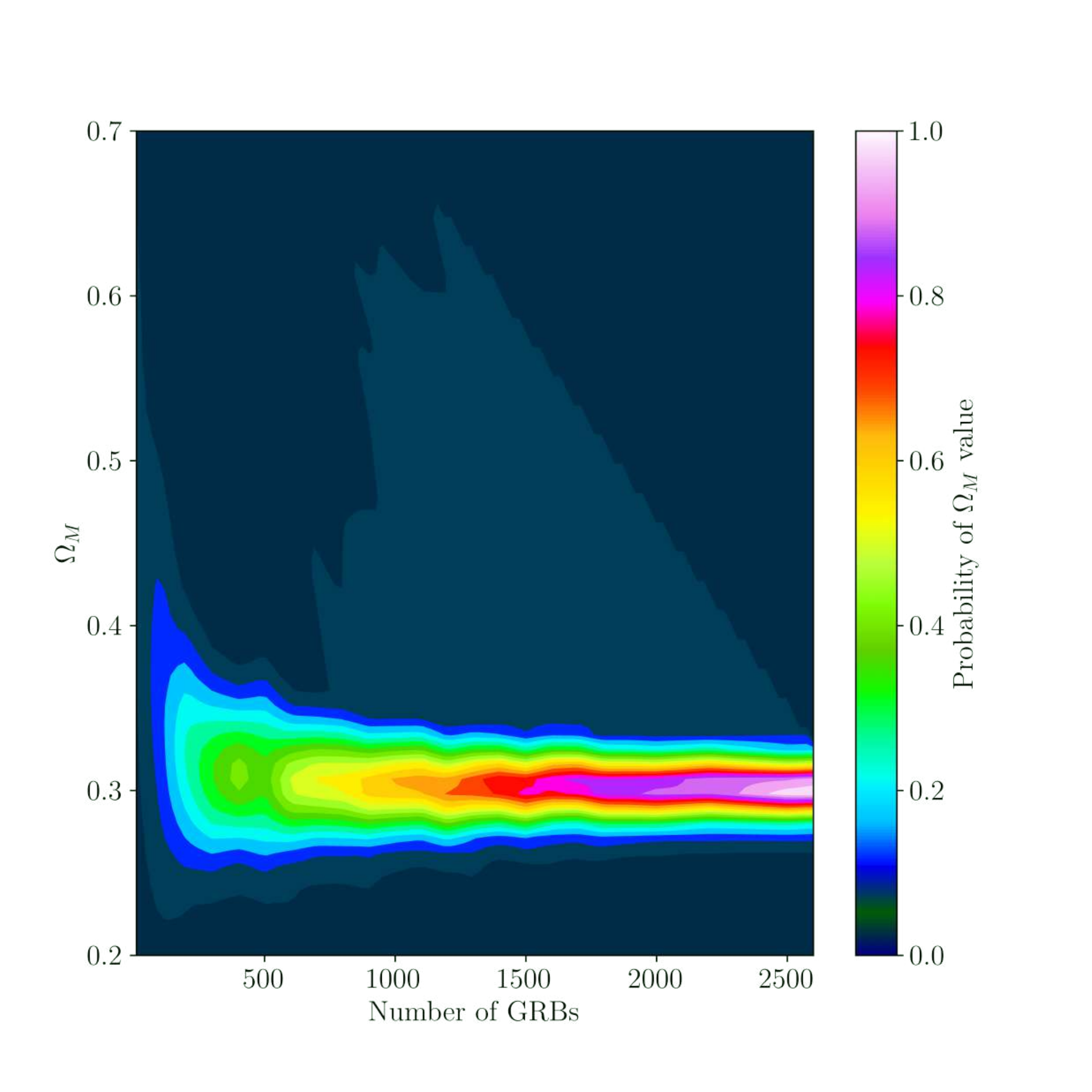}\label{fig:FULL OPT probmap noev n=2}}
\caption{Upper left panel: the mean values of $\Omega_{\text{M}}$ vs. the numbers of GRBs obeying the optical fundamental plane to converge upon a value of $\Omega_{\text{M}}$ using GRBs as the standalone probe by considering the observed error bars. The upper right panel: the same as the left panel, but considering the error bars divided by 2. The bottom left and right panels shows the corresponding probability distributions of the upper left and right panels, respectively.}
\label{fig:sim results for Full OPT}
\end{figure*}

\begin{figure*}
    \centering
    {\includegraphics[width=8.5cm]{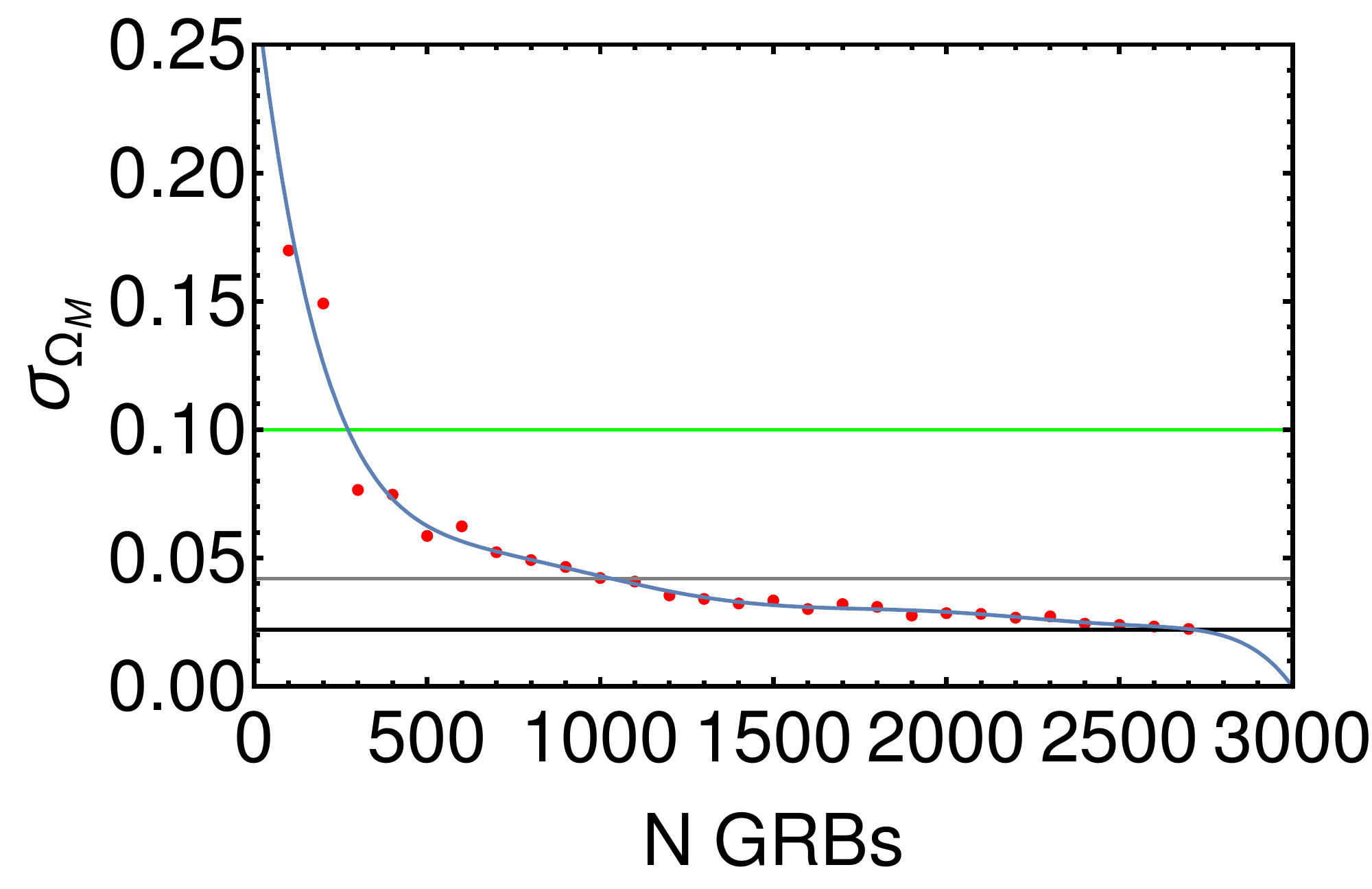}
    \label{fig:FULL OPT cont noev n=1 betoule}}
   {\includegraphics[width=8.5cm]{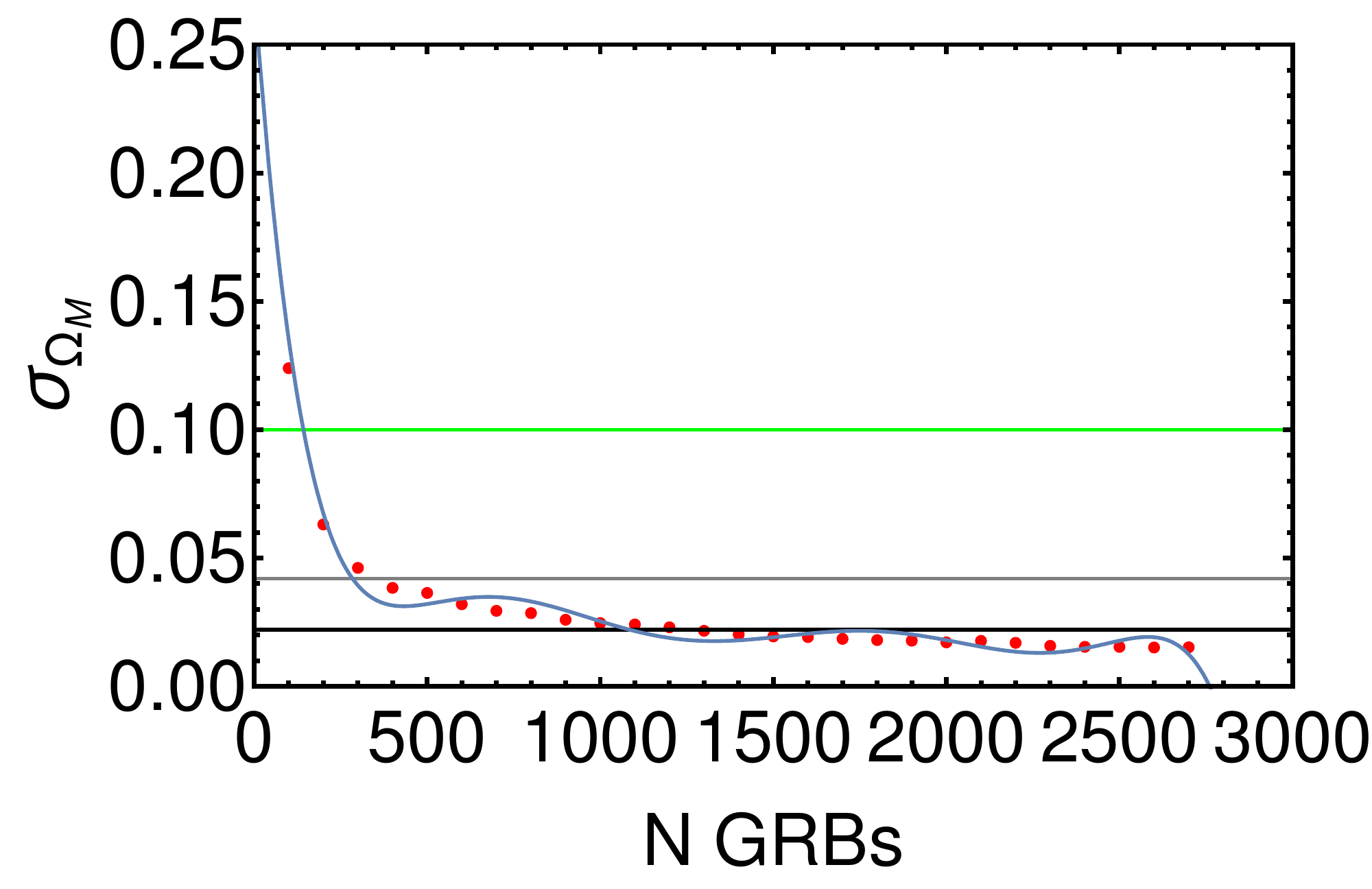}
    \label{fig:FULL OPT cont noev n=2 betoule}}
\caption{The plots show the number of simulated GRBs versus the error on $\Omega_{\text{M}}$ derived by the simulations starting from the full optical sample. On the left panel we have the undivided errors for the physical quantities related to the GRBs, while on the right panel we have divided these error by 2. The green, grey and black lines identify the \citet{2011ApJS..192....1C,2014A&A...568A..22B} and \citet{2018ApJ...859..101S} errors on $\Omega_{\text{M}}$, respectively.}
\label{fig: opt conts}
\end{figure*}

We now perform the same simulations and analysis on the full optical GRB sample. In Fig. \ref{fig:sim results for Full OPT}, the upper panels show the convergence plots, the lower panels show the probability maps, and Fig. \ref{fig: opt conts} shows for which number of simulated optical GRBs we reach the \cite{2011ApJS..192....1C}, \cite{2014A&A...568A..22B} and \cite{2018ApJ...859..101S} limits. These plots are surprising yet beneficial; despite a smaller sample size and increased $\sigma_{\text{int}}$ on the optical fundamental plane, simulations using the full optical sample of 45 GRBs as a base produce clearly and abundantly more precise values for $\Omega_{\text{M}}$ than what the X-ray sample achieves. In fact, the optical sample with errors undivided trumps the X-ray sample with halved errors. The reason for this behaviour is still unclear and additional investigation alongside a larger sample is needed for a deeper explanation. The comparisons between the two wavelengths' results are showcased in Table \ref{tab:compare opt xray sims}. The optical sample proves itself to be $47.6\%$ more precise than the X-ray sample when we consider n=1 and $44.4\%$ for n=2 for 2700 simulated GRBs. Similarly, if we consider the cases with evolution the optical sample has a decrease in the scatter in determining $\Omega_{\text{M}}$ of $44.2\%$ compared to the X-ray sample. 

\begin{figure*}
    \centering
    \subfloat[X-RAY | Evolution-Corrected Probability Map]
    {\includegraphics[width=8.65cm]{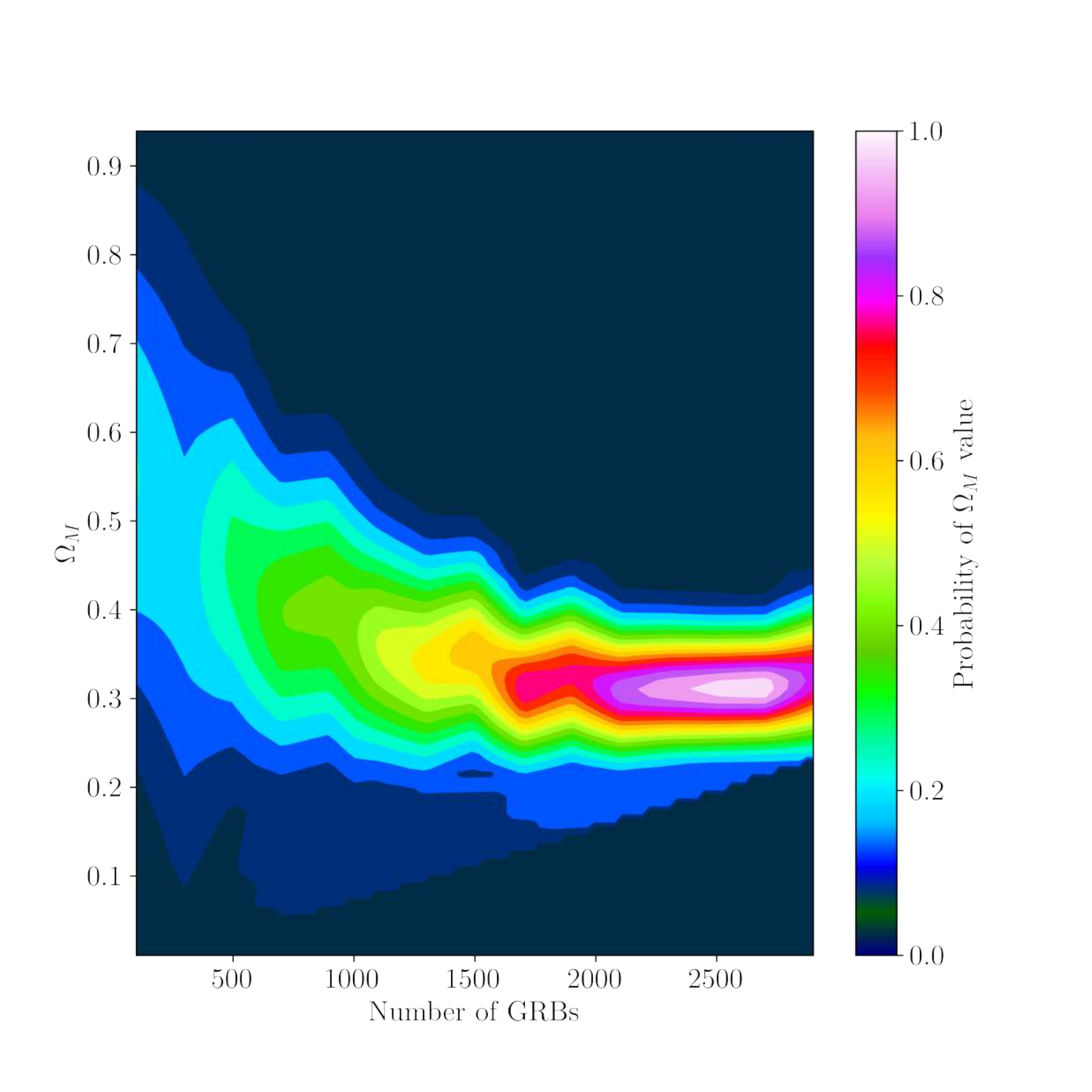}\label{fig:FULL PLAT probmap EV}}
    \subfloat[OPTICAL | Evolution-Corrected Probability Map]
    {\includegraphics[width=8.65cm]{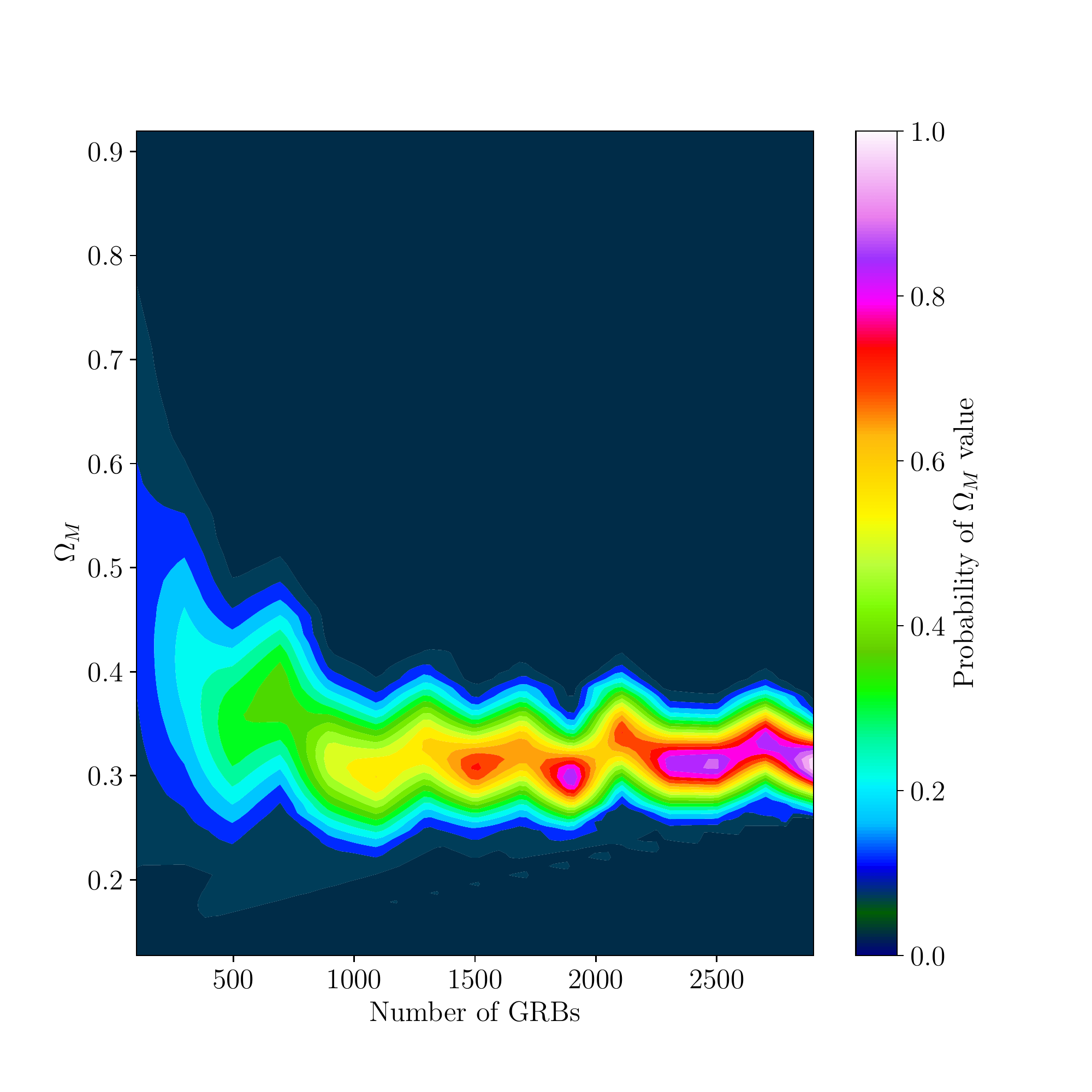}\label{fig:FULL OPT probmap EV}}
\caption{The probability map for the X-ray (left panel) and optical (right) fundamental planes, both corrected for evolution.}
\label{fig: full probmaps ev}
\end{figure*}

Furthermore, we find out from Fig. \ref{fig: opt conts} that we only require 1031 optical GRBs to reach the \cite{2014A&A...568A..22B} precision limit (left panel), and 284 when errors are halved (right panel). The \cite{2018ApJ...859..101S} limit is reached with 2718 and with 1086 GRBs for the undivided errors and divided by two errors scenarios, respectively (left and right panel of Fig. \ref{fig: opt conts}).
We have extended this analysis by performing redshift evolution and selection bias corrections on the simulated GRBs in both wavelengths as was done in Sec.\S \ref{sec:EP_method}. However, as seen in Table \ref{tab:compare opt xray sims}, this correction results in less precise derivations of $\Omega_{\text{M}}$. The behavior of the simulation results in general become less uniform in terms of convergence, as shown in Fig. \ref{fig: full probmaps ev}. This is because we now must also consider the larger error bars on the evolutionary function slope for each variable.

Whereas before the optical samples provided the most precision, we find that after bias corrections, the X-ray GRB data remains the most reliable. This is because the slope of the optical evolutionary functions have larger uncertainties than their X-ray counterparts, given that the OPT sample is slightly smaller with a fundamental plane having a higher $\sigmaint$.

\begin{table*}
\caption{The first column is the X-ray and optical samples used, respectively, with “n" the number by which the sample errors are divided before entering the simulations. The second column is the number of GRBs. The third column shows the most probable value of $\Omega_{\text{M}}$. The forth column is the standard deviation of the probability density functions correspondent to the numbers of simulated GRBs in the second column. \label{tab:compare opt xray sims}}
    \begin{tabular}{c|c|c|c}
        \hline
        Sample & \# GRBs & Most Probable $\Omega_{\text{M}}$ & $\sigma_{\text{pdf}}$ \\\hline
        X-ray n = 1 & 2700 & $0.308 \pm 0.042$ & 0.037 \\
        Optical n = 1 & 2700 & $0.299 \pm 0.022$ & 0.018\\
        X-ray n = 2 & 2600 & $0.300 \pm 0.027$ & 0.022 \\
        Optical n = 2 & 2600 & $0.301 \pm 0.015$ & 0.012\\
        X-ray (EV) n = 1 & 2700 & $0.312 \pm 0.052$ & 0.043\\ Optical (EV) n = 1 & 2900 & $0.311 \pm 0.029$ & 0.023\\
        \hline
    \end{tabular}
\end{table*}

Discussion on the methodology we employed to determine the error estimate on the MCMC sampler is drawn out in Appendix Sec. \S \ref{MCMC error}.

\subsection{Simulated trimmed Samples as Increasingly Precise Cosmological Tools} \label{sec:simulations_trimmed}

We now explore ways of trimming the X-ray and optical samples to derive smaller error bars on the value of $\Omega_{\text{M}}$ than the ones derived by the full samples. The overall methodologies can indeed be repeated for future works as our observed sample sizes increase. First we use the 3D fundamental plane in X-ray as defined in Sec.\S \ref{sec:PLATtrim+SNe} by the 10 GRBs that hold the intrinsic scatter of the plane near zero as a base for simulations.

\begin{figure*}
  \centering
  \subfloat[X-RAY | Simulation Results for PLATtrim = 10 with Halved Errors]
  {\includegraphics[width=8.5cm]{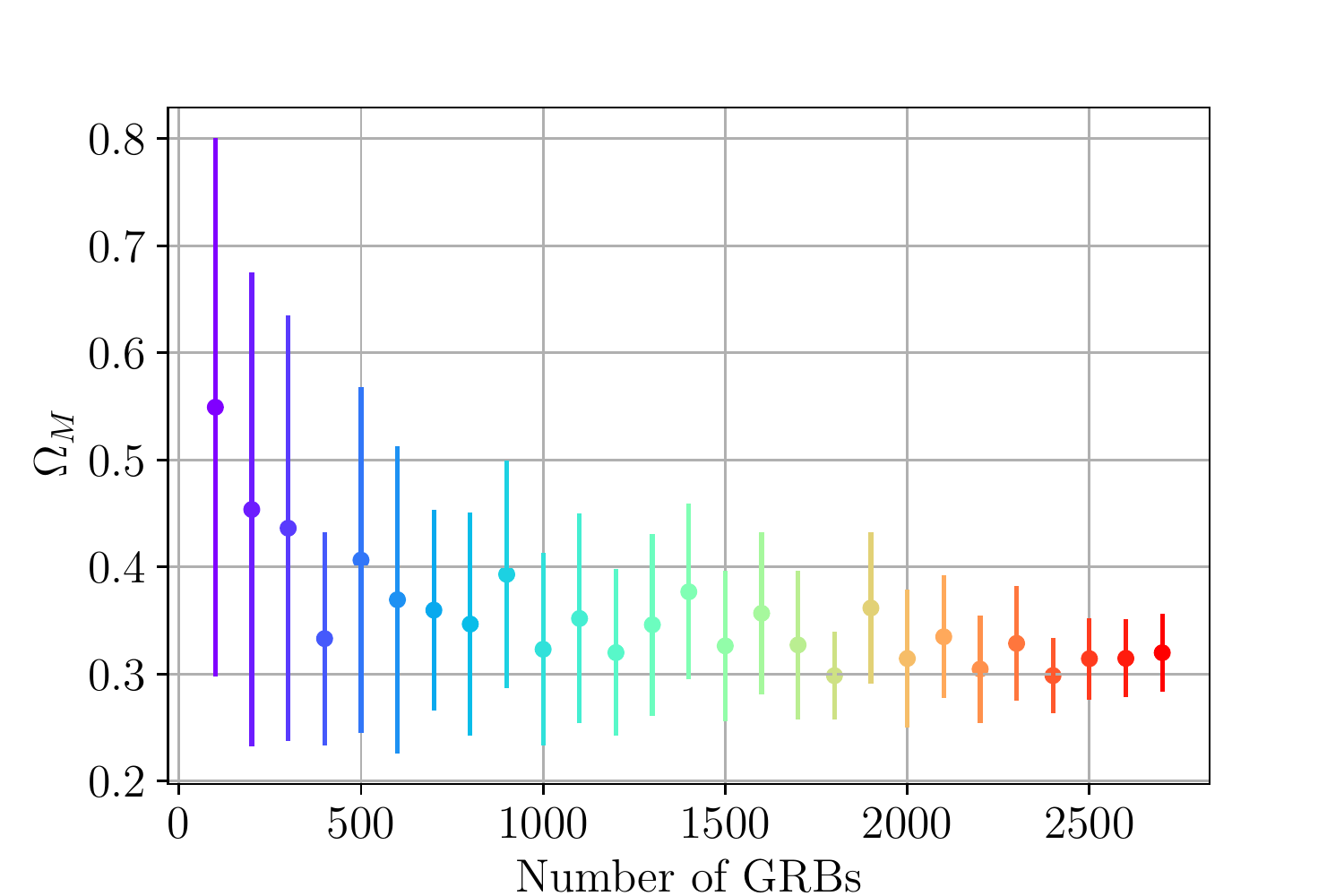}\label{fig:PLATtrim10 n=2 conv}}
  \subfloat[X-RAY | Simulation Results for PLATTtrim = 20 with Halved Errors]
  {\includegraphics[width=8.5cm]{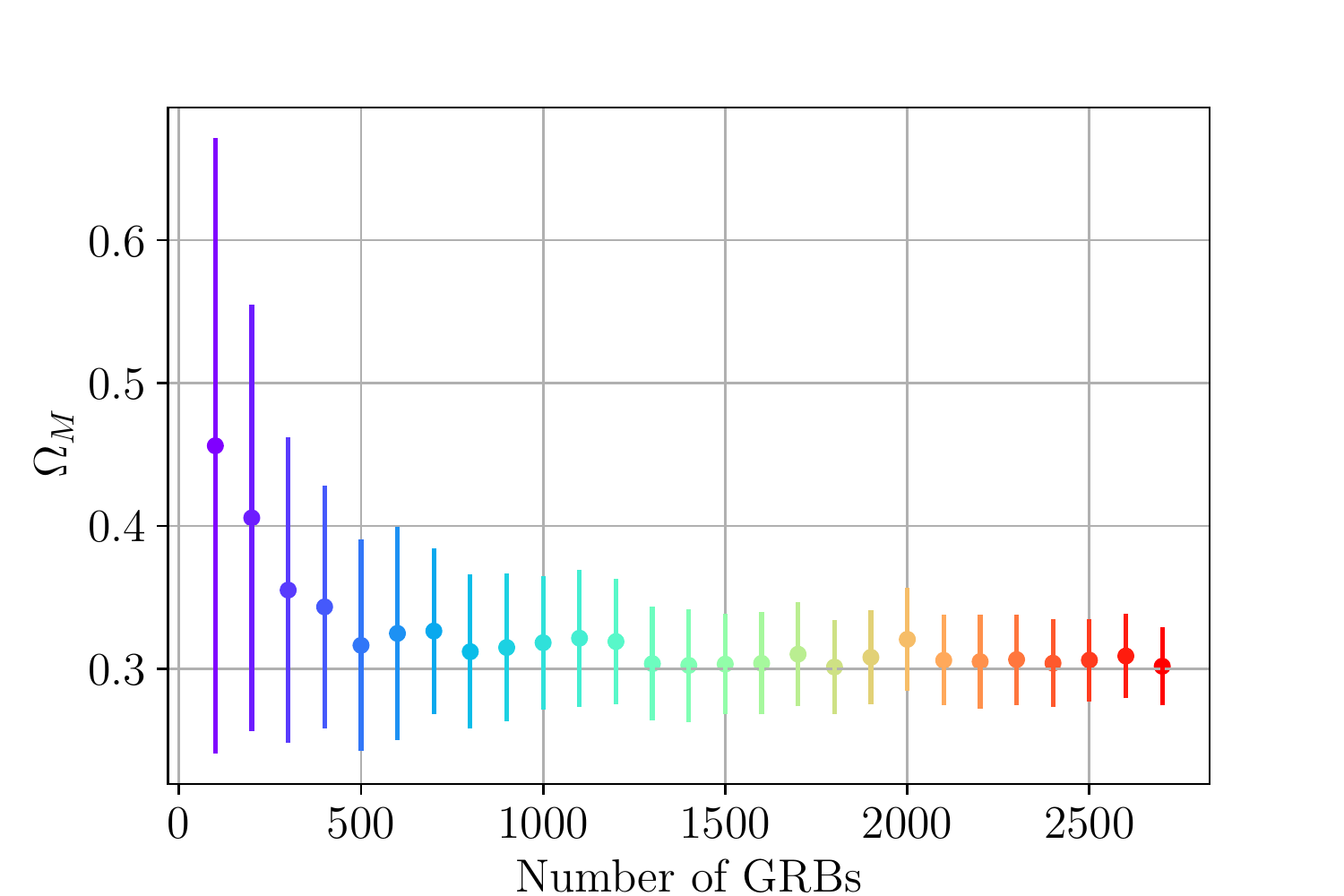}\label{fig:PLATtrim20 n=2 noev conv}}
  \vspace{0.1cm}
  \subfloat[X-RAY | Probability Map for PLATtrim = 10 with Halved Errors]
    {\includegraphics[width=8.65cm]{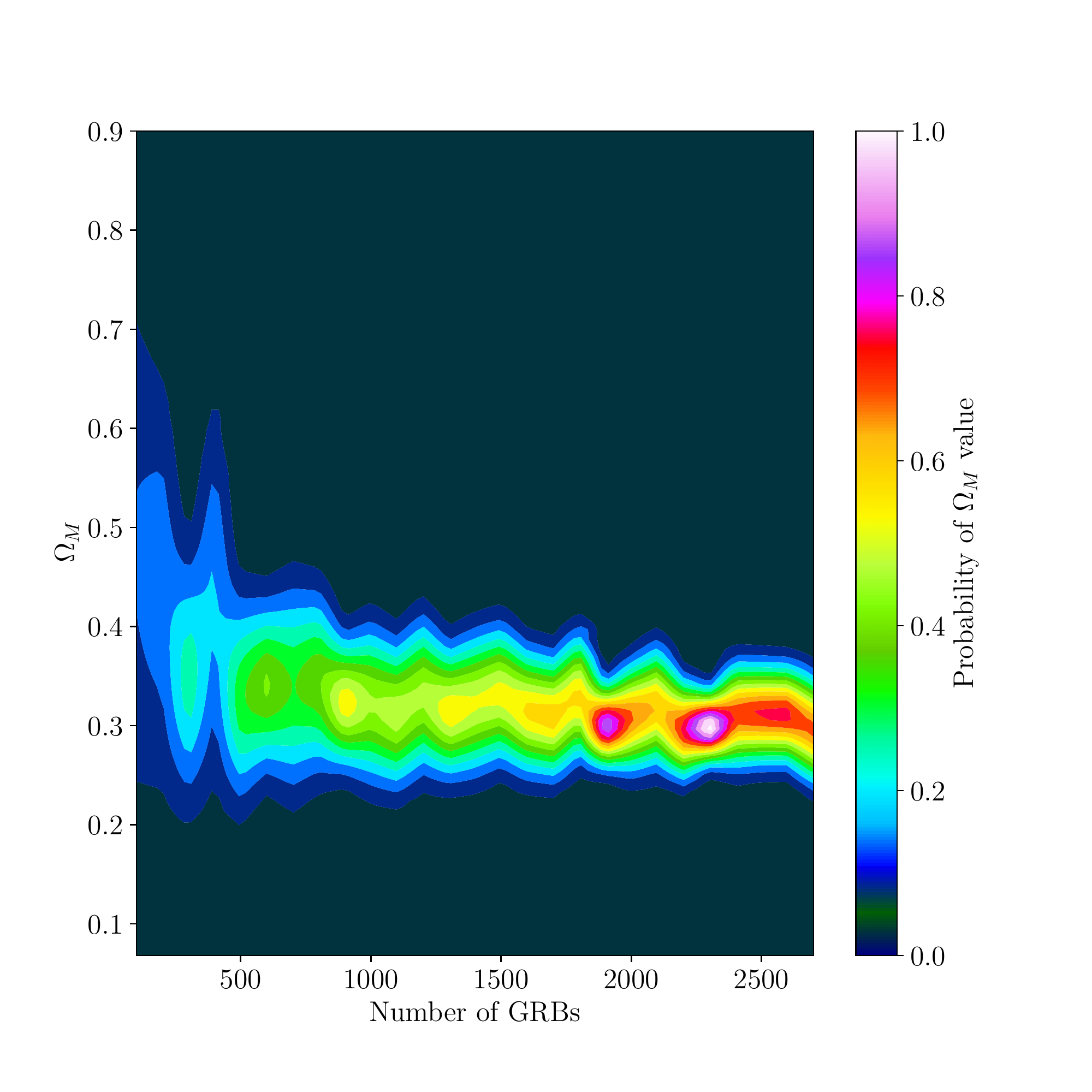}\label{fig:PLATtrim10 n=2 noev probmap}}
  \subfloat[X-RAY | Probability Map for PLATtrim = 20 with Halved Errors]
  {\includegraphics[width=8.65cm]{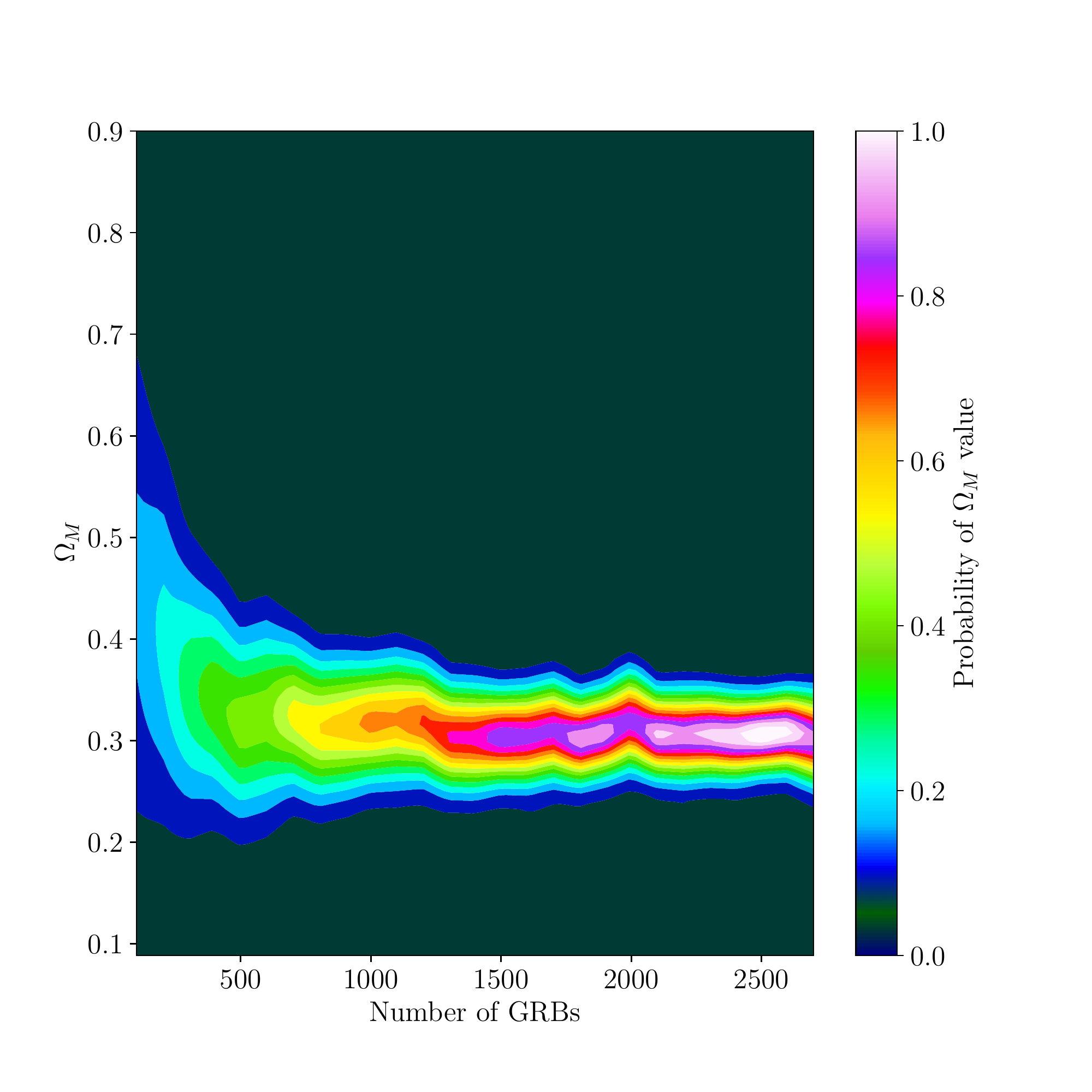}\label{fig:PLATtrim20 n=2 noev probmap}}
  \caption{Upper left panel: the mean values of $\Omega_{\text{M}}$ vs. the numbers of GRBs obeying the optical fundamental plane simulated with 10 GRBs. The upper right panel: the same as the left panel, but considering the plane simulated with 20 GRBs instead of 10. The bottom left and right panels shows the corresponding probability distributions of the upper left and right panels, respectively.}
  \label{fig:plat trim convs}
\end{figure*}

\begin{figure*}
  \centering
  {\includegraphics[width=8.6cm]{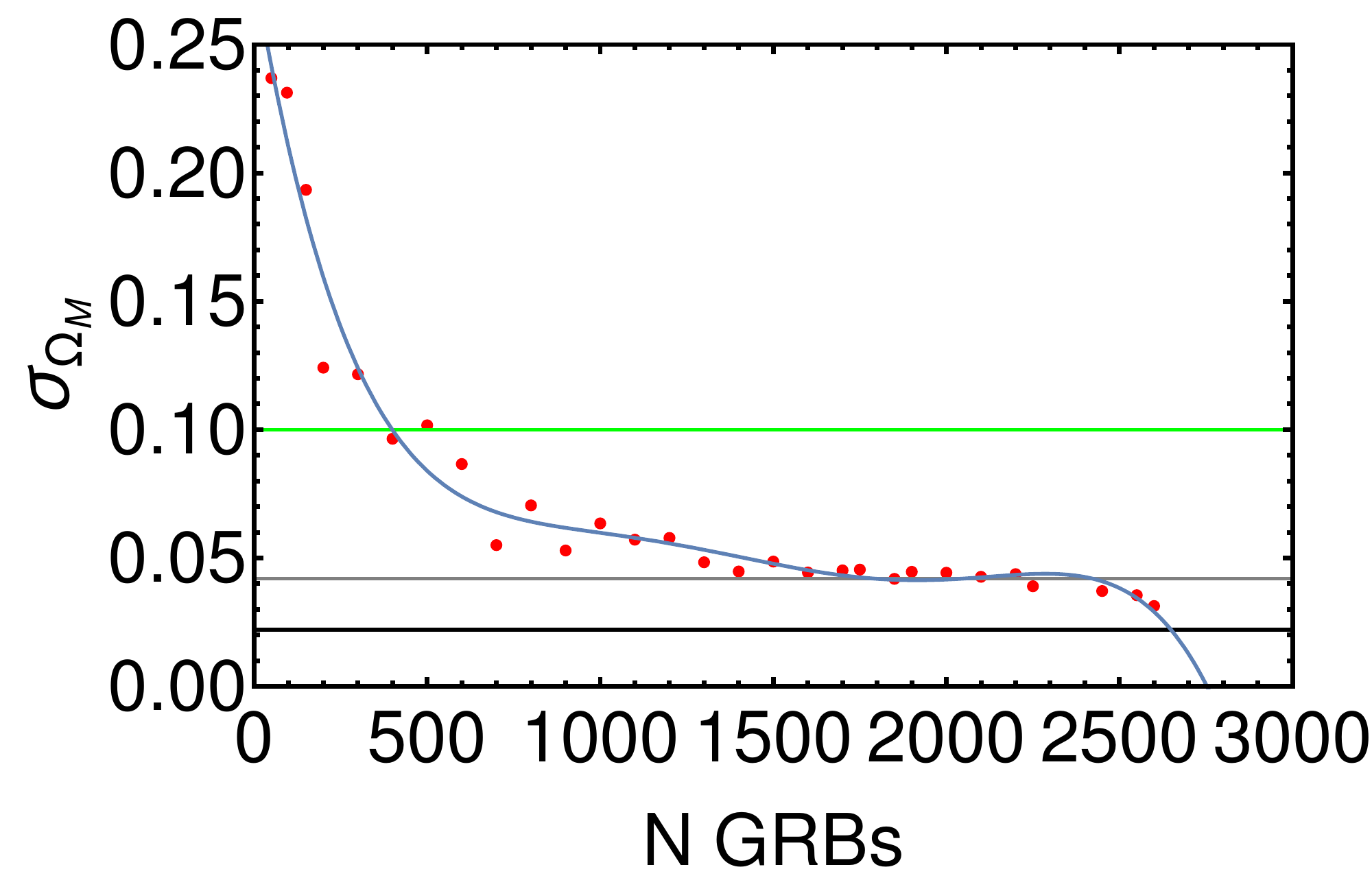}}
  \hspace{0.4cm}
  {\includegraphics[width=8.6cm]{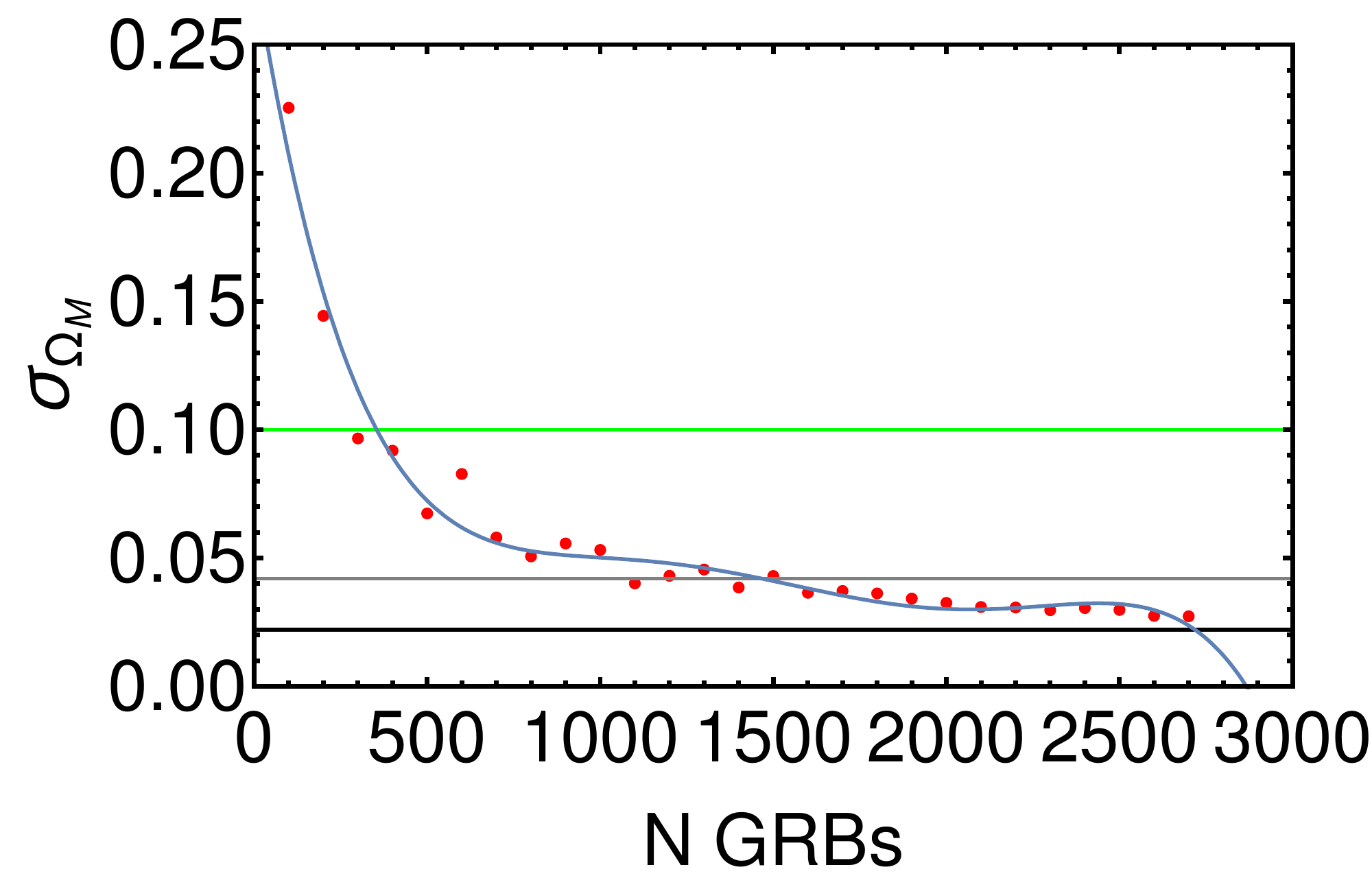}}
  \caption{The plots show the number of simulated GRBs versus the error on $\Omega_{\text{M}}$ derived by the simulations. On the left panel we start from the PLAT sample trimmed with 10 GRBs and halved errors, while on the right panel we start from the PLAT sample trimmed with 20 GRBs and halved errors. The green, grey and black lines identify the \citet{2011ApJS..192....1C,2014A&A...568A..22B} and \citet{2018ApJ...859..101S} errors on $\Omega_{\text{M}}$, respectively. \label{fig:trim10_vs_trim20}}
\end{figure*}

By looking at the initial simulation results for a PLATtrim set of 10, we note a less smooth convergence than in the previous section where the full samples were instead used; fluctuating values are evident in Fig. \ref{fig:plat trim convs}. Although less uniform, the advantage to this model is that we do achieve smaller error bars than could be done with the full samples. In fact, it is by this trim in X-ray that we reach the smallest uncertainty yet with the X-ray sample for only 2300 GRBs (left panel of Fig. \ref{fig:trim10_vs_trim20}). All comparisons of the trimmed samples to the full samples in determining $\Omega_{\text{M}}$ are detailed in Table \ref{tab:compare trims}.

In addition to trimming the samples by those that hold the intrinsic scatter of the fundamental plane near-zero, we also choose to analyze an alternative method (called a posteriori) to the PLAT and OPT sample trimming to improve our results even further. We now run an array of simulations for varying PLATtrim and OPTtrim selections, looking for the number of GRBs used as a base for our simulations that optimizes our current sample. The criteria for which we consider a trimmed sample to be optimized follows from the computation of the smallest standard deviation on $\Omega_{\text{M}}$ for a given number of simulated GRBs. Although more time (and processor) expensive to test, this a posteriori sample trimming has potential to decrease the computed uncertainty significantly. We find that drawing from the full 50 X-ray and 45 optical GRBs, trimmed samples containing 20 and 25 GRBs, respectively, optimize our calculations. The plots from which we determined these values can be found in Appendix Sec. \S \ref{retro choices}.

Fig. \ref{fig:plat trim convs} compares the different trimmings and confirms our initial inference; although the trim done a priori with 10 GRBs (Fig. \ref{fig:PLATtrim10 n=2 noev probmap}) reduces uncertainties better than a posteriori trim with 20 GRBs, fluctuations in the a priori convergence exist already for n = 1. Therefore, when we choose to test n = 2, our a posteriori trimming choice (Fig. \ref{fig:PLATtrim20 n=2 noev probmap}) is more reliable. For details see Table \ref{tab:compare trims}. The best estimates from these trims are depicted in Fig. \ref{fig:trim10_vs_trim20}. For both trims, the \cite{2011ApJS..192....1C}, \cite{2014A&A...568A..22B} and \cite{2018ApJ...859..101S} error limits are reached with a smaller number of GRBs. All error limit results can be seen in Table \ref{tab: numgrbs needed}.

\begin{figure*}
  \centering
  \subfloat[OPTICAL | Simulation Results for OPTtrim = 25 with Undivided Errors]{\includegraphics[width=8.5cm]{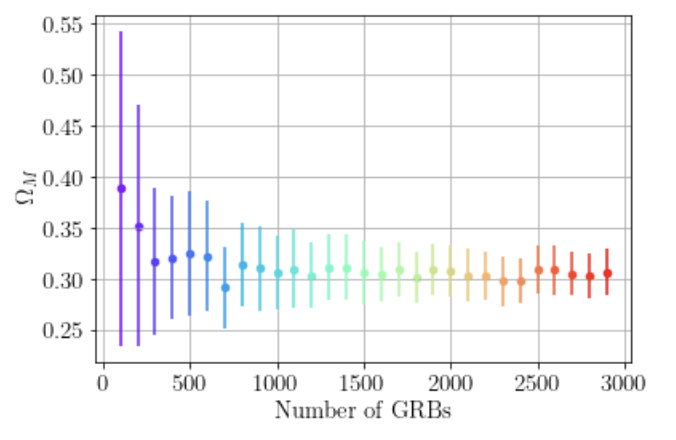}\label{fig:OPTtrim25 n=1 noev conv}}
  \hspace{0.4cm}
  \subfloat[OPTICAL | Simulation Results for OPTtrim = 25 with Halved Errors]{\includegraphics[width=8.5cm]{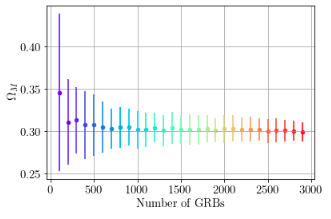}\label{fig:OPTtrim25 n=2 noev conv}}
    \vspace{0.5cm}
    \subfloat[OPTICAL | Probability Map for OPTtrim = 25 with Undivided Errors]{\includegraphics[width=8.65cm]{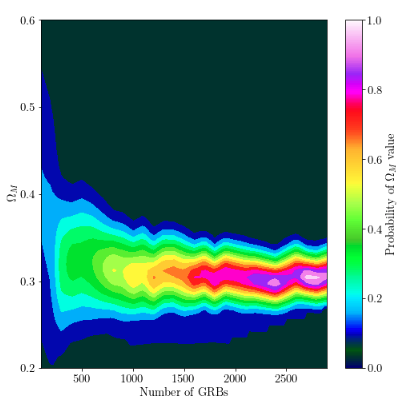}\label{fig:OPTtrim25 n=1 noev probmap clip}}
    \hspace{0.4cm}
  \subfloat[OPTICAL | Probability Map for OPTtrim = 25 with Halved Errors]{\includegraphics[width=8.65cm]{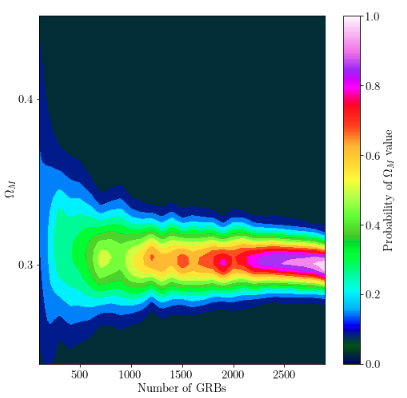}\label{fig:OPTtrim25 n=2 noev probmap clip}}
  \caption{Upper left panel: the mean values of $\Omega_{\text{M}}$ vs. the numbers of GRBs obeying the optical fundamental plane simulated with 25 GRBs. The upper right panel: the same as the left panel, but considering the error bars divided by 2. The bottom left and right panels shows the corresponding probability distributions of the upper left and right panels, respectively.}
  \label{fig:opttrim convs}
\end{figure*}

\begin{table*}
\caption{The first column shows the samples used, the second shows the number of GRBs used, with “n" the number by which the sample errors are divided before entering the simulations. A trimmed sample “a priori" refers to the sample of the 10 closest GRBs to their respective fundamental planes that yield an intrinsic scatter near-zero. The uncertainties reported are standard deviations. These results do not take into account redshift evolution or selection biases. \label{tab:compare trims}}
    \begin{tabular}{|c|c|c|}
        \hline
        Sample & \# GRBs & $\Omega_{\text{M}}$\\
        \hline
        PLAT; n = 1 & 2700 & $0.308 \pm 0.042$\\
        PLAT; n = 2 & 2600 & $0.300 \pm 0.027$ \\
        OPT; n = 1 & 2700 & $0.299 \pm 0.022$\\
        OPT; n = 2 & 2600 & $0.301 \pm 0.015$ \\
        PLATtrim (a priori); n = 1 & 2400 & $0.299 \pm 0.035$\\ 
        PLATtrim (a posteriori); n = 1 & 2400 & $0.300 \pm 0.042$ \\
        OPTtrim (a priori); n = 1 & 2900 & $0.306 \pm 0.024$ \\
        OPTtrim (a posteriori); n = 1 & 2700 & $0.305 \pm 0.021$ \\
        PLATtrim (a priori); n = 2 & 2300 &  $0.299 \pm 0.026$\\
        PLATtrim (a posteriori); n = 2 & 2700 & $0.302 \pm 0.027$ \\
        OPTtrim (a priori); n = 2 & 2600 & $0.301 \pm 0.016$ \\ 
        OPTtrim (a posteriori); n = 2 & 2600 & $0.301 \pm 0.014$\\
        OPTtrim (a priori); n = 2 & 2900 & $0.300 \pm 0.015$\\ OPTtrim (a posteriori); n = 2 & 2900 & $0.299 \pm 0.012$\\
        \hline
    \end{tabular}
\end{table*}

\begin{figure}
  \centering
  {\includegraphics[width=8.6cm]{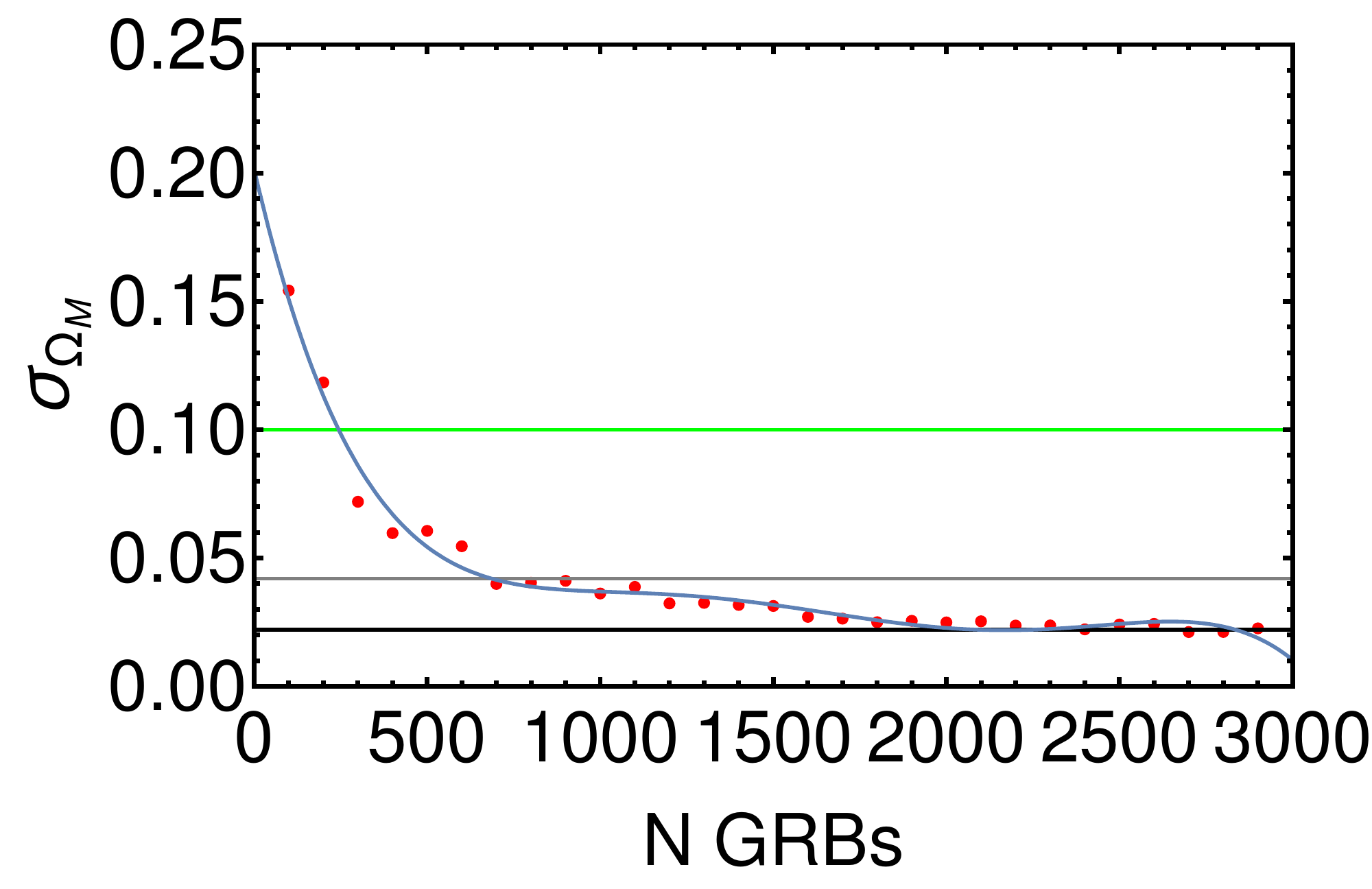}}
  \hspace{0.4cm}
  {\includegraphics[width=8.6cm]{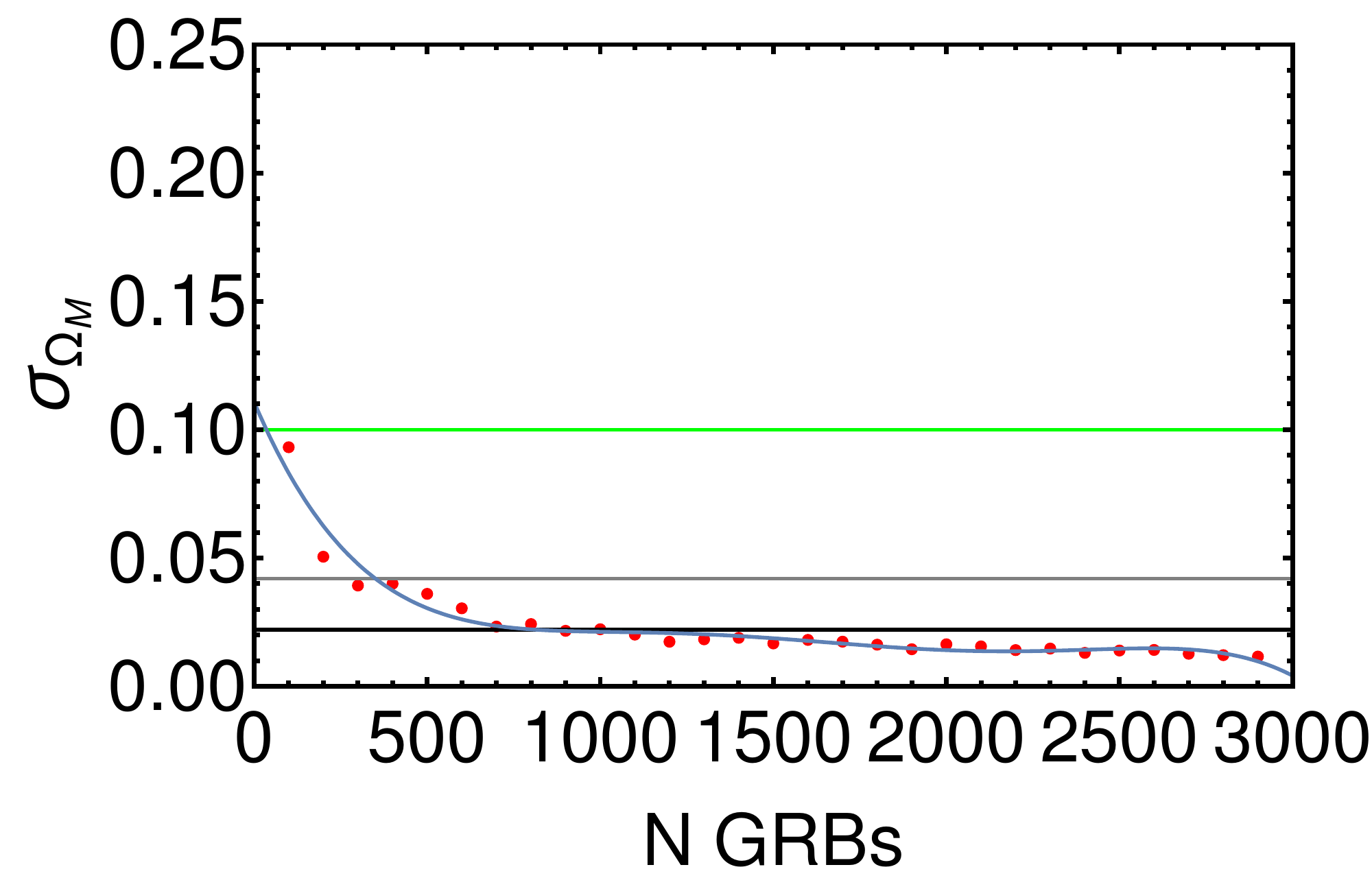}}
  \caption{The plots show the number of simulated GRBs versus the error on $\Omega_{\text{M}}$ derived by the simulations. On the left panel we start from the optical sample trimmed with 25 GRBs (a posteriori) and undivided errors, while on the right panel we start from the same sample but with halved errors. The green, grey and black lines identify the \citet{2011ApJS..192....1C,2014A&A...568A..22B} and \citet{2018ApJ...859..101S} errors on $\Omega_{\text{M}}$, respectively.  \label{fig:trim25_opt_error_limit}}
\end{figure}

We find the smallest error bars, considering the optical sample, for an a posteriori-decided OPTtrim of 25 GRBs in both error division cases. And in fact, out of both wavelengths and trimming methodologies, this sample yields our best results yet in both error division scenarios and for three error limit definitions we considered in comparison with the literature (Fig. \ref{fig:opttrim convs}). We also note that in Table \ref{tab:compare trims}, for all instances of the optical GRB sample, we not only fall below the \cite{2014A&A...568A..22B} standard deviation limit, but also near or even below the \cite{2018ApJ...859..101S} error limit (as again determined by SNe Ia only) of $\sigma = 0.022$. \cite{2018ApJ...859..101S} arrived at such a high precision using a large sample of 1048 SNe Ia, and the data we produce in the correspondent extrapolation (Fig. \ref{fig:trim25_opt_error_limit}) suggests that, in the case of halved errors, we only need 36, 350 and 822 GRBs with plateaus to reach the \cite{2011ApJS..192....1C}, \cite{2014A&A...568A..22B} and \cite{2018ApJ...859..101S} limits, respectively. It is remarkable that the   \cite{2011ApJS..192....1C} limit is already reachable now.

Taking this a posteriori approach to trimming, we re-run our previous computations involving the trimmed samples in both wavelengths including SNe Ia data. These results are displayed in Table \ref{tab:new sne+baos}. We see no noticeable difference in either the values of $\Omega_{\text{M}}$ themselves, or in their uncertainties. We performed the same a posteriori trimming on the EV-corrected data, and similarly re-performed our computations considering the most efficacious cut of both wavelengths with SNe Ia. Therefore, we can conclude that the EV-corrected simulations presented in Table \ref{tab:new sne+baos} below are the most precise derivations of the matter density of the Universe we can possibly achieve today given that the errors on $\Omega_{\text{M}}$ remain the same, but the treatment allow us to correct for selection biases and redshift evolution.

\begin{table*}
\caption{The first column shows the sample, while the successive ones the numbers of GRBs needed for the limits set by \citet{2011ApJS..192....1C,2014A&A...568A..22B}, and \citet{2018ApJ...859..101S} for the full error bars and the halved ones. \label{tab: numgrbs needed}. We put an hyphen when the limit is not reached.}
    \begin{tabular}{|l|c|c|c|c|c|c|}
        \hline
        \multicolumn{7}{|c|}{Number of GRBs with Plateaus Needed} \\
        \hline
        \multicolumn{1}{|c|}{GRB} & \multicolumn{2}{c|}{Conley (2011)} & \multicolumn{2}{c|}{Betoule (2014)} & \multicolumn{2}{c|}{Scolnic (2018)}\\
        \cline{2-7}
        \multicolumn{1}{|c|}{Sample} & n = 1 & n = 2 & n = 1 & n = 2 & n = 1 & n = 2 \\
        \hline
        PLAT & 789 & 357 & 2653 & 1452 & - & 2724 \\
        OPT & 271 & 142 & 1031 & 284 & 2718 & 1086 \\
        PLATtrim (a priori) & 847 & 399 & 2705 & 1788 & 2839 & 2649 \\
        OPTtrim (a priori)  & 330 & 112 & 829 & 393 & 2870 & 1513 \\
        PLATtrim (a posteriori) & 646 & 354 & 2699 & 1466 & - & 2719 \\
        OPTtrim (a posteriori) & 244 & 36 & 685 & 350 & 2104 & 822 \\
        \hline
    \end{tabular}
\end{table*}

\begin{table*}
\caption{The first column shows the samples used, while the second shows the results on $\Omega_{\text{M}}$ with their errors, which are variances on the \emph{emcee} chain, corresponding to the $68\%$ confidence limit. The asterisk on the trimmed GRB sample indicates that it has been trimmed a posteriori. All fits are considering redshift evolution effects. \label{tab:new sne+baos}}
\begin{tabular}{c|c|c|c}
\hline
Sample & $\Omega_{\text{M}}$ \\\hline
PLATtrim$^{*}$+SNe Ia & $0.299 \pm 0.009$ \\
PLATtrim$^{*}$+SNe Ia (EV) & $0.299 \pm 0.009$\\ OPTtrim$^{*}$+SNe Ia & $0.299 \pm 0.009$ \\
OPTtrim$^{*}$+SNe Ia (EV) & $0.299 \pm 0.009$\\
\hline
\vspace{0.4cm}
\end{tabular}
\end{table*}

\section{Future Deep-Space Surveys and their Detection Power} \label{sec:future_surveys}

Extrapolating minimum numbers of GRBs dependent upon error becomes useful in predicting which precision on $\Omega_{\text{M}}$ we can achieve in the future as current and future satellite missions observe increasingly detailed and more numerous data. One of these planned launches, studying X-ray emissions from GRBs, is the Space Variable Objects Monitor (SVOM, \cite{wei2016deep}). The proposed launch date is June 2023, and it is to act as a pathfinder for a later mission, that is, the Transient High Energy Sources and Early Universe Surveyor (THESEUS, \cite{2018AdSpR..62..191A}). SVOM is expected to detect around 80 GRBs per year, and for a planned three year mission, this means it should gather $\sim 240$ GRBs throughout the course of its lifetime with $\sim 1-2$ triggers per week expected \citep{2018MmSAI..89..266C}.

SVOM's successor, THESEUS, has a very tentative launch date in 2037. Although the THESEUS mission has not been selected under Phase A study by ESA as a candidate M5 mission \citep{2021arXiv210409531A}, there is the intent and the effort by the community to apply for other future funding schemes and opportunities. The number of expected GRBs triggers per year can reach up to 1000 \citep{2017xru..conf..250A}. So one can expect up to three triggers per day \citep{2018MmSAI..89..157F}. THESEUS holds a hypothesized rate of GRB observation between  $300\text{-}700$ GRBs yr$^{\text{-}1}$ \citep{2018AdSpR..62..191A}. 


\begin{table*}
\caption{The first column shows all of the samples used. The three rows associated with each sample definition show the number of GRBs needed that have plateau phases, the total number of GRBs needed once the chosen sample proportionality out of the total samples are considered, and lastly, the year in which that number of GRBs is achieved with observed data. These rows are repeated for samples with the errors undivided, the errors halved, the errors undivided considering LCR, and finally, the errors halved including LCR. The overarching columns three and four give these estimates considering the precision reached by \citet{2011ApJS..192....1C} and \citet{2014A&A...568A..22B}, respectively. A sample ``+ ML'' implies that machine learning techniques are employed to double the initial sample size by redshift inference.}\label{Tab:Time Estimates conley,betoule}
{\centering
\hspace*{-0.6cm}\scalebox{0.85}{\begin{tabular}{|c|c|c|c|c|c|c|c|c|c|}
\hline
\multicolumn{1}{|c|}{GRB} & & \multicolumn{4}{c|}{Conley (2011)} & \multicolumn{4}{c|}{Betoule (2014)} \\
\cline{3-10}
\multicolumn{1}{|c|}{Sample} & & n = 1 & n = 2 & n = 1 ($47.5\%$ LCR) & n = 2 ($47.5\%$ LCR) & n = 1 & n = 2 & n = 1 ($47.5\%$ LCR) & n = 2 ($47.5\%$ LCR) \\
\hline
& \# GRBs (Plateau) & 789 & 357 & 374 & 169 & 2653 & 1452 & 1260 & 689 \\  
PLAT & \# GRBs (Total) & 16789 & 7596 & 7975 & 3608 & 56455 & 30898 & 26816 & 14676 \\ 
& Year Achieved & 2060 & 2044 & 2045 & 2037 & 2129 & 2085 & 2078 & 2056 \\ \hline 

& \# GRBs (Total) & 8394& 3798& 3987& 1804 & 28227 &15449 & 13408 & 7338 \\ 
PLAT + ML & Year Achieved & 2045 & 2037 & 2038 & 2026 & 2080 & 2058 & 2054 & 2043\\ \hline

& \# GRBs (Plateau) & 271 & 142 & 128 & 67 & 1031 & 284 & 489 & 134 \\ 
OPT & \# GRBs (Total) & 4582 & 2401 & 2176 & 1140 & 17435 & 4802 & 8281 & 2281 \\ 
& Year Achieved & 2046 & 2038 & 2038 & 2026 & 2093 & 2047 & 2060 & 2038 \\\hline 

& \# GRBs (Total) & 2291 & 1200 & 1088 & 570 & 8717 & 2401 & 4140 & 1140 \\ 
OPT + ML & Year Achieved & 2038 & 2027 & 2025 & Now & 2061 & 2038 & 2045 & 2026\\ \hline

& \# GRBs (Plateau) & 847 & 399 & 402 & 189 & 2705 & 1788 & 1284 & 849 \\ 
PLATtrim  (10) & \# GRBs (Total) & 18024 & 8490 & 8561 & 4033& 57562 & 38048 & 27342 & 18073 \\ 
& Year Achieved & 2062 & 2045 & 2046 & 2038 & 2131 &2097 & 2078 & 2062\\\hline 

& \# GRBs (Total) &9012 & 4245 &4280 & 2016 & 28781 &19024 & 13671 & 9036 \\ 
PLATtrim (10) + ML & Year Achieved & 2046 & 2038 & 2038 & 2027 & 2081 & 2064 & 2054 & 2046 \\\hline 

& \# GRBs (Plateau) & 330 & 112 & 156 & 53 & 829 & 393 & 393 & 186 \\ 
OPTtrim (10) & \# GRBs (Total) & 5580 & 1894 & 2650 & 899 & 14019 & 6646 & 6659 & 3156 \\ 
& Year Achieved & 2050 & 2037 & 2039 & 2022 & 2081 & 2054 & 2054 & 2041 \\\hline 

& \# GRBs (Total) & 2790 & 947 & 1325 & 449 & 7009 & 3323 & 3329 & 1578 \\ 
OPTtrim (10) + ML & Year Achieved & 2040 & 2023 & 2029 & Now & 2055 & 2042 & 2042 & 2032 \\\hline 

& \# GRBs (Plateau) & 646 & 354 & 306 & 168& 2699 & 1466 & 1282 & 696 \\ 
PLATtrim (20) & \# GRBs (Total) & 13746 & 7533 & 6529 & 3578 & 57434 & 31196 & 27281 & 14818  \\ 
& Year Achieved &2055 & 2044 & 2045 & 2037 & 2131  &2085 & 2078 & 2057 \\\hline 

& \# GRBs (Total) & 6873 & 3766 &3264& 1789 & 28717 & 15598 & 13640 & 7409 \\ 
PLATtrim (20) + ML & Year Achieved & 2043 & 2037 & 2036 & 2025 & 2081 &2058 & 2054 & 2044 \\ \hline

& \# GRBs (Plateau) & 244 & 36 & 115 & 17 & 685 & 350 & 325 & 166 \\ 
OPTtrim (25) & \# GRBs (Total) & 4126 & 608 & 1959 & 289 & 11584 & 5918 & 5502 & 2811 \\ 
& Year Achieved & 2045 & 2018 & 2037 & Now & 2072 & 2051 & 2050 & 2040 \\\hline 

& \# GRBs (Total) & 2063 & 304 & 979 & 144 & 5792 & 2959 & 2751 & 1405\\ 
OPTtrim (25) + ML & Year Achieved & 2037 & Now & 2024 & Now & 2051 & 2040 & 2040 & 2030 \\ 

\hline
\end{tabular}}}
\end{table*}

We now use the calculated number of GRBs necessary to obtain ideal precision to predict, given the observing capability of both the present and the aforementioned future deep space survey missions, to estimate the length of time until X-ray and optical GRB emissions can be used in practice as standalone standard candles with limits on the errors on $\Omega_{\text{M}}$ comparable with the ones obtained using SNe Ia. Although the THESEUS team states that the mission will last for 3.45 years, we work under the assumption that its lifetime will be prolonged, as it has happened for many satellite missions. Specifically, we estimate its lifetime to endure as long as the Konus-Wind mission, that has been in service for 27 years, or the Chandra X-ray Observatory, that is now almost 23 years old.
Thus, our computation related to the lifetime of THESEUS for simplicity is posed to be 27 years.

Given these estimates, the number of GRBs expected to be detected by THESEUS throughout the course of its mission is then 18900. By THESEUS, SVOM, and current rates of detection by Swift, we calculate the year in which the number of detected GRBs needed is achieved. The number of needed GRBs taken from the simulated data is listed as the number of GRBs with observed plateaus (\# GRBs (Plateau)) in Tables \ref{Tab:Time Estimates conley,betoule} for the estimates to reach the limits of \citet{2011ApJS..192....1C} and \citet{2014A&A...568A..22B} and Table \ref{tab:Time Estimates scolnic} for reaching the limit of \cite{2018ApJ...859..101S}. However, we must also consider that our PLAT sample of GRBs is taken from the full 1064 GRBs presenting X-rays observed by Swift-XRT up until August 2019, and that our OPT sample is taken from the full 761 GRBs with optical observations from January 1997 - December 2018, including the ones without redshift measurements and detected plateaus. We must maintain these proportionalities when considering a minimum number of GRBs because our simulated sample must have an observable plateau emission phase and a detected redshift. So, more explicitly, we need to multiply the number of GRBs observable in X-rays by a factor of $\frac{1064}{50}$ (where 50 is the full PLAT sample) to compute how many GRBs with X-ray plateaus and with observed redshift will belong to the PLAT sample, and similarly by $\frac{761}{45}$ (where 45 is the full OPT sample) for the GRBs possessing an observed optical plateau with redshift. Multiplying by these ratios ensures that we account for these requirements. In addition, we need to consider the ratio of observations of the new missions.
For clarity, we here summarize the assumptions and the rates underlying the forecasts of this section in bullet points:
\begin{itemize}
\item  Lifetime of Theseus = 27 years, with an estimated launch in 2037.
\item Lifetime of SVOM = 15 years, with an estimated launch in 2023.
\item Total number of GRBs observed throughout the lifetime of THESEUS =
18900.
\item Total GRBs observed throughout the lifetime of SVOM = 1350.
\item Current rates of detection of Swift = 89.45 $\text{yr}^{-1}$.
\item For X-ray wavelengths, we assume that the ratio (0.82) of the total number of GRBs observed by Swift-XRT (1064) from 2005 January until 2019 August compared to the total number of GRBs observed by Swift (1305) will be the same ratio of GRBs observed by the ECLAIRs (4-120 keV) on board of SVOM and the X-Gamma rays Imaging Spectrometer (XGIS, 2-20 MeV) and Soft X-ray Imager (SXI, 0.3-5 keV) on board of THESEUS.
\item For optical wavelengths, we assume that the ratio (0.39) of the total number of optical afterglows (761) observed from 1997 up to December 2018, compared to the total number of GRBs by all missions (1942), is again the same as the one observed by the Infrared Telescope (IRT, 0.7-1.8 $\mu$m) on THESEUS, and as the Visible Telescope (VT, 540-600 nm) on board of SVOM.
\item For optical wavelengths, we assume that additional ground-based instruments will also be operative as they are at this time. This is because, even if some of them will stop operating, we expect to have at least the same number and with the same capabilities as today from the ground based observations. We actually expect to have more, but we remain conservative in this estimate.
\end{itemize}

Keeping these assumptions and rates in mind,
we define in Tables \ref{Tab:Time Estimates conley,betoule} and \ref{tab:Time Estimates scolnic} a second column describing the total number of GRBs needed to achieve the error limit, considering both GRBs with and without plateau phases. We then calculate the year in which these simulated GRBs are predicted to be observed considering four types of GRB samples:

\begin{enumerate}
    \item \ all errors on the measured quantities undivided with no statistical reconstructions.
    \item \ all errors halved with no statistical reconstructions.
    \item \ all errors undivided initially, but by current LCR capabilities, $47.5 \%$ of the sample's errors are already halved.
    \item \ all errors halved initially, and by current LCR capabilities, $47.5 \%$ of the sample's errors are halved again.
\end{enumerate}

\begin{table*}
\caption{The first column shows all of the different optical GRB samples used. The three rows associated with each sample show the number of GRBs needed that have 1) plateau phases, 2) the total number of GRBs needed, and 3) the year in which that number of GRBs is achieved with observed data. These rows are repeated for samples with the errors undivided, the errors halved, the errors undivided considering LCR, and finally, the errors halved including LCR. These estimates are given considering the precision reached by \citet{2018ApJ...859..101S}. A sample ``+ ML'' implies that machine learning techniques are employed to double the initial sample size by redshift inference.}\label{tab:Time Estimates scolnic}
\begin{center}
\hspace*{-1cm}\begin{tabular}{|l|l|c|c|c|c|}
\hline
\multicolumn{1}{|c|}{GRB} &\multicolumn{1}{c|}{} & \multicolumn{4}{c|}{Scolnic (2018)}\\
\cline{3-6}
\multicolumn{1}{|c|}{Sample} & & n = 1 & n = 2 & n = 1 ($47.5\%$ LCR) & n = 2 (47.5\% LCR) \\
\hline
& \# GRBs (Plateau) & 2718 & 1086 & 1291 & 515 \\ 
OPT & \# GRBs (Total) & 45964 & 18365 & 21833 & 8723 \\ 
& Year Achieved & 2197 & 2097 & 2140 & 2061 \\ 
\hline
& \# GRBs (Total) & 22982 & 9182 & 10916 & 4361 \\ 
OPT + ML & Year Achieved & 2113 & 2063 & 2069 & 2046 \\ 

\hline
& \# GRBs (Plateau) & 2870 & 1513 & 1363 & 718 \\ 
OPTtrim (10) & \# GRBs (Total) & 48534 & 23054 & 12153 & 19840 \\ 
 & Year Achieved & 2207 & 2123 & 2114 & 2074 \\ 
\hline
& \# GRBs (Total) & 24267 & 12793 & 11527 & 6076 \\ 
OPTtrim (10) + ML & Year Achieved & 2118 & 2076 & 2072 & 2052 \\

\hline
& \# GRBs (Plateau) & 2104 & 822 & 999 & 390\\ 
OPTtrim (25) & \# GRBs (Total) & 35580 & 13900 & 16900 & 6602 \\ 
& Year Achieved & 2159 & 2080 & 2091 & 2054 \\ 
\hline
& \# GRBs (Total)& 17790 & 6950 & 8450 & 3301 \\ 
OPTtrim (25) + ML & Year Achieved & 2094 & 2055 & 2060 & 2042 \\ 
\hline
\end{tabular}
\end{center}
\end{table*}

The year achieved is, of course, closer to the present for a simulated number of GRBs whose likelihood errors have been divided by two. 
As for the \cite{2014A&A...568A..22B} limit, the earliest year in which we fall below the cutoff given current capabilities is in 2038 with the full OPT sample and 2056 for the full PLAT X-ray sample. It is in the decade or so that we expect to have enough observational data to use GRBs as standalone standard candles that produce errors on cosmological parameters less than or equivalent to those produced by SNe Ia measures. This time estimate is reduced, however, if we consider the applicability of current machine learning (ML) techniques including the LCR to effectively reduce the errorbars on the measurement of the fundamental plane relation and the redshift inference to double the PLAT sample of GRBs in X-ray that we can use for these calculations. Because such a small number of observed GRBs have had their redshift recorded ($\simeq \frac{1}{3}$), ML techniques have been developed by \cite{2019arXiv190705074D} to estimate the missing redshifts. Indeed, it is possible to determine the redshift of GRBs possessing the plateau emission by completely non parametric models or semi-parametric models or a combination of the two. Considering also this research line, we find that we reach the full number of observed GRBs in X-ray necessary to use them as standard candles by 2026 for the \cite{2011ApJS..192....1C} limit, and by 2043 for the \cite{2014A&A...568A..22B} limit. The limit found by \cite{2014A&A...568A..22B} with the optical trimmed a posteriori (20 GRB) sample can be achieved much earlier, in 2025 (in just 3 years from now), using the ML techniques.

Furthermore, because of the particular efficacy of the OPTtrim sample in the constraint of $\Omega_{\text{M}}$, we can estimate similarly the time frame in which the most recent and precise \cite{2018ApJ...859..101S} limit is achieved. From Table \ref{tab:Time Estimates scolnic}, considering both current statistical LCR and the redshift inference through the ML applications, GRBs will be just as ideal probes as SNe Ia reaching the \cite{2018ApJ...859..101S} limit in 2042, if we instead consider the limit of \cite{2014A&A...568A..22B} we are able to reach it with LCR and redshift inference in just 4 years, in 2026.
Both estimates are performed with the OPT trimmed a posteriori (25 GRBs).

We here point out that the value of $\Omega_{\text{M}}$ can be measured as a guide. This forecast, although based on the flat $\Lambda$CDM model, can be generalized.
 In principle, we can use the fundamental plane for any cosmological model, and we don’t need to restrict ourselves to a specific number of parameters. We can indeed use this data sample to constrain extended theories of gravity or more exotic cosmological models as well.
The methodology that we have demonstrated is very general, and can be applied to any future cosmological model. The forecast in this particular paper is restricted to the flat $\Lambda$CDM model, and we are interested in constraining only $\Omega_{\text{M}}$, as a benchmark quantity. Any additional forecasts are interesting, but beyond the scope of the current paper.

\section{Conclusions and Insights} \label{conclusions}

In this paper, we have defined a subsample of 222 GRBs with redshift measurements and LC plateaus from all 1064 GRBs detected by Swift-XRT, and in combination with SNe Ia measurements, we have calculated the matter content of the Universe today to be $\Omega_{\text{M}} = 0.299 \pm 0.009$. We arrive at this precision using GRB emission data in optical wavelengths as well, marking a significant step in the recognition and usability of optical plateaus studies as cosmological tools. We have applied a procedure according to which a subsample of GRBs closer to the plane has been chosen, so that we could reduce the scatter of the fundamental plane at the smallest possible. The Sec. \S \ref{sec:subample_test} results wherein we compare the distributions of the fluxes and times of the PLAT, OPT, PLATtrim and OPTtrim samples assert that picking a smaller number of GRBs does not bias the sample. The KS test shows that the hypothesis that the two samples are drawn by the same distribution cannot be rejected at the $5\%$ level for all cases, but for the optical sample for $T^{*}_a$. We also stress here that the cosmological computations performed do not suffer of the circularity problem when we add GRBs and SNe Ia together GRBs are not calibrated over SNe Ia and do not assume any a priori cosmological models. We found that by trimming the X-ray sample to a sample of 10 (a priori) or 20 GRBs (a posteriori) in combination with SNe Ia leads to the same error as with the full PLAT sample. The results were as follows: $\Omega_{\text{M}}=0.299 \pm 0.009$.

For both GRB samples we account for redshift evolutionary effects using the EP methodology, thus achieving the smallest intrinsic scatter on the X-ray 3D fundamental plane in the literature to date, yielding a $44.4\%$ reduction with respect to the one computed without considering these corrections. Employing this correction and retrieving the same value of $\Omega_{\text{M}}$ as stated above, we can be sure that this is the most precise determination of the density content yet. As was noted and is described in detail within the Appendix Sec. \S \ref{MCMC error}, all MCMC sampler errors have been quantified and determined to be an order of magnitude less than the order on which we compare the $\Omega_{\text{M}}$ symmetrized errors and standard deviations. This is the first time in which GRBs have been attempted to be used as standalone standard candles through simulations, with an inferred precision higher than the one obtained by SNe Ia samples alone once a sufficient number of simulated GRBs are accounted for. Because we do not currently have enough GRB data to achieve this goal, we run MCMC simulations to define a probabilistic minimum number needed to satisfy them.

Simulating first using the full PLAT and OPT samples as the base, we find that a minimum of 150 GRBs is needed to provide closed-contours around an $\Omega_{\text{M}}$ value in our prior interval. However, to arrive at a desirable error on this value comparable to the limit set by SNe Ia, we instead simulate using as a base the PLATtrim and OPTtrim samples of 10 GRBs, because they yield intrinsic scatters near-zero. For instance, the PLATtrim gives a $\sigmaintxtrim = 0.05 \pm 0.05$ (right upper panel of Fig. \ref{fig:PLAT_GRB+SNE}), which is significantly less, 86\% than that of the full PLAT, $\sigmaintxfull = 0.36 \pm 0.04$ (left upper panel of Fig. \ref{fig:PLAT_GRB+SNE}), or even that of the evolution-corrected samples; PLAT (EV) yields $\sigmaintxfullev = 0.20 \pm 0.06$ (44 \% less than the PLAT), and PLATtrim (EV) gives $\sigmaintxtrimev = 0.13 \pm 0.09$ (64\% less than the PLAT). 

We aim to seek how many GRBs are needed to yield a value of $\Omega_{\text{M}}$ with a very high precision. For undivided errors on PLATtrim, the probability of $\Omega_{\text{M}} = 0.299 \pm 0.035$ approaches for 2400 GRBs (Table \ref{tab:compare trims}); in addition only 2705 X-ray GRBs are needed to reach the \cite{2014A&A...568A..22B} standard deviation limit (Table \ref{Tab:Time Estimates conley,betoule}). Instead, with undivided errors on OPTtrim, only 829 optical GRBs are needed. Because we have to consider the proportionality of the sample, these estimates must be increased to account for the presence of the redshift and the plateau in the sample. Therefore, in the attempt to decrease the GRBs numbers further, we explore alternative trimming techniques. By such, we trim both the optical and X-ray samples a posteriori and find more compelling uncertainties on the fundamental planes. Specifically, dividing the likelihood errors by two on PLATtrim (a posteriori) we find a most-probable $\Omega_{\text{M}} = 0.302 \pm 0.027$ for 2700 GRBs, and with halved errors on OPTtrim (a posteriori), we see a most-probable $\Omega_{\text{M}} = 0.299 \pm 0.021$ for 2700 GRBs (see Table \ref{tab:compare trims}).
It is clear that the OPTtrim reaches an error bar on the $\Omega_{\text{M}}$ which is 22 \% smaller compared to the one achieved by the PLAT with the same number of simulated GRBs. This allows to safely state that the OPTtrim is more efficacious than the PLATtrim.

We achieve the \cite{2014A&A...568A..22B} limit with 1452 X-ray GRBs or 284 optical GRBs (see Table \ref{Tab:Time Estimates conley,betoule}).
We finally test for the more recent \cite{2018ApJ...859..101S} error cutoff of $\sigma = 0.022$, and find that the OPTtrim (a posteriori with 25 GRBs) sample can achieve this limit with 822 GRBs with plateau and halved error bars (see Table \ref{tab:Time Estimates scolnic}). Because of this, we look into deep-space survey missions that will be collecting this GRB data in coming years. By surveys such as SVOM, THESEUS, and the continued use of Swift, it is estimated that a number of 22486 GRBs will be gathered in their lifetimes. By factoring in the ability of machine learning techniques to derive redshifts and the successful halving of errors by the statistical reconstruction of GRB LCs (currently we can achieve this for $47.5\%$ of a sample), we predict to arrive at the \cite{2018ApJ...859..101S} limit by 2042, and to the \cite{2014A&A...568A..22B} limit with only 134 optical GRBs by 2026. This will be the time frame in which there will be enough observational data to effectively use GRBs as standalone standard candles, with SNe Ia comparable error deeming GRBs ideal cosmological probes.

These results are interesting because, as the definition of GRBs as standard candles becomes more and more reliable, the addition of these astrophysical objects to SNe Ia data will soon give the most precise derivation of $\Omega_{\text{M}}$ ever achieved. Furthermore, as it has been done previously \citep{Dainotti_2021_sub2}, we will also arrive at the most precise values for the dark energy parameter $w$, and for the Hubble Constant $H_{0}$. The inclusion of a larger X-ray PLAT sample or a platinum sample in optical to trim would decrease the intrinsic scatter of the plane even further and potentially lower our number of estimated GRBs. Not only this, but larger samples would allow for more constraining results on the redshift evolutionary effects. This would again increase the accuracies of our measurements. 
Furthermore, as we keep investigating the physics of GRB progenitors, we will be able to better define GRB classes, samples, and, thus, continue to improve the precision on our results as well as shed more and more light on the mechanism behind GRBs.

\section*{Acknowledgements}
This work is supported by JSPS Grants-in-Aid for Scientific Research “KAKENHI” (A: Grant Number JP19H00693). This work was supported in part by a RIKEN pioneering project ``Evolution of Matter in the Universe (r-EMU)''. This work made use of data supplied by the UK Swift Science Data Centre at the University of Leicester. V. Nielson acknowledges the support by the United States Department of Energy in funding the Science Undergraduate Laboratory Internship (SULI) program. We are grateful to S. Savastano for the very preliminary help in writing the python codes for the simulations, and to N. Hornby, B. De Simone, A. Lenart, S. Young, N. O'Shea, D. Levine, N. Osborn, D. Zhou, and Z. Kania for running some of the computations here shown. We are also grateful to A. Lenart and D. Levine for running the Efron \& Petrosian method on the optical and X-ray data. We also acknowledge the use of the RIKEN Hokusai BigWaterfall cluster for some of the simulations.

\section{Data availability}
The data underlying this article will be shared upon a reasonable request to the corresponding author.

\bibliographystyle{mnras}
\bibliography{newtables}

\begin{thebibliography}{}
\makeatletter
\relax
\def\mn@urlcharsother{\let\do\@makeother \do\$\do\&\do\#\do\^\do\_\do\%\do\~}
\def\mn@doi{\begingroup\mn@urlcharsother \@ifnextchar [ {\mn@doi@}
  {\mn@doi@[]}}
\def\mn@doi@[#1]#2{\def\@tempa{#1}\ifx\@tempa\@empty \href
  {http://dx.doi.org/#2} {doi:#2}\else \href {http://dx.doi.org/#2} {#1}\fi
  \endgroup}
\def\mn@eprint#1#2{\mn@eprint@#1:#2::\@nil}
\def\mn@eprint@arXiv#1{\href {http://arxiv.org/abs/#1} {{\tt arXiv:#1}}}
\def\mn@eprint@dblp#1{\href {http://dblp.uni-trier.de/rec/bibtex/#1.xml}
  {dblp:#1}}
\def\mn@eprint@#1:#2:#3:#4\@nil{\def\@tempa {#1}\def\@tempb {#2}\def\@tempc
  {#3}\ifx \@tempc \@empty \let \@tempc \@tempb \let \@tempb \@tempa \fi \ifx
  \@tempb \@empty \def\@tempb {arXiv}\fi \@ifundefined
  {mn@eprint@\@tempb}{\@tempb:\@tempc}{\expandafter \expandafter \csname
  mn@eprint@\@tempb\endcsname \expandafter{\@tempc}}}

\bibitem[\protect\citeauthoryear{{Abbott} et~al.,}{{Abbott}
  et~al.}{2018}]{2018MNRAS.480.3879A}
{Abbott} T.~M.~C.,  et~al., 2018, \mn@doi [\mnras] {10.1093/mnras/sty1939},
  \href {https://ui.adsabs.harvard.edu/abs/2018MNRAS.480.3879A} {480, 3879}

\bibitem[\protect\citeauthoryear{{Ajello} et~al.,}{{Ajello}
  et~al.}{2019}]{2019ApJ...878...52A}
{Ajello} M.,  et~al., 2019, \mn@doi [\apj] {10.3847/1538-4357/ab1d4e}, \href
  {https://ui.adsabs.harvard.edu/abs/2019ApJ...878...52A} {878, 52}

\bibitem[\protect\citeauthoryear{{Amati} et~al.,}{{Amati}
  et~al.}{2002a}]{2002MmSAI..73.1178A}
{Amati} L.,  et~al., 2002a, \memsai, \href
  {https://ui.adsabs.harvard.edu/abs/2002MmSAI..73.1178A} {73, 1178}

\bibitem[\protect\citeauthoryear{{Amati} et~al.,}{{Amati}
  et~al.}{2002b}]{2002A&A...390...81A}
{Amati} L.,  et~al., 2002b, \mn@doi [\aap] {10.1051/0004-6361:20020722}, \href
  {https://ui.adsabs.harvard.edu/abs/2002A&A...390...81A} {390, 81}

\bibitem[\protect\citeauthoryear{{Amati}, {Guidorzi}, {Frontera}, {Della
  Valle}, {Finelli}, {Landi}  \& {Montanari}}{{Amati}
  et~al.}{2008}]{2008MNRAS.391..577A}
{Amati} L.,  {Guidorzi} C.,  {Frontera} F.,  {Della Valle} M.,  {Finelli} F.,
  {Landi} R.,   {Montanari} E.,  2008, \mn@doi [\mnras]
  {10.1111/j.1365-2966.2008.13943.x}, \href
  {https://ui.adsabs.harvard.edu/abs/2008MNRAS.391..577A} {391, 577}

\bibitem[\protect\citeauthoryear{{Amati}, {O'Brien}, {Goetz}, {Tenzer}  \&
  {Bozzo}}{{Amati} et~al.}{2017}]{2017xru..conf..250A}
{Amati} L.,  {O'Brien} P.,  {Goetz} D.,  {Tenzer} C.,   {Bozzo} E.,  2017, in
  Proceedings of the conference~held 6-9~June 2017 in~Rome I.,  ed., The X-ray
  Universe 2017. p.~250

\bibitem[\protect\citeauthoryear{{Amati} et~al.,}{{Amati}
  et~al.}{2018}]{2018AdSpR..62..191A}
{Amati} L.,  et~al., 2018, \mn@doi [Advances in Space Research]
  {10.1016/j.asr.2018.03.010}, \href
  {https://ui.adsabs.harvard.edu/abs/2018AdSpR..62..191A} {62, 191}

\bibitem[\protect\citeauthoryear{Amati, D’Agostino, Luongo, Muccino  \&
  Tantalo}{Amati et~al.}{2019}]{Amati_2019}
Amati L.,  D’Agostino R.,  Luongo O.,  Muccino M.,   Tantalo M.,  2019,
  \mn@doi [Monthly Notices of the Royal Astronomical Society: Letters]
  {10.1093/mnrasl/slz056}, 486, L46–L51

\bibitem[\protect\citeauthoryear{{Amati} et~al.,}{{Amati}
  et~al.}{2021}]{2021arXiv210409531A}
{Amati} L.,  et~al., 2021, arXiv e-prints, \href
  {https://ui.adsabs.harvard.edu/abs/2021arXiv210409531A} {p. arXiv:2104.09531}

\bibitem[\protect\citeauthoryear{Bargiacchi, Risaliti, Benetti, Capozziello,
  Lusso, Saccardi  \& Signorini}{Bargiacchi et~al.}{2021}]{Bargiacchi}
Bargiacchi G.,  Risaliti G.,  Benetti M.,  Capozziello S.,  Lusso E.,  Saccardi
  A.,   Signorini M.,  2021, \mn@doi [Astron. Astrophys.]
  {10.1051/0004-6361/202140386}, 649, A65

\bibitem[\protect\citeauthoryear{{Beaton} \& {Carnegie-Chicago Hubble Program
  Team}}{{Beaton} \& {Carnegie-Chicago Hubble Program
  Team}}{2018}]{2018AAS...23135105B}
{Beaton} R.~L.,  {Carnegie-Chicago Hubble Program Team} 2018, in American
  Astronomical Society Meeting Abstracts \#231. p. 351.05

\bibitem[\protect\citeauthoryear{{Beniamini}, {Duque}, {Daigne}  \&
  {Mochkovitch}}{{Beniamini} et~al.}{2020}]{2020MNRAS.492.2847B}
{Beniamini} P.,  {Duque} R.,  {Daigne} F.,   {Mochkovitch} R.,  2020, \mn@doi
  [\mnras] {10.1093/mnras/staa070}, \href
  {https://ui.adsabs.harvard.edu/abs/2020MNRAS.492.2847B} {492, 2847}

\bibitem[\protect\citeauthoryear{{Bernardini}}{{Bernardini}}{2015}]{2015JHEAp...7...64B}
{Bernardini} M.~G.,  2015, \mn@doi [Journal of High Energy Astrophysics]
  {10.1016/j.jheap.2015.05.003}, \href
  {https://ui.adsabs.harvard.edu/abs/2015JHEAp...7...64B} {7, 64}

\bibitem[\protect\citeauthoryear{{Betoule} et~al.,}{{Betoule}
  et~al.}{2014}]{2014A&A...568A..22B}
{Betoule} M.,  et~al., 2014, \mn@doi [\aap] {10.1051/0004-6361/201423413},
  \href {https://ui.adsabs.harvard.edu/abs/2014A&A...568A..22B} {568, A22}

\bibitem[\protect\citeauthoryear{{Birrer} et~al.,}{{Birrer}
  et~al.}{2020}]{2020A&A...643A.165B}
{Birrer} S.,  et~al., 2020, \mn@doi [\aap] {10.1051/0004-6361/202038861}, \href
  {https://ui.adsabs.harvard.edu/abs/2020A&A...643A.165B} {643, A165}

\bibitem[\protect\citeauthoryear{{Bloom}, {Frail}  \& {Sari}}{{Bloom}
  et~al.}{2001}]{2001AJ....121.2879B}
{Bloom} J.~S.,  {Frail} D.~A.,   {Sari} R.,  2001, \mn@doi [\aj]
  {10.1086/321093}, \href
  {https://ui.adsabs.harvard.edu/abs/2001AJ....121.2879B} {121, 2879}

\bibitem[\protect\citeauthoryear{{Boella}, {Butler}, {Perola}, {Piro}, {Scarsi}
   \& {Bleeker}}{{Boella} et~al.}{1997}]{1997A&AS..122..299B}
{Boella} G.,  {Butler} R.~C.,  {Perola} G.~C.,  {Piro} L.,  {Scarsi} L.,
  {Bleeker} J.~A.~M.,  1997, \mn@doi [\aaps] {10.1051/aas:1997136}, \href
  {https://ui.adsabs.harvard.edu/abs/1997A&AS..122..299B} {122, 299}

\bibitem[\protect\citeauthoryear{{Bromberg}, {Nakar}, {Piran}  \&
  {Sari}}{{Bromberg} et~al.}{2013}]{2013ApJ...764..179B}
{Bromberg} O.,  {Nakar} E.,  {Piran} T.,   {Sari} R.,  2013, \mn@doi [\apj]
  {10.1088/0004-637X/764/2/179}, \href
  {https://ui.adsabs.harvard.edu/abs/2013ApJ...764..179B} {764, 179}

\bibitem[\protect\citeauthoryear{{Campana}, {Guidorzi}, {Tagliaferri},
  {Chincarini}, {Moretti}, {Rizzuto}  \& {Romano}}{{Campana}
  et~al.}{2007}]{2007A&A...472..395C}
{Campana} S.,  {Guidorzi} C.,  {Tagliaferri} G.,  {Chincarini} G.,  {Moretti}
  A.,  {Rizzuto} D.,   {Romano} P.,  2007, \mn@doi [\aap]
  {10.1051/0004-6361:20066984}, \href
  {https://ui.adsabs.harvard.edu/abs/2007A&A...472..395C} {472, 395}

\bibitem[\protect\citeauthoryear{{Cannizzo} \& {Gehrels}}{{Cannizzo} \&
  {Gehrels}}{2009}]{2009AAS...21361002C}
{Cannizzo} J.~K.,  {Gehrels} N.,  2009, in American Astronomical Society
  Meeting Abstracts \#213. p. 610.02

\bibitem[\protect\citeauthoryear{{Cannizzo} \& {Gehrels}}{{Cannizzo} \&
  {Gehrels}}{2010}]{2010HEAD...11.1404C}
{Cannizzo} J.~K.,  {Gehrels} N.,  2010, in AAS/High Energy Astrophysics
  Division \#11. p. 14.04

\bibitem[\protect\citeauthoryear{{Cao}, {Khadka}  \& {Ratra}}{{Cao}
  et~al.}{2021a}]{2021arXiv211014840C}
{Cao} S.,  {Khadka} N.,   {Ratra} B.,  2021a, arXiv e-prints, \href
  {https://ui.adsabs.harvard.edu/abs/2021arXiv211014840C} {p. arXiv:2110.14840}

\bibitem[\protect\citeauthoryear{{Cao}, {Ryan}  \& {Ratra}}{{Cao}
  et~al.}{2021b}]{2021MNRAS.504..300C}
{Cao} S.,  {Ryan} J.,   {Ratra} B.,  2021b, \mn@doi [\mnras]
  {10.1093/mnras/stab942}, \href
  {https://ui.adsabs.harvard.edu/abs/2021MNRAS.504..300C} {504, 300}

\bibitem[\protect\citeauthoryear{{Cao}, {Dainotti}  \& {Ratra}}{{Cao}
  et~al.}{2022}]{2022arXiv220105245C}
{Cao} S.,  {Dainotti} M.,   {Ratra} B.,  2022, arXiv e-prints, \href
  {https://ui.adsabs.harvard.edu/abs/2022arXiv220105245C} {p. arXiv:2201.05245}

\bibitem[\protect\citeauthoryear{Capozziello \& Izzo}{Capozziello \&
  Izzo}{2008}]{Izzo2}
Capozziello S.,  Izzo L.,  2008, \mn@doi [Astron. Astrophys.]
  {10.1051/0004-6361:200810337}, 490, 31

\bibitem[\protect\citeauthoryear{Capozziello \& Izzo}{Capozziello \&
  Izzo}{2010}]{Izzo}
Capozziello S.,  Izzo L.,  2010, \mn@doi [Astron. Astrophys.]
  {10.1051/0004-6361/201014522}, 519, A73

\bibitem[\protect\citeauthoryear{{Capozziello}, {Benetti}  \&
  {Spallicci}}{{Capozziello} et~al.}{2020}]{2020FoPh...50..893C}
{Capozziello} S.,  {Benetti} M.,   {Spallicci} A. D.~A.~M.,  2020, \mn@doi
  [Foundations of Physics] {10.1007/s10701-020-00356-2}, \href
  {https://ui.adsabs.harvard.edu/abs/2020FoPh...50..893C} {50, 893}

\bibitem[\protect\citeauthoryear{{Cardone}, {Capozziello}  \&
  {Dainotti}}{{Cardone} et~al.}{2009}]{2009MNRAS.400..775C}
{Cardone} V.~F.,  {Capozziello} S.,   {Dainotti} M.~G.,  2009, \mn@doi [\mnras]
  {10.1111/j.1365-2966.2009.15456.x}, \href
  {https://ui.adsabs.harvard.edu/abs/2009MNRAS.400..775C} {400, 775}

\bibitem[\protect\citeauthoryear{{Cardone}, {Dainotti}, {Capozziello}  \&
  {Willingale}}{{Cardone} et~al.}{2010}]{2010MNRAS.408.1181C}
{Cardone} V.~F.,  {Dainotti} M.~G.,  {Capozziello} S.,   {Willingale} R.,
  2010, \mn@doi [\mnras] {10.1111/j.1365-2966.2010.17197.x}, \href
  {https://ui.adsabs.harvard.edu/abs/2010MNRAS.408.1181C} {408, 1181}

\bibitem[\protect\citeauthoryear{{Carroll}}{{Carroll}}{2001}]{2001LRR.....4....1C}
{Carroll} S.~M.,  2001, \mn@doi [Living Reviews in Relativity]
  {10.12942/lrr-2001-1}, \href
  {https://ui.adsabs.harvard.edu/abs/2001LRR.....4....1C} {4, 1}

\bibitem[\protect\citeauthoryear{{Chen}, {Kumar}  \& {Ratra}}{{Chen}
  et~al.}{2017}]{2017ApJ...835...86C}
{Chen} Y.,  {Kumar} S.,   {Ratra} B.,  2017, \mn@doi [\apj]
  {10.3847/1538-4357/835/1/86}, \href
  {https://ui.adsabs.harvard.edu/abs/2017ApJ...835...86C} {835, 86}

\bibitem[\protect\citeauthoryear{{Collazzi}, {Schaefer}, {Goldstein}  \&
  {Preece}}{{Collazzi} et~al.}{2012}]{2012ApJ...747...39C}
{Collazzi} A.~C.,  {Schaefer} B.~E.,  {Goldstein} A.,   {Preece} R.~D.,  2012,
  \mn@doi [\apj] {10.1088/0004-637X/747/1/39}, \href
  {https://ui.adsabs.harvard.edu/abs/2012ApJ...747...39C} {747, 39}

\bibitem[\protect\citeauthoryear{{Conley} et~al.,}{{Conley}
  et~al.}{2011}]{2011ApJS..192....1C}
{Conley} A.,  et~al., 2011, \mn@doi [\apjs] {10.1088/0067-0049/192/1/1}, \href
  {https://ui.adsabs.harvard.edu/abs/2011ApJS..192....1C} {192, 1}

\bibitem[\protect\citeauthoryear{{Cordier}, {G{\"o}tz}, {Motch}  \& {SVOM
  Collaboration}}{{Cordier} et~al.}{2018}]{2018MmSAI..89..266C}
{Cordier} B.,  {G{\"o}tz} D.,  {Motch} C.,   {SVOM Collaboration} 2018,
  \memsai, \href {https://ui.adsabs.harvard.edu/abs/2018MmSAI..89..266C} {89,
  266}

\bibitem[\protect\citeauthoryear{{Cucchiara} et~al.,}{{Cucchiara}
  et~al.}{2011}]{2011ApJ...736....7C}
{Cucchiara} A.,  et~al., 2011, \mn@doi [\apj] {10.1088/0004-637X/736/1/7},
  \href {https://ui.adsabs.harvard.edu/abs/2011ApJ...736....7C} {736, 7}

\bibitem[\protect\citeauthoryear{{D'Agostini}}{{D'Agostini}}{2005}]{2005physics..11182D}
{D'Agostini} G.,  2005, arXiv e-prints, \href
  {https://ui.adsabs.harvard.edu/abs/2005physics..11182D} {p. physics/0511182}

\bibitem[\protect\citeauthoryear{{Dainotti}}{{Dainotti}}{2019}]{2019gbcc.book.....D}
{Dainotti} M.,  2019, {Gamma-ray Burst Correlations; Current status and open
  questions}.
Institute of Physics Publishing, \mn@doi{10.1088/2053-2563/aae15c}

\bibitem[\protect\citeauthoryear{{Dainotti} \& {Amati}}{{Dainotti} \&
  {Amati}}{2018}]{2018PASP..130e1001D}
{Dainotti} M.~G.,  {Amati} L.,  2018, \mn@doi [\pasp]
  {10.1088/1538-3873/aaa8d7}, \href
  {https://ui.adsabs.harvard.edu/abs/2018PASP..130e1001D} {130, 051001}

\bibitem[\protect\citeauthoryear{{Dainotti} \& {Del Vecchio}}{{Dainotti} \&
  {Del Vecchio}}{2017}]{2017NewAR..77...23D}
{Dainotti} M.~G.,  {Del Vecchio} R.,  2017, \mn@doi [\nar]
  {10.1016/j.newar.2017.04.001}, \href
  {https://ui.adsabs.harvard.edu/abs/2017NewAR..77...23D} {77, 23}

\bibitem[\protect\citeauthoryear{{Dainotti}, {Cardone}  \&
  {Capozziello}}{{Dainotti} et~al.}{2008}]{2008MNRAS.391L..79D}
{Dainotti} M.~G.,  {Cardone} V.~F.,   {Capozziello} S.,  2008, \mn@doi [\mnras]
  {10.1111/j.1745-3933.2008.00560.x}, \href
  {https://ui.adsabs.harvard.edu/abs/2008MNRAS.391L..79D} {391, L79}

\bibitem[\protect\citeauthoryear{{Dainotti}, {Ostrowski}  \&
  {Willingale}}{{Dainotti} et~al.}{2011a}]{2011MNRAS.418.2202D}
{Dainotti} M.~G.,  {Ostrowski} M.,   {Willingale} R.,  2011a, \mn@doi [\mnras]
  {10.1111/j.1365-2966.2011.19433.x}, \href
  {https://ui.adsabs.harvard.edu/abs/2011MNRAS.418.2202D} {418, 2202}

\bibitem[\protect\citeauthoryear{{Dainotti}, {Cardone}, {Capozziello},
  {Ostrowski}  \& {Willingale}}{{Dainotti} et~al.}{2011b}]{2011ApJ...730..135D}
{Dainotti} M.~G.,  {Cardone} F.~V.,  {Capozziello} S.,  {Ostrowski} M.,
  {Willingale} R.,  2011b, \mn@doi [\apj] {10.1088/0004-637X/730/2/135}, \href
  {https://ui.adsabs.harvard.edu/abs/2011ApJ...730..135D} {730, 135}

\bibitem[\protect\citeauthoryear{{Dainotti}, {Cardone}, {Piedipalumbo}  \&
  {Capozziello}}{{Dainotti} et~al.}{2013a}]{2013MNRAS.436...82D}
{Dainotti} M.~G.,  {Cardone} V.~F.,  {Piedipalumbo} E.,   {Capozziello} S.,
  2013a, \mn@doi [\mnras] {10.1093/mnras/stt1516}, \href
  {https://ui.adsabs.harvard.edu/abs/2013MNRAS.436...82D} {436, 82}

\bibitem[\protect\citeauthoryear{{Dainotti}, {Petrosian}, {Singal}  \&
  {Ostrowski}}{{Dainotti} et~al.}{2013b}]{2013ApJ...774..157D}
{Dainotti} M.~G.,  {Petrosian} V.,  {Singal} J.,   {Ostrowski} M.,  2013b,
  \mn@doi [\apj] {10.1088/0004-637X/774/2/157}, \href
  {https://ui.adsabs.harvard.edu/abs/2013ApJ...774..157D} {774, 157}

\bibitem[\protect\citeauthoryear{{Dainotti}, {Petrosian}, {Willingale},
  {O'Brien}, {Ostrowski}  \& {Nagataki}}{{Dainotti}
  et~al.}{2015a}]{2015MNRAS.451.3898D}
{Dainotti} M.,  {Petrosian} V.,  {Willingale} R.,  {O'Brien} P.,  {Ostrowski}
  M.,   {Nagataki} S.,  2015a, \mn@doi [\mnras] {10.1093/mnras/stv1229}, \href
  {https://ui.adsabs.harvard.edu/abs/2015MNRAS.451.3898D} {451, 3898}

\bibitem[\protect\citeauthoryear{{Dainotti}, {Del Vecchio}, {Shigehiro}  \&
  {Capozziello}}{{Dainotti} et~al.}{2015b}]{2015ApJ...800...31D}
{Dainotti} M.~G.,  {Del Vecchio} R.,  {Shigehiro} N.,   {Capozziello} S.,
  2015b, \mn@doi [\apj] {10.1088/0004-637X/800/1/31}, \href
  {https://ui.adsabs.harvard.edu/abs/2015ApJ...800...31D} {800, 31}

\bibitem[\protect\citeauthoryear{{Dainotti}, {Postnikov}, {Hernandez}  \&
  {Ostrowski}}{{Dainotti} et~al.}{2016}]{2016ApJ...825L..20D}
{Dainotti} M.~G.,  {Postnikov} S.,  {Hernandez} X.,   {Ostrowski} M.,  2016,
  \mn@doi [\apjl] {10.3847/2041-8205/825/2/L20}, \href
  {https://ui.adsabs.harvard.edu/abs/2016ApJ...825L..20D} {825, L20}

\bibitem[\protect\citeauthoryear{{Dainotti}, {Willingale}, {Capozziello},
  {Cardone}  \& {Ostrowski}}{{Dainotti} et~al.}{2017a}]{2011AIPC.1358..113D}
{Dainotti} M.~G.,  {Willingale} R.,  {Capozziello} S.,  {Cardone} V.~F.,
  {Ostrowski} M.,  2017a, \mn@doi [AIP Conference Proceedings]
  {10.1063/1.3621750}, \href
  {https://ui.adsabs.harvard.edu/abs/2011AIPC.1358..113D} {}

\bibitem[\protect\citeauthoryear{{Dainotti}, {Nagataki}, {Maeda}, {Postnikov}
  \& {Pian}}{{Dainotti} et~al.}{2017b}]{2017A&A...600A..98D}
{Dainotti} M.~G.,  {Nagataki} S.,  {Maeda} K.,  {Postnikov} S.,   {Pian} E.,
  2017b, \mn@doi [\aap] {10.1051/0004-6361/201628384}, \href
  {https://ui.adsabs.harvard.edu/abs/2017A&A...600A..98D} {600, A98}

\bibitem[\protect\citeauthoryear{{Dainotti}, {Hernandez}, {Postnikov},
  {Nagataki}, {O'brien}, {Willingale}  \& {Striegel}}{{Dainotti}
  et~al.}{2017c}]{2017ApJ...848...88D}
{Dainotti} M.~G.,  {Hernandez} X.,  {Postnikov} S.,  {Nagataki} S.,  {O'brien}
  P.,  {Willingale} R.,   {Striegel} S.,  2017c, \mn@doi [\apj]
  {10.3847/1538-4357/aa8a6b}, \href
  {https://ui.adsabs.harvard.edu/abs/2017ApJ...848...88D} {848, 88}

\bibitem[\protect\citeauthoryear{{Dainotti}, {Del Vecchio}  \&
  {Tarnopolski}}{{Dainotti} et~al.}{2018}]{2018AdAst2018E...1D}
{Dainotti} M.~G.,  {Del Vecchio} R.,   {Tarnopolski} M.,  2018, \mn@doi
  [Advances in Astronomy] {10.1155/2018/4969503}, \href
  {https://ui.adsabs.harvard.edu/abs/2018AdAst2018E...1D} {2018, 4969503}

\bibitem[\protect\citeauthoryear{{Dainotti} et~al.,}{{Dainotti}
  et~al.}{2019}]{2019arXiv190705074D}
{Dainotti} M.,  et~al., 2019, arXiv e-prints, \href
  {https://ui.adsabs.harvard.edu/abs/2019arXiv190705074D} {p. arXiv:1907.05074}

\bibitem[\protect\citeauthoryear{{Dainotti}, {Lenart}, {Sarracino}, {Nagataki},
  {Capozziello}  \& {Fraija}}{{Dainotti} et~al.}{2020a}]{2020ApJ...904...97D}
{Dainotti} M.~G.,  {Lenart} A.~{\L}.,  {Sarracino} G.,  {Nagataki} S.,
  {Capozziello} S.,   {Fraija} N.,  2020a, \mn@doi [\apj]
  {10.3847/1538-4357/abbe8a}, \href
  {https://ui.adsabs.harvard.edu/abs/2020ApJ...904...97D} {904, 97}

\bibitem[\protect\citeauthoryear{{Dainotti} et~al.,}{{Dainotti}
  et~al.}{2020b}]{2020ApJ...905L..26D}
{Dainotti} M.~G.,  et~al., 2020b, \mn@doi [\apjl] {10.3847/2041-8213/abcda9},
  \href {https://ui.adsabs.harvard.edu/abs/2020ApJ...905L..26D} {905, L26}

\bibitem[\protect\citeauthoryear{{Dainotti}, {Lenart}, {Sarracino},
  {Fernandez}, {Nagataki}, {Fraija}, {Capozziello}  \& M.}{{Dainotti}
  et~al.}{2021a}]{Dainotti_2021_sub2}
{Dainotti} M.~G.,  {Lenart} A.~L.,  {Sarracino} G.,  {Fernandez} J.,
  {Nagataki} S.,  {Fraija} N.,  {Capozziello} S.,   M. B.,  2021a, Submitted to
  MNRAS

\bibitem[\protect\citeauthoryear{{Dainotti}, {Levine}, {Fraija}  \&
  {Chandra}}{{Dainotti} et~al.}{2021b}]{2021Galax...9...95D}
{Dainotti} M.,  {Levine} D.,  {Fraija} N.,   {Chandra} P.,  2021b, \mn@doi
  [Galaxies] {10.3390/galaxies9040095}, \href
  {https://ui.adsabs.harvard.edu/abs/2021Galax...9...95D} {9, 95}

\bibitem[\protect\citeauthoryear{{Dainotti}, {Lenart}, {Fraija}, {Nagataki},
  {Warren}, {De Simone}, {Srinivasaragavan}  \& {Mata}}{{Dainotti}
  et~al.}{2021c}]{2021PASJ...73..970D}
{Dainotti} M.~G.,  {Lenart} A.~{\L}.,  {Fraija} N.,  {Nagataki} S.,  {Warren}
  D.~C.,  {De Simone} B.,  {Srinivasaragavan} G.,   {Mata} A.,  2021c, \mn@doi
  [\pasj] {10.1093/pasj/psab057}, \href
  {https://ui.adsabs.harvard.edu/abs/2021PASJ...73..970D} {73, 970}

\bibitem[\protect\citeauthoryear{{Dainotti} et~al.,}{{Dainotti}
  et~al.}{2021d}]{2021ApJS..255...13D}
{Dainotti} M.~G.,  et~al., 2021d, \mn@doi [\apjs] {10.3847/1538-4365/abfe17},
  \href {https://ui.adsabs.harvard.edu/abs/2021ApJS..255...13D} {255, 13}

\bibitem[\protect\citeauthoryear{{Dainotti}, {De Simone}, {Schiavone},
  {Montani}, {Rinaldi}  \& {Lambiase}}{{Dainotti}
  et~al.}{2021e}]{2021ApJ...912..150D}
{Dainotti} M.~G.,  {De Simone} B.,  {Schiavone} T.,  {Montani} G.,  {Rinaldi}
  E.,   {Lambiase} G.,  2021e, \mn@doi [\apj] {10.3847/1538-4357/abeb73}, \href
  {https://ui.adsabs.harvard.edu/abs/2021ApJ...912..150D} {912, 150}

\bibitem[\protect\citeauthoryear{{Dainotti}, {Bargiacchi}, {Lenart},
  {Capozziello}, {Colgain}, {Solomon}, {Stojkovic}  \&
  {Sheikh-Jabbari}}{{Dainotti} et~al.}{2022a}]{ep2022}
{Dainotti} M.~G.,  {Bargiacchi} G.,  {Lenart} A.~L.,  {Capozziello} S.,
  {Colgain} E.~O.,  {Solomon} R.,  {Stojkovic} D.,   {Sheikh-Jabbari} M.,
  2022a

\bibitem[\protect\citeauthoryear{{Dainotti} et~al.,}{{Dainotti}
  et~al.}{2022b}]{2022arXiv220312908D}
{Dainotti} M.~G.,  et~al., 2022b, arXiv e-prints, \href
  {https://ui.adsabs.harvard.edu/abs/2022arXiv220312908D} {p. arXiv:2203.12908}

\bibitem[\protect\citeauthoryear{{Dainotti}, {De Simone}, {Schiavone},
  {Montani}, {Rinaldi}, {Lambiase}, {Bogdan}  \& {Ugale}}{{Dainotti}
  et~al.}{2022c}]{2022Galax..10...24D}
{Dainotti} M.~G.,  {De Simone} B.,  {Schiavone} T.,  {Montani} G.,  {Rinaldi}
  E.,  {Lambiase} G.,  {Bogdan} M.,   {Ugale} S.,  2022c, \mn@doi [Galaxies]
  {doi:10.3390/galaxies10010024}, 10, 24

\bibitem[\protect\citeauthoryear{{Del Vecchio}, {Dainotti}  \&
  {Ostrowski}}{{Del Vecchio} et~al.}{2016}]{2016ApJ...828...36D}
{Del Vecchio} R.,  {Dainotti} M.~G.,   {Ostrowski} M.,  2016, \mn@doi [\apj]
  {10.3847/0004-637X/828/1/36}, \href
  {https://ui.adsabs.harvard.edu/abs/2016ApJ...828...36D} {828, 36}

\bibitem[\protect\citeauthoryear{{Demianski}, {Piedipalumbo}, {Rubano}  \&
  {Scudellaro}}{{Demianski} et~al.}{2012}]{2012MNRAS.426.1396D}
{Demianski} M.,  {Piedipalumbo} E.,  {Rubano} C.,   {Scudellaro} P.,  2012,
  \mn@doi [\mnras] {10.1111/j.1365-2966.2012.21568.x}, \href
  {https://ui.adsabs.harvard.edu/abs/2012MNRAS.426.1396D} {426, 1396}

\bibitem[\protect\citeauthoryear{{Demianski}, {Piedipalumbo}, {Sawant}  \&
  {Amati}}{{Demianski} et~al.}{2017a}]{2017A&A...598A.112D}
{Demianski} M.,  {Piedipalumbo} E.,  {Sawant} D.,   {Amati} L.,  2017a, \mn@doi
  [\aap] {10.1051/0004-6361/201628909}, \href
  {https://ui.adsabs.harvard.edu/abs/2017A&A...598A.112D} {598, A112}

\bibitem[\protect\citeauthoryear{{Demianski}, {Piedipalumbo}, {Sawant}  \&
  {Amati}}{{Demianski} et~al.}{2017b}]{2017A&A...598A.113D}
{Demianski} M.,  {Piedipalumbo} E.,  {Sawant} D.,   {Amati} L.,  2017b, \mn@doi
  [\aap] {10.1051/0004-6361/201628911}, \href
  {https://ui.adsabs.harvard.edu/abs/2017A&A...598A.113D} {598, A113}

\bibitem[\protect\citeauthoryear{{Eddington}}{{Eddington}}{1913}]{1913MNRAS..73..359E}
{Eddington} A.~S.,  1913, \mn@doi [\mnras] {10.1093/mnras/73.5.359}, \href
  {https://ui.adsabs.harvard.edu/abs/1913MNRAS..73..359E} {73, 359}

\bibitem[\protect\citeauthoryear{{Efron} \& {Petrosian}}{{Efron} \&
  {Petrosian}}{1992}]{1992ApJ...399..345E}
{Efron} B.,  {Petrosian} V.,  1992, \mn@doi [\apj] {10.1086/171931}, \href
  {https://ui.adsabs.harvard.edu/abs/1992ApJ...399..345E} {399, 345}

\bibitem[\protect\citeauthoryear{{Efstathiou}}{{Efstathiou}}{2020}]{2020arXiv200710716E}
{Efstathiou} G.,  2020, arXiv e-prints, \href
  {https://ui.adsabs.harvard.edu/abs/2020arXiv200710716E} {p. arXiv:2007.10716}

\bibitem[\protect\citeauthoryear{{Evans} et~al.,}{{Evans}
  et~al.}{2009}]{Evans2009}
{Evans} P.~A.,  et~al., 2009, \mn@doi [Monthly Notices of the Royal
  Astronomical Society] {10.1111/j.1365-2966.2009.14913.x}, \href
  {https://ui.adsabs.harvard.edu/abs/2009MNRAS.397.1177E} {397, 1177}

\bibitem[\protect\citeauthoryear{{Fana Dirirsa} et~al.,}{{Fana Dirirsa}
  et~al.}{2019}]{2019ApJ...887...13F}
{Fana Dirirsa} F.,  et~al., 2019, \mn@doi [\apj] {10.3847/1538-4357/ab4e11},
  \href {https://ui.adsabs.harvard.edu/abs/2019ApJ...887...13F} {887, 13}

\bibitem[\protect\citeauthoryear{{Foreman-Mackey}, {Hogg}, {Lang}  \&
  {Goodman}}{{Foreman-Mackey} et~al.}{2013}]{2013PASP..125..306F}
{Foreman-Mackey} D.,  {Hogg} D.~W.,  {Lang} D.,   {Goodman} J.,  2013, \mn@doi
  [\pasp] {10.1086/670067}, \href
  {https://ui.adsabs.harvard.edu/abs/2013PASP..125..306F} {125, 306}

\bibitem[\protect\citeauthoryear{{Freedman}}{{Freedman}}{2021}]{2021ApJ...919...16F}
{Freedman} W.~L.,  2021, \mn@doi [\apj] {10.3847/1538-4357/ac0e95}, \href
  {https://ui.adsabs.harvard.edu/abs/2021ApJ...919...16F} {919, 16}

\bibitem[\protect\citeauthoryear{{Freedman} et~al.,}{{Freedman}
  et~al.}{2020}]{2020ApJ...891...57F}
{Freedman} W.~L.,  et~al., 2020, \mn@doi [\apj] {10.3847/1538-4357/ab7339},
  \href {https://ui.adsabs.harvard.edu/abs/2020ApJ...891...57F} {891, 57}

\bibitem[\protect\citeauthoryear{{Frontera} et~al.,}{{Frontera}
  et~al.}{2018}]{2018MmSAI..89..157F}
{Frontera} F.,  et~al., 2018, \memsai, \href
  {https://ui.adsabs.harvard.edu/abs/2018MmSAI..89..157F} {89, 157}

\bibitem[\protect\citeauthoryear{{Gendre} et~al.,}{{Gendre}
  et~al.}{2013}]{2013arXiv1308.1001G}
{Gendre} B.,  et~al., 2013, arXiv e-prints, \href
  {https://ui.adsabs.harvard.edu/abs/2013arXiv1308.1001G} {p. arXiv:1308.1001}

\bibitem[\protect\citeauthoryear{{Ghirlanda}, {Ghisellini}  \&
  {Lazzati}}{{Ghirlanda} et~al.}{2004}]{2004ApJ...616..331G}
{Ghirlanda} G.,  {Ghisellini} G.,   {Lazzati} D.,  2004, \mn@doi [\apj]
  {10.1086/424913}, \href
  {https://ui.adsabs.harvard.edu/abs/2004ApJ...616..331G} {616, 331}

\bibitem[\protect\citeauthoryear{Ghirlanda, Nava, Ghisellini  \&
  Firmani}{Ghirlanda et~al.}{2007}]{Ghirlanda_2007}
Ghirlanda G.,  Nava L.,  Ghisellini G.,   Firmani C.,  2007, \mn@doi [Astronomy
  \& Astrophysics] {10.1051/0004-6361:20077119}, 466, 127–136

\bibitem[\protect\citeauthoryear{{Gompertz}, {van der Horst}, {O'Brien}, {Wynn}
   \& {Wiersema}}{{Gompertz} et~al.}{2015}]{2015MNRAS.448..629G}
{Gompertz} B.~P.,  {van der Horst} A.~J.,  {O'Brien} P.~T.,  {Wynn} G.~A.,
  {Wiersema} K.,  2015, \mn@doi [\mnras] {10.1093/mnras/stu2752}, \href
  {https://ui.adsabs.harvard.edu/abs/2015MNRAS.448..629G} {448, 629}

\bibitem[\protect\citeauthoryear{{Grupe}, {Nousek}, {Veres}, {Zhang}  \&
  {Gehrels}}{{Grupe} et~al.}{2013}]{2013ApJS..209...20G}
{Grupe} D.,  {Nousek} J.~A.,  {Veres} P.,  {Zhang} B.-B.,   {Gehrels} N.,
  2013, \mn@doi [\apjs] {10.1088/0067-0049/209/2/20}, \href
  {https://ui.adsabs.harvard.edu/abs/2013ApJS..209...20G} {209, 20}

\bibitem[\protect\citeauthoryear{{Hu}, {Wang}  \& {Dai}}{{Hu}
  et~al.}{2021}]{2021MNRAS.507..730H}
{Hu} J.~P.,  {Wang} F.~Y.,   {Dai} Z.~G.,  2021, \mn@doi [\mnras]
  {10.1093/mnras/stab2180}, \href
  {https://ui.adsabs.harvard.edu/abs/2021MNRAS.507..730H} {507, 730}

\bibitem[\protect\citeauthoryear{Kendall}{Kendall}{1938}]{10.1093/biomet/30.1-2.81}
Kendall M.~G.,  1938, \mn@doi [Biometrika] {10.1093/biomet/30.1-2.81}, 30, 81

\bibitem[\protect\citeauthoryear{{Khadka} \& {Ratra}}{{Khadka} \&
  {Ratra}}{2020}]{2020MNRAS.499..391K}
{Khadka} N.,  {Ratra} B.,  2020, \mn@doi [\mnras] {10.1093/mnras/staa2779},
  \href {https://ui.adsabs.harvard.edu/abs/2020MNRAS.499..391K} {499, 391}

\bibitem[\protect\citeauthoryear{{Khadka} \& {Ratra}}{{Khadka} \&
  {Ratra}}{2021a}]{2021arXiv210707600K}
{Khadka} N.,  {Ratra} B.,  2021a, arXiv e-prints, \href
  {https://ui.adsabs.harvard.edu/abs/2021arXiv210707600K} {p. arXiv:2107.07600}

\bibitem[\protect\citeauthoryear{{Khadka} \& {Ratra}}{{Khadka} \&
  {Ratra}}{2021b}]{2021MNRAS.502.6140K}
{Khadka} N.,  {Ratra} B.,  2021b, \mn@doi [\mnras] {10.1093/mnras/stab486},
  \href {https://ui.adsabs.harvard.edu/abs/2021MNRAS.502.6140K} {502, 6140}

\bibitem[\protect\citeauthoryear{{Khadka}, {Luongo}, {Muccino}  \&
  {Ratra}}{{Khadka} et~al.}{2021}]{2021JCAP...09..042K}
{Khadka} N.,  {Luongo} O.,  {Muccino} M.,   {Ratra} B.,  2021, \mn@doi [\jcap]
  {10.1088/1475-7516/2021/09/042}, \href
  {https://ui.adsabs.harvard.edu/abs/2021JCAP...09..042K} {2021, 042}

\bibitem[\protect\citeauthoryear{{Khetan} et~al.,}{{Khetan}
  et~al.}{2021}]{2021A&A...647A..72K}
{Khetan} N.,  et~al., 2021, \mn@doi [\aap] {10.1051/0004-6361/202039196}, \href
  {https://ui.adsabs.harvard.edu/abs/2021A&A...647A..72K} {647, A72}

\bibitem[\protect\citeauthoryear{{Knust} et~al.,}{{Knust}
  et~al.}{2017}]{2017A&A...607A..84K}
{Knust} F.,  et~al., 2017, \mn@doi [\aap] {10.1051/0004-6361/201730578}, \href
  {https://ui.adsabs.harvard.edu/abs/2017A&A...607A..84K} {607, A84}

\bibitem[\protect\citeauthoryear{{Kocevski} \& {Butler}}{{Kocevski} \&
  {Butler}}{2007}]{2007AAS...211.1801K}
{Kocevski} D.,  {Butler} N.,  2007, in American Astronomical Society Meeting
  Abstracts. p. 18.01

\bibitem[\protect\citeauthoryear{{Kodama}, {Yonetoku}, {Murakami}, {Tanabe},
  {Tsutsui}  \& {Nakamura}}{{Kodama} et~al.}{2008}]{2008MNRAS.391L...1K}
{Kodama} Y.,  {Yonetoku} D.,  {Murakami} T.,  {Tanabe} S.,  {Tsutsui} R.,
  {Nakamura} T.,  2008, \mn@doi [\mnras] {10.1111/j.1745-3933.2008.00508.x},
  \href {https://ui.adsabs.harvard.edu/abs/2008MNRAS.391L...1K} {391, L1}

\bibitem[\protect\citeauthoryear{{Kouveliotou}, {Meegan}, {Fishman}, {Bhat},
  {Briggs}, {Koshut}, {Paciesas}  \& {Pendleton}}{{Kouveliotou}
  et~al.}{1993}]{1993ApJ...413L.101K}
{Kouveliotou} C.,  {Meegan} C.~A.,  {Fishman} G.~J.,  {Bhat} N.~P.,  {Briggs}
  M.~S.,  {Koshut} T.~M.,  {Paciesas} W.~S.,   {Pendleton} G.~N.,  1993,
  \mn@doi [\apjl] {10.1086/186969}, \href
  {https://ui.adsabs.harvard.edu/abs/1993ApJ...413L.101K} {413, L101}

\bibitem[\protect\citeauthoryear{Lamb}{Lamb}{2003}]{Lamb_2003}
Lamb D.~Q.,  2003, \mn@doi [AIP Conference Proceedings] {10.1063/1.1579395}

\bibitem[\protect\citeauthoryear{{Levan}}{{Levan}}{2017}]{2017hst..prop15349L}
{Levan} A.,  2017, {From the longest GRBs to the brightest supernovae}, HST
  Proposal

\bibitem[\protect\citeauthoryear{{Levan} et~al.,}{{Levan}
  et~al.}{2007}]{2007MNRAS.378.1439L}
{Levan} A.~J.,  et~al., 2007, \mn@doi [\mnras]
  {10.1111/j.1365-2966.2007.11879.x}, \href
  {https://ui.adsabs.harvard.edu/abs/2007MNRAS.378.1439L} {378, 1439}

\bibitem[\protect\citeauthoryear{{Leventis}, {Wijers}  \& {van der
  Horst}}{{Leventis} et~al.}{2014}]{2014MNRAS.437.2448L}
{Leventis} K.,  {Wijers} R.~A.~M.~J.,   {van der Horst} A.~J.,  2014, \mn@doi
  [\mnras] {10.1093/mnras/stt2055}, \href
  {https://ui.adsabs.harvard.edu/abs/2014MNRAS.437.2448L} {437, 2448}

\bibitem[\protect\citeauthoryear{{Levine}, {Dainotti}, {Zvonarek}, {Fraija},
  {Warren}, {Chandra}  \& {Lloyd-Ronning}}{{Levine}
  et~al.}{2022}]{2022ApJ...925...15L}
{Levine} D.,  {Dainotti} M.,  {Zvonarek} K.~J.,  {Fraija} N.,  {Warren} D.~C.,
  {Chandra} P.,   {Lloyd-Ronning} N.,  2022, \mn@doi [\apj]
  {10.3847/1538-4357/ac4221}, \href
  {https://ui.adsabs.harvard.edu/abs/2022ApJ...925...15L} {925, 15}

\bibitem[\protect\citeauthoryear{{Li}, {Wu}, {Lei}, {Dai}, {Liang}  \&
  {Ryde}}{{Li} et~al.}{2018}]{2018ApJS..236...26L}
{Li} L.,  {Wu} X.-F.,  {Lei} W.-H.,  {Dai} Z.-G.,  {Liang} E.-W.,   {Ryde} F.,
  2018, \mn@doi [\apjs] {10.3847/1538-4365/aabaf3}, \href
  {https://ui.adsabs.harvard.edu/abs/2018ApJS..236...26L} {236, 26}

\bibitem[\protect\citeauthoryear{{Liang} \& {Zhang}}{{Liang} \&
  {Zhang}}{2005}]{2005ApJ...633..611L}
{Liang} E.,  {Zhang} B.,  2005, \mn@doi [\apj] {10.1086/491594}, \href
  {https://ui.adsabs.harvard.edu/abs/2005ApJ...633..611L} {633, 611}

\bibitem[\protect\citeauthoryear{{Lin} \& {Ishak}}{{Lin} \&
  {Ishak}}{2021}]{2021JCAP...05..009L}
{Lin} W.,  {Ishak} M.,  2021, \mn@doi [\jcap] {10.1088/1475-7516/2021/05/009},
  \href {https://ui.adsabs.harvard.edu/abs/2021JCAP...05..009L} {2021, 009}

\bibitem[\protect\citeauthoryear{{Lloyd}}{{Lloyd}}{2000}]{2000AAS...197.2506L}
{Lloyd} N.~M.,  2000, in American Astronomical Society Meeting Abstracts. p.
  25.06

\bibitem[\protect\citeauthoryear{{L{\"u}}, {Zhang}, {Liang}, {Zhang}  \&
  {Sakamoto}}{{L{\"u}} et~al.}{2014}]{2014MNRAS.442.1922L}
{L{\"u}} H.-J.,  {Zhang} B.,  {Liang} E.-W.,  {Zhang} B.-B.,   {Sakamoto} T.,
  2014, \mn@doi [\mnras] {10.1093/mnras/stu982}, \href
  {https://ui.adsabs.harvard.edu/abs/2014MNRAS.442.1922L} {442, 1922}

\bibitem[\protect\citeauthoryear{{L{\"u}}, {Zhang}, {Lei}, {Li}  \&
  {Lasky}}{{L{\"u}} et~al.}{2015}]{2015ApJ...805...89L}
{L{\"u}} H.-J.,  {Zhang} B.,  {Lei} W.-H.,  {Li} Y.,   {Lasky} P.~D.,  2015,
  \mn@doi [\apj] {10.1088/0004-637X/805/2/89}, \href
  {https://ui.adsabs.harvard.edu/abs/2015ApJ...805...89L} {805, 89}

\bibitem[\protect\citeauthoryear{{Luongo} \& {Muccino}}{{Luongo} \&
  {Muccino}}{2020}]{2020A&A...641A.174L}
{Luongo} O.,  {Muccino} M.,  2020, \mn@doi [\aap]
  {10.1051/0004-6361/202038264}, \href
  {https://ui.adsabs.harvard.edu/abs/2020A&A...641A.174L} {641, A174}

\bibitem[\protect\citeauthoryear{Lusso et~al.}{Lusso et~al.}{2020}]{Lusso}
Lusso E.,  et~al., 2020, \mn@doi [Astron. Astrophys.]
  {10.1051/0004-6361/202038899}, 642, A150

\bibitem[\protect\citeauthoryear{{MacFadyen} \& {Woosley}}{{MacFadyen} \&
  {Woosley}}{1999}]{1999ApJ...524..262M}
{MacFadyen} A.~I.,  {Woosley} S.~E.,  1999, \mn@doi [\apj] {10.1086/307790},
  \href {https://ui.adsabs.harvard.edu/abs/1999ApJ...524..262M} {524, 262}

\bibitem[\protect\citeauthoryear{{MacFadyen}, {Woosley}  \&
  {Heger}}{{MacFadyen} et~al.}{2001}]{2001ApJ...550..410M}
{MacFadyen} A.~I.,  {Woosley} S.~E.,   {Heger} A.,  2001, \mn@doi [\apj]
  {10.1086/319698}, \href
  {https://ui.adsabs.harvard.edu/abs/2001ApJ...550..410M} {550, 410}

\bibitem[\protect\citeauthoryear{{Malmquist}}{{Malmquist}}{1922}]{1922MeLuF.100....1M}
{Malmquist} K.~G.,  1922, Meddelanden fran Lunds Astronomiska Observatorium
  Serie I, \href {https://ui.adsabs.harvard.edu/abs/1922MeLuF.100....1M} {100,
  1}

\bibitem[\protect\citeauthoryear{{Mazets} et~al.,}{{Mazets}
  et~al.}{1981}]{1981Ap&SS..80....3M}
{Mazets} E.~P.,  et~al., 1981, \mn@doi [\apss] {10.1007/BF00649140}, \href
  {https://ui.adsabs.harvard.edu/abs/1981Ap&SS..80....3M} {80, 3}

\bibitem[\protect\citeauthoryear{{Norris} \& {Bonnell}}{{Norris} \&
  {Bonnell}}{2006}]{2006ApJ...643..266N}
{Norris} J.~P.,  {Bonnell} J.~T.,  2006, \mn@doi [\apj] {10.1086/502796}, \href
  {https://ui.adsabs.harvard.edu/abs/2006ApJ...643..266N} {643, 266}

\bibitem[\protect\citeauthoryear{{Norris}, {Gehrels}  \& {Scargle}}{{Norris}
  et~al.}{2010}]{2010ApJ...717..411N}
{Norris} J.~P.,  {Gehrels} N.,   {Scargle} J.~D.,  2010, \mn@doi [\apj]
  {10.1088/0004-637X/717/1/411}, \href
  {https://ui.adsabs.harvard.edu/abs/2010ApJ...717..411N} {717, 411}

\bibitem[\protect\citeauthoryear{{O'Brien} \& {Willingale}}{{O'Brien} \&
  {Willingale}}{2007}]{2007Ap&SS.311..167O}
{O'Brien} P.~T.,  {Willingale} R.,  2007, \mn@doi [\apss]
  {10.1007/s10509-007-9551-3}, \href
  {https://ui.adsabs.harvard.edu/abs/2007Ap&SS.311..167O} {311, 167}

\bibitem[\protect\citeauthoryear{{Oates} et~al.,}{{Oates}
  et~al.}{2015}]{2015MNRAS.453.4121O}
{Oates} S.~R.,  et~al., 2015, \mn@doi [\mnras] {10.1093/mnras/stv1956}, \href
  {https://ui.adsabs.harvard.edu/abs/2015MNRAS.453.4121O} {453, 4121}

\bibitem[\protect\citeauthoryear{{Oates} et~al.,}{{Oates}
  et~al.}{2017}]{2017Galax...5....4O}
{Oates} S.,  et~al., 2017, \mn@doi [Galaxies] {10.3390/galaxies5010004}, \href
  {https://ui.adsabs.harvard.edu/abs/2017Galax...5....4O} {5, 4}

\bibitem[\protect\citeauthoryear{{Paczy{\'n}ski}}{{Paczy{\'n}ski}}{1998}]{1998ApJ...494L..45P}
{Paczy{\'n}ski} B.,  1998, \mn@doi [\apjl] {10.1086/311148}, \href
  {https://ui.adsabs.harvard.edu/abs/1998ApJ...494L..45P} {494, L45}

\bibitem[\protect\citeauthoryear{{Petrosian}, {Kitanidis}  \&
  {Kocevski}}{{Petrosian} et~al.}{2015}]{2015ApJ...806...44P}
{Petrosian} V.,  {Kitanidis} E.,   {Kocevski} D.,  2015, \mn@doi [\apj]
  {10.1088/0004-637X/806/1/44}, \href
  {https://ui.adsabs.harvard.edu/abs/2015ApJ...806...44P} {806, 44}

\bibitem[\protect\citeauthoryear{{Piro}, {Ricci}, {Wieringa}, {Bannister},
  {Troja}, {Gendre}, {Fiore}  \& {Piranomonte}}{{Piro}
  et~al.}{2014}]{2014atnf.prop.6334P}
{Piro} L.,  {Ricci} R.,  {Wieringa} M.,  {Bannister} K.,  {Troja} E.,  {Gendre}
  B.,  {Fiore} F.,   {Piranomonte} S.,  2014, {ATCA observations of the new
  class of ultralong GRBs: a local proxy of popIII explosions?}, ATNF Proposal

\bibitem[\protect\citeauthoryear{{Postnikov}, {Dainotti}, {Hernandez}  \&
  {Capozziello}}{{Postnikov} et~al.}{2014}]{2014ApJ...783..126P}
{Postnikov} S.,  {Dainotti} M.~G.,  {Hernandez} X.,   {Capozziello} S.,  2014,
  \mn@doi [\apj] {10.1088/0004-637X/783/2/126}, \href
  {https://ui.adsabs.harvard.edu/abs/2014ApJ...783..126P} {783, 126}

\bibitem[\protect\citeauthoryear{{Rea}, {Gull{\'o}n}, {Pons}, {Perna},
  {Dainotti}, {Miralles}  \& {Torres}}{{Rea}
  et~al.}{2015}]{2015ApJ...813...92R}
{Rea} N.,  {Gull{\'o}n} M.,  {Pons} J.~A.,  {Perna} R.,  {Dainotti} M.~G.,
  {Miralles} J.~A.,   {Torres} D.~F.,  2015, \mn@doi [\apj]
  {10.1088/0004-637X/813/2/92}, \href
  {https://ui.adsabs.harvard.edu/abs/2015ApJ...813...92R} {813, 92}

\bibitem[\protect\citeauthoryear{{Riess}, {Anderson}, {Breuval}, {Casertano},
  {Macri}, {Scolnic}  \& {Yuan}}{{Riess} et~al.}{2021}]{2021jwst.prop.1685R}
{Riess} A.,  {Anderson} R.~I.,  {Breuval} L.,  {Casertano} S.,  {Macri} L.~M.,
  {Scolnic} D.,   {Yuan} W.,  2021, {Uncrowding the Cepheids for an Improved
  Determination of the Hubble Constant}, JWST Proposal. Cycle 1

\bibitem[\protect\citeauthoryear{{Rodney} et~al.,}{{Rodney}
  et~al.}{2015}]{2015AJ....150..156R}
{Rodney} S.~A.,  et~al., 2015, \mn@doi [\aj] {10.1088/0004-6256/150/5/156},
  \href {https://ui.adsabs.harvard.edu/abs/2015AJ....150..156R} {150, 156}

\bibitem[\protect\citeauthoryear{{Rowlinson}, {O'Brien}, {Metzger}, {Tanvir}
  \& {Levan}}{{Rowlinson} et~al.}{2013}]{2013MNRAS.430.1061R}
{Rowlinson} A.,  {O'Brien} P.~T.,  {Metzger} B.~D.,  {Tanvir} N.~R.,   {Levan}
  A.~J.,  2013, \mn@doi [\mnras] {10.1093/mnras/sts683}, \href
  {https://ui.adsabs.harvard.edu/abs/2013MNRAS.430.1061R} {430, 1061}

\bibitem[\protect\citeauthoryear{{Rowlinson}, {Gompertz}, {Dainotti},
  {O'Brien}, {Wijers}  \& {van der Horst}}{{Rowlinson}
  et~al.}{2014}]{2014MNRAS.443.1779R}
{Rowlinson} A.,  {Gompertz} B.~P.,  {Dainotti} M.,  {O'Brien} P.~T.,  {Wijers}
  R.~A.~M.~J.,   {van der Horst} A.~J.,  2014, \mn@doi [\mnras]
  {10.1093/mnras/stu1277}, \href
  {https://ui.adsabs.harvard.edu/abs/2014MNRAS.443.1779R} {443, 1779}

\bibitem[\protect\citeauthoryear{{Rowlinson}, {Patruno}  \&
  {O'Brien}}{{Rowlinson} et~al.}{2017}]{2017MNRAS.472.1152R}
{Rowlinson} A.,  {Patruno} A.,   {O'Brien} P.~T.,  2017, \mn@doi [\mnras]
  {10.1093/mnras/stx2023}, \href
  {https://ui.adsabs.harvard.edu/abs/2017MNRAS.472.1152R} {472, 1152}

\bibitem[\protect\citeauthoryear{{Sakamoto} et~al.,}{{Sakamoto}
  et~al.}{2007}]{2007AAS...210.1004S}
{Sakamoto} T.,  et~al., 2007, in American Astronomical Society Meeting
  Abstracts. p. 10.04

\bibitem[\protect\citeauthoryear{{Scolnic} et~al.,}{{Scolnic}
  et~al.}{2018}]{2018ApJ...859..101S}
{Scolnic} D.~M.,  et~al., 2018, \mn@doi [\apj] {10.3847/1538-4357/aab9bb},
  \href {https://ui.adsabs.harvard.edu/abs/2018ApJ...859..101S} {859, 101}

\bibitem[\protect\citeauthoryear{{Si} et~al.,}{{Si}
  et~al.}{2018}]{2018ApJ...863...50S}
{Si} S.-K.,  et~al., 2018, \mn@doi [\apj] {10.3847/1538-4357/aad08a}, \href
  {https://ui.adsabs.harvard.edu/abs/2018ApJ...863...50S} {863, 50}

\bibitem[\protect\citeauthoryear{{Srinivasaragavan}, {Dainotti}, {Fraija},
  {Hernandez}, {Nagataki}, {Lenart}, {Bowden}  \& {Wagner}}{{Srinivasaragavan}
  et~al.}{2020}]{2020ApJ...903...18S}
{Srinivasaragavan} G.~P.,  {Dainotti} M.~G.,  {Fraija} N.,  {Hernandez} X.,
  {Nagataki} S.,  {Lenart} A.,  {Bowden} L.,   {Wagner} R.,  2020, \mn@doi
  [\apj] {10.3847/1538-4357/abb702}, \href
  {https://ui.adsabs.harvard.edu/abs/2020ApJ...903...18S} {903, 18}

\bibitem[\protect\citeauthoryear{{Stratta}, {Dainotti}, {Dall'Osso},
  {Hernandez}  \& {De Cesare}}{{Stratta} et~al.}{2018}]{2018ApJ...869..155S}
{Stratta} G.,  {Dainotti} M.~G.,  {Dall'Osso} S.,  {Hernandez} X.,   {De
  Cesare} G.,  2018, \mn@doi [\apj] {10.3847/1538-4357/aadd8f}, \href
  {https://ui.adsabs.harvard.edu/abs/2018ApJ...869..155S} {869, 155}

\bibitem[\protect\citeauthoryear{{Tang}, {Huang}, {Geng}  \& {Zhang}}{{Tang}
  et~al.}{2019}]{2019ApJS..245....1T}
{Tang} C.-H.,  {Huang} Y.-F.,  {Geng} J.-J.,   {Zhang} Z.-B.,  2019, \mn@doi
  [\apjs] {10.3847/1538-4365/ab4711}, \href
  {https://ui.adsabs.harvard.edu/abs/2019ApJS..245....1T} {245, 1}

\bibitem[\protect\citeauthoryear{{Torrado} \& {Lewis}}{{Torrado} \&
  {Lewis}}{2019}]{2019ascl.soft10019T}
{Torrado} J.,  {Lewis} A.,  2019, Astrophysics Source Code Library, \href
  {https://ui.adsabs.harvard.edu/abs/2019ascl.soft10019T} {p. ascl:1910.019}

\bibitem[\protect\citeauthoryear{{Torrado} \& {Lewis}}{{Torrado} \&
  {Lewis}}{2021}]{2021JCAP...05..057T}
{Torrado} J.,  {Lewis} A.,  2021, \mn@doi [\jcap]
  {10.1088/1475-7516/2021/05/057}, \href
  {https://ui.adsabs.harvard.edu/abs/2021JCAP...05..057T} {2021, 057}

\bibitem[\protect\citeauthoryear{{Van Eerten}}{{Van
  Eerten}}{2014a}]{2014MNRAS.442.3495V}
{Van Eerten} H.,  2014a, \mn@doi [\mnras] {10.1093/mnras/stu1025}, \href
  {https://ui.adsabs.harvard.edu/abs/2014MNRAS.442.3495V} {442, 3495}

\bibitem[\protect\citeauthoryear{{Van Eerten}}{{Van
  Eerten}}{2014b}]{2014MNRAS.445.2414V}
{Van Eerten} H.~J.,  2014b, \mn@doi [\mnras] {10.1093/mnras/stu1921}, \href
  {https://ui.adsabs.harvard.edu/abs/2014MNRAS.445.2414V} {445, 2414}

\bibitem[\protect\citeauthoryear{{Varela} et~al.,}{{Varela}
  et~al.}{2016}]{2016A&A...589A..37V}
{Varela} K.,  et~al., 2016, \mn@doi [\aap] {10.1051/0004-6361/201526260}, \href
  {https://ui.adsabs.harvard.edu/abs/2016A&A...589A..37V} {589, A37}

\bibitem[\protect\citeauthoryear{{Wang}, {Wang}, {Cheng}  \& {Dai}}{{Wang}
  et~al.}{2016}]{2016A&A...585A..68W}
{Wang} J.~S.,  {Wang} F.~Y.,  {Cheng} K.~S.,   {Dai} Z.~G.,  2016, \mn@doi
  [\aap] {10.1051/0004-6361/201526485}, \href
  {https://ui.adsabs.harvard.edu/abs/2016A&A...585A..68W} {585, A68}

\bibitem[\protect\citeauthoryear{Wang, Zhang, Liang, Lu, Lin, Li  \& Lin}{Wang
  et~al.}{2018}]{Wang_2018}
Wang X.-G.,  Zhang B.,  Liang E.-W.,  Lu R.-J.,  Lin D.-B.,  Li J.,   Lin L.,
  2018, \mn@doi [The Astrophysical Journal] {10.3847/1538-4357/aabc13}, 859,
  160

\bibitem[\protect\citeauthoryear{{Wang}, {Hu}, {Zhang}  \& {Dai}}{{Wang}
  et~al.}{2021a}]{2021arXiv210614155W}
{Wang} F.~Y.,  {Hu} J.~P.,  {Zhang} G.~Q.,   {Dai} Z.~G.,  2021a, arXiv
  e-prints, \href {https://ui.adsabs.harvard.edu/abs/2021arXiv210614155W} {p.
  arXiv:2106.14155}

\bibitem[\protect\citeauthoryear{{Wang} et~al.,}{{Wang}
  et~al.}{2021b}]{2021ApJ...907L...1W}
{Wang} F.,  et~al., 2021b, \mn@doi [\apjl] {10.3847/2041-8213/abd8c6}, \href
  {https://ui.adsabs.harvard.edu/abs/2021ApJ...907L...1W} {907, L1}

\bibitem[\protect\citeauthoryear{Wei et~al.,}{Wei et~al.}{2016}]{wei2016deep}
Wei J.,  et~al., 2016, arXiv e-prints, \href
  {https://ui.adsabs.harvard.edu/abs/2016arXiv161006892W} {p. arXiv:1610.06892}

\bibitem[\protect\citeauthoryear{{Willingale} et~al.,}{{Willingale}
  et~al.}{2007}]{2007ApJ...662.1093W}
{Willingale} R.,  et~al., 2007, \mn@doi [\apj] {10.1086/517989}, \href
  {https://ui.adsabs.harvard.edu/abs/2007ApJ...662.1093W} {662, 1093}

\bibitem[\protect\citeauthoryear{{Woosley}}{{Woosley}}{1993}]{1993ApJ...405..273W}
{Woosley} S.~E.,  1993, \mn@doi [\apj] {10.1086/172359}, \href
  {https://ui.adsabs.harvard.edu/abs/1993ApJ...405..273W} {405, 273}

\bibitem[\protect\citeauthoryear{{Xu}, {Tang}, {Geng}, {Wang}, {Wang},
  {Kuerban}  \& {Huang}}{{Xu} et~al.}{2020}]{2020arXiv201205627X}
{Xu} F.,  {Tang} C.-H.,  {Geng} J.-J.,  {Wang} F.-Y.,  {Wang} Y.-Y.,  {Kuerban}
  A.,   {Huang} Y.-F.,  2020, arXiv e-prints, \href
  {https://ui.adsabs.harvard.edu/abs/2020arXiv201205627X} {p. arXiv:2012.05627}

\bibitem[\protect\citeauthoryear{Yonetoku, Murakami, Nakamura, Yamazaki, Inoue
  \& Ioka}{Yonetoku et~al.}{2004}]{Yonetoku_2004}
Yonetoku D.,  Murakami T.,  Nakamura T.,  Yamazaki R.,  Inoue A.~K.,   Ioka K.,
   2004, \mn@doi [The Astrophysical Journal] {10.1086/421285}, 609, 935–951

\bibitem[\protect\citeauthoryear{{Yu}, {Zhu}, {Li}, {L{\"u}}  \& {Zou}}{{Yu}
  et~al.}{2017}]{2017ApJ...840...12Y}
{Yu} Y.-W.,  {Zhu} J.-P.,  {Li} S.-Z.,  {L{\"u}} H.-J.,   {Zou} Y.-C.,  2017,
  \mn@doi [\apj] {10.3847/1538-4357/aa6c27}, \href
  {https://ui.adsabs.harvard.edu/abs/2017ApJ...840...12Y} {840, 12}

\bibitem[\protect\citeauthoryear{{Zhang}, {Fan}, {Dyks}, {Kobayashi},
  {M{\'e}sz{\'a}ros}, {Burrows}, {Nousek}  \& {Gehrels}}{{Zhang}
  et~al.}{2006}]{2006ApJ...642..354Z}
{Zhang} B.,  {Fan} Y.~Z.,  {Dyks} J.,  {Kobayashi} S.,  {M{\'e}sz{\'a}ros} P.,
  {Burrows} D.~N.,  {Nousek} J.~A.,   {Gehrels} N.,  2006, \mn@doi [\apj]
  {10.1086/500723}, \href
  {https://ui.adsabs.harvard.edu/abs/2006ApJ...642..354Z} {642, 354}

\bibitem[\protect\citeauthoryear{{Zhang} et~al.,}{{Zhang}
  et~al.}{2009}]{2009ApJ...703.1696Z}
{Zhang} B.,  et~al., 2009, \mn@doi [\apj] {10.1088/0004-637X/703/2/1696}, \href
  {https://ui.adsabs.harvard.edu/abs/2009ApJ...703.1696Z} {703, 1696}

\bibitem[\protect\citeauthoryear{{Zhao}, {Zhang}, {Gao}, {Lan}, {L{\"u}}  \&
  {Zhang}}{{Zhao} et~al.}{2019}]{2019ApJ...883...97Z}
{Zhao} L.,  {Zhang} B.,  {Gao} H.,  {Lan} L.,  {L{\"u}} H.,   {Zhang} B.,
  2019, \mn@doi [\apj] {10.3847/1538-4357/ab38c4}, \href
  {https://ui.adsabs.harvard.edu/abs/2019ApJ...883...97Z} {883, 97}

\makeatother
\end{thebibliography}

\begin{appendix}

\section{Monte Carlo Markov Chain Sampling Error Prediction} \label{MCMC error}

\begin{figure*}[h!]
  \includegraphics[width=8.5cm]{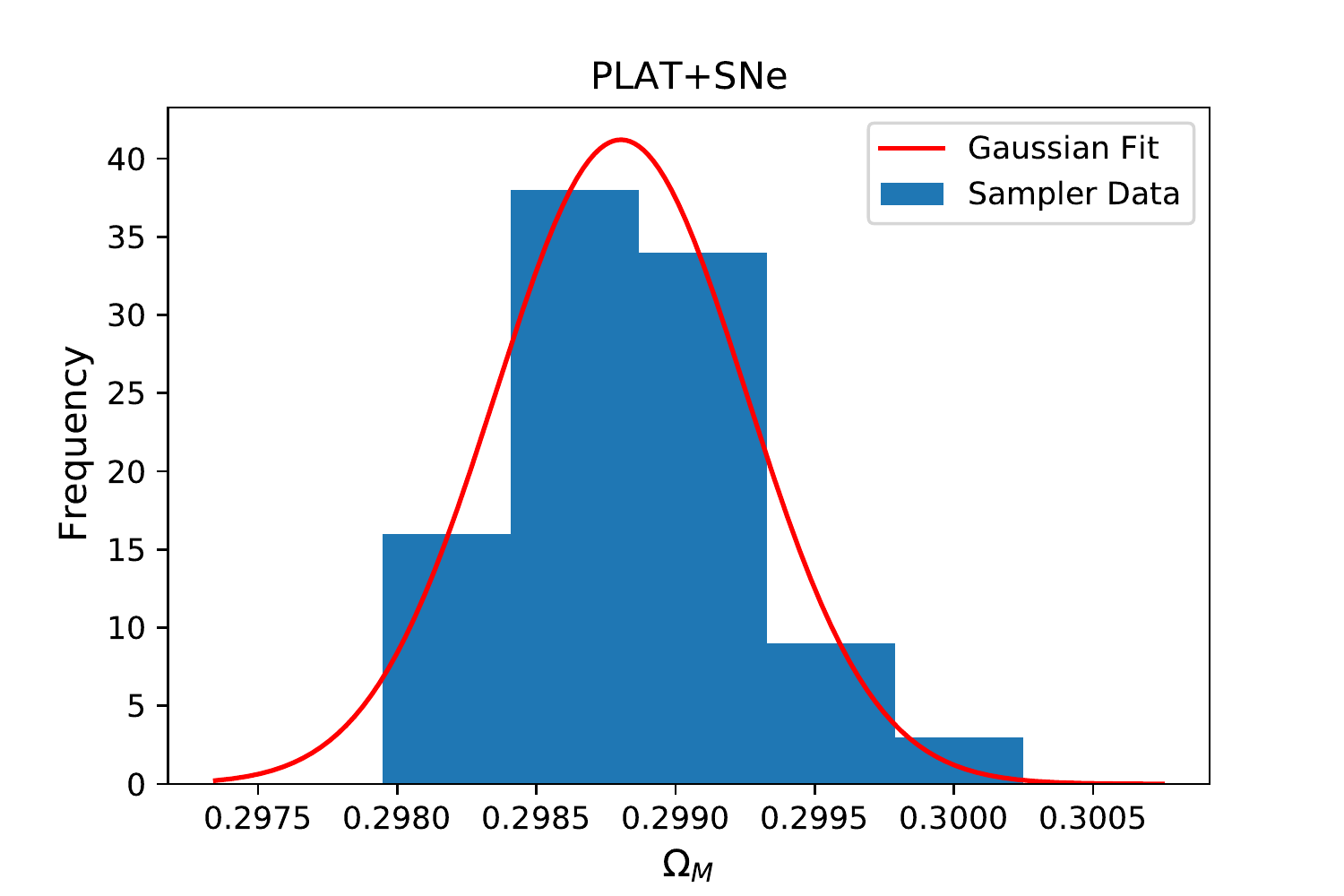}
  \includegraphics[width=8.5cm]{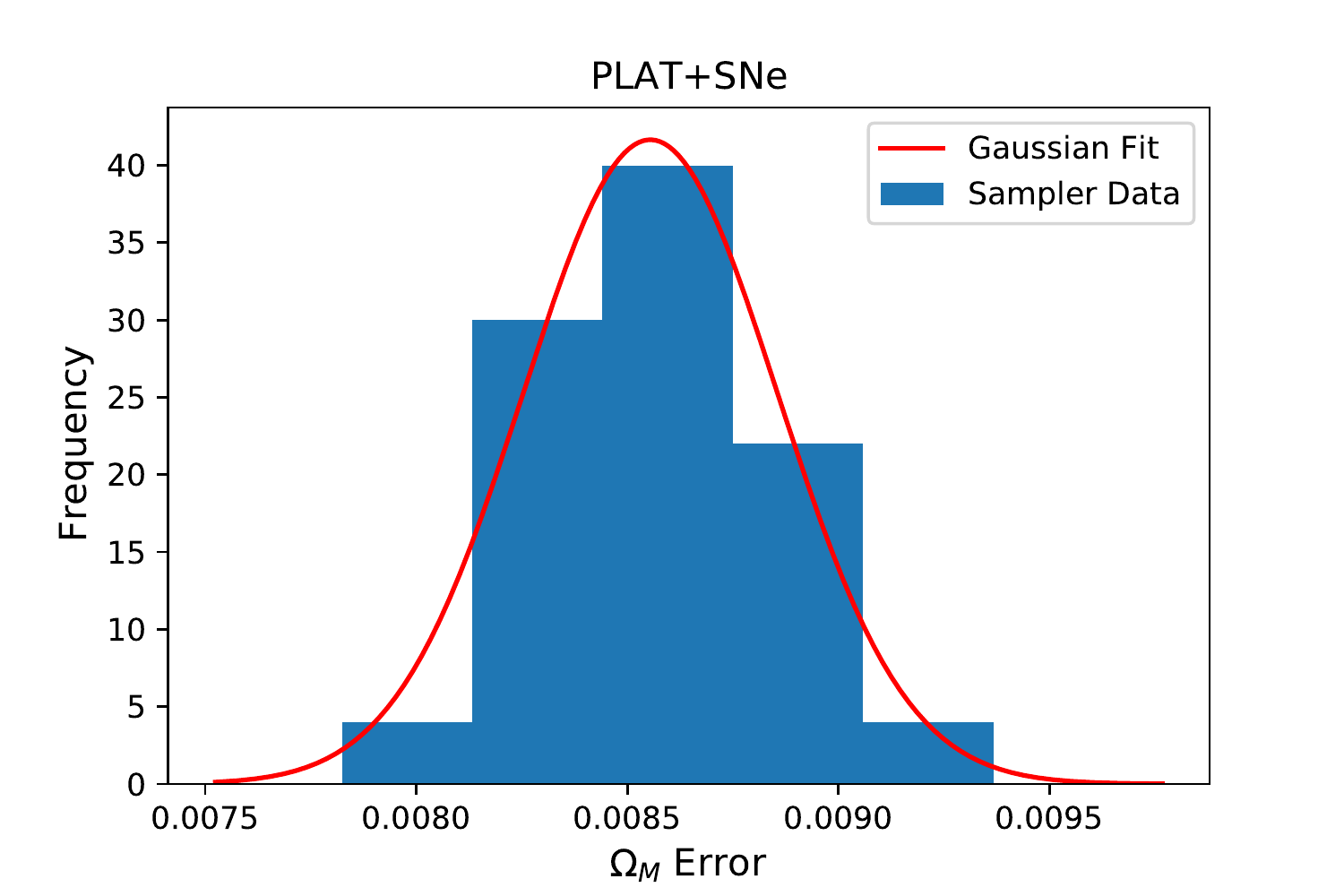}
  \includegraphics[width=8.5cm]{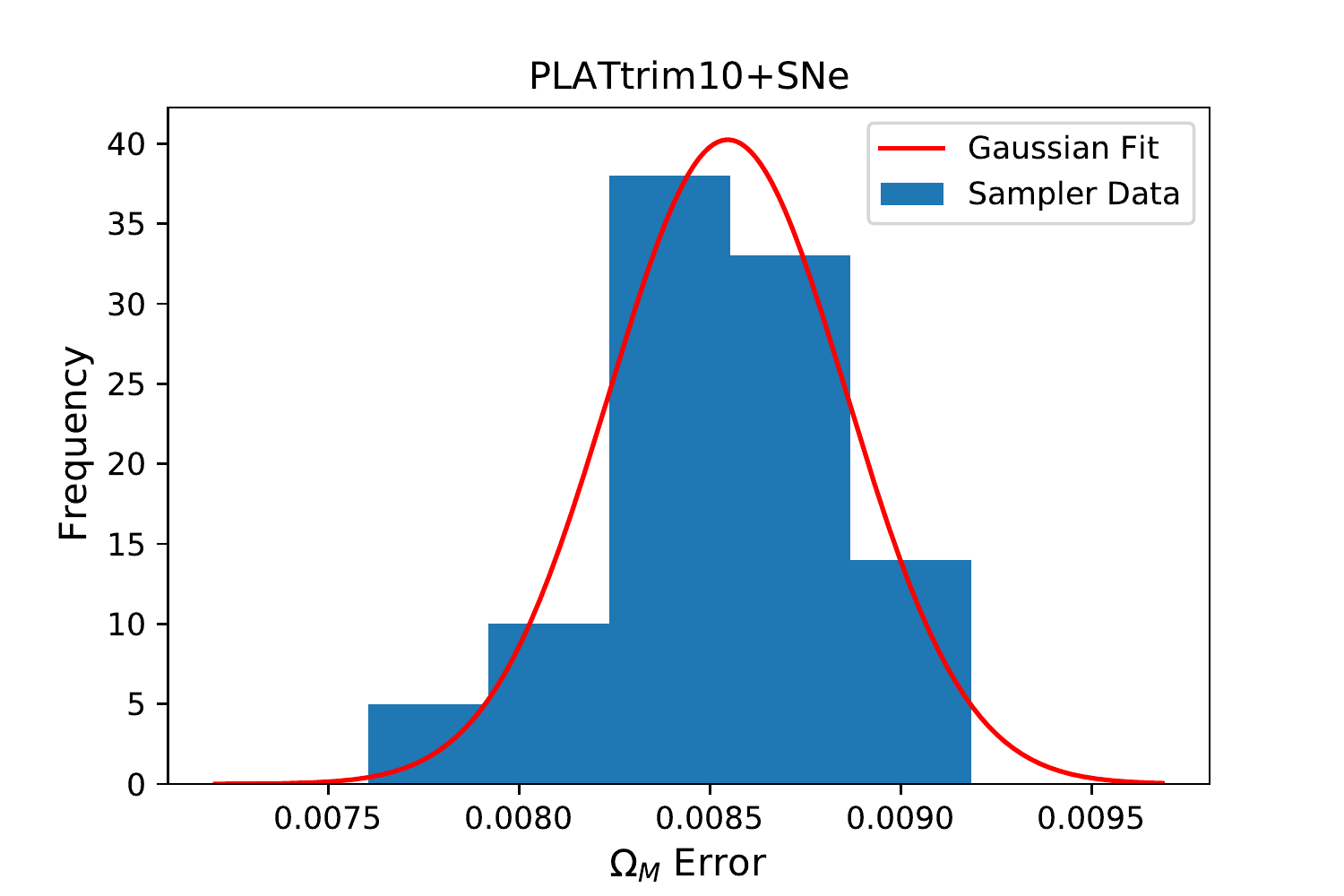}
  \includegraphics[width=8.5cm]{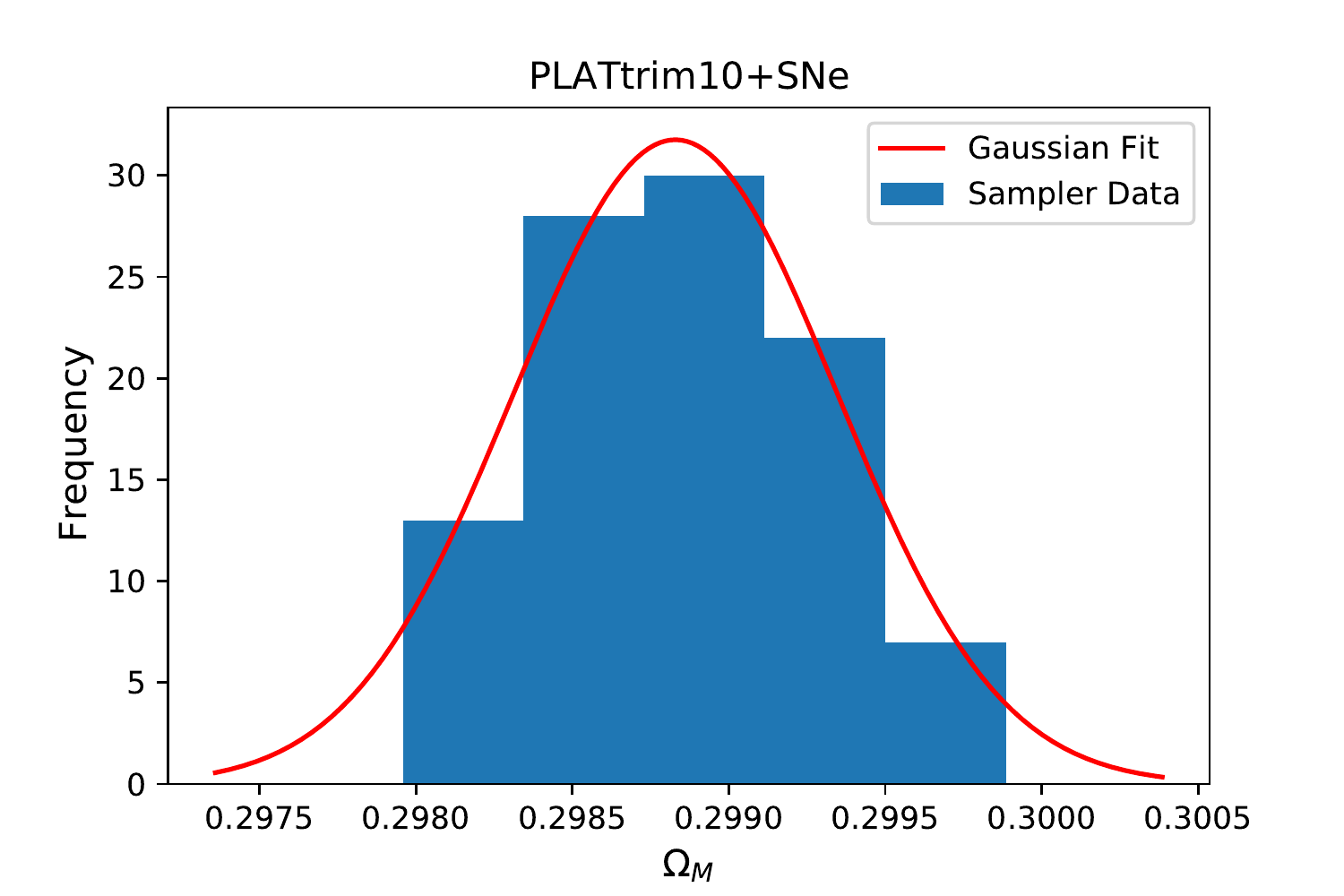}
\caption{The two uppermost panels depict the MCMC sampler results for $\Omega_{\text{M}}$ (upper left) and its error (upper right) considering PLAT+SNe Ia probes together. The bottom panels show the same considering instead PLATtrim+SNe Ia probes. The PLATtrim sample in this Figure refers the apriori trim taken of the 10 GRBs closest to the fundamental plane.}\label{fig:loops}
\end{figure*}

In this section of the Appendix, we aim to quantify the numerical uncertainty that the MCMC sampling is inducing on the error on $\Omega_{\text{M}}$. This will allow us to state that the errors provided on $\Omega_{\text{M}}$ in the text are at least one order of magnitude larger than the uncertainty on the MCMC sampling, thus confirming the reliability of the precision reached. We measure the numerical error introduced through \emph{cobaya} MCMC sampling by monitoring the chain convergence using the Gelman-Rubin statistic (G-R). But even placing restrictions on both the G-R and its defined confidence level leaves some undefined uncertainties on the sampler-derived $\Omega_{\text{M}}$ and its error. Therefore, to make our results even more reliable, we loop all sampler runs 100 times until a Gaussian distribution of the inferred errors is produced to quantify those. From such, we derive the mean of both the $\Omega_{\text{M}}$ and its error, and these are the values which we report for all samples and combinations of samples. Two instances of this method are visualized in Fig. \ref{fig:loops}. To have a precise estimate of the $\sigma$ associated with the error on $\Omega_{\text{M}}$, we fit these distributions with a Gaussian curve and we superimpose the fit with a red line. We note that the error obtained by our simulation is smaller than the $\sigma=0.0005$ of the errors of the distributions. This means that this fluctuation is one order of magnitude less than the uncertainties we are comparing in the main text.

When simulating additional GRBs, we control the numerical error innate to \emph{emcee} by periodically computing the integrated auto-correlation time $\tau$ during sampling to manage the number steps taken. Because \emph{emcee} uses parallel chains to reduce the variance, we can stop the sampling once the chains are longer than $\simeq 50 \tau \,\text{or}\, 100 \tau$. This method ensures that we generate the minimum number of samples to effectively reduce the relative error on our target integral. It is by these operations that we have confidence in not only our results, but also in their comparability between each other those by SNe Ia.

\section{Choosing A Posteriori Trim Values} \label{retro choices}

\begin{figure*}[h]
\centering
    \begin{tabular}{cc}
    \includegraphics[width=0.45\textwidth]{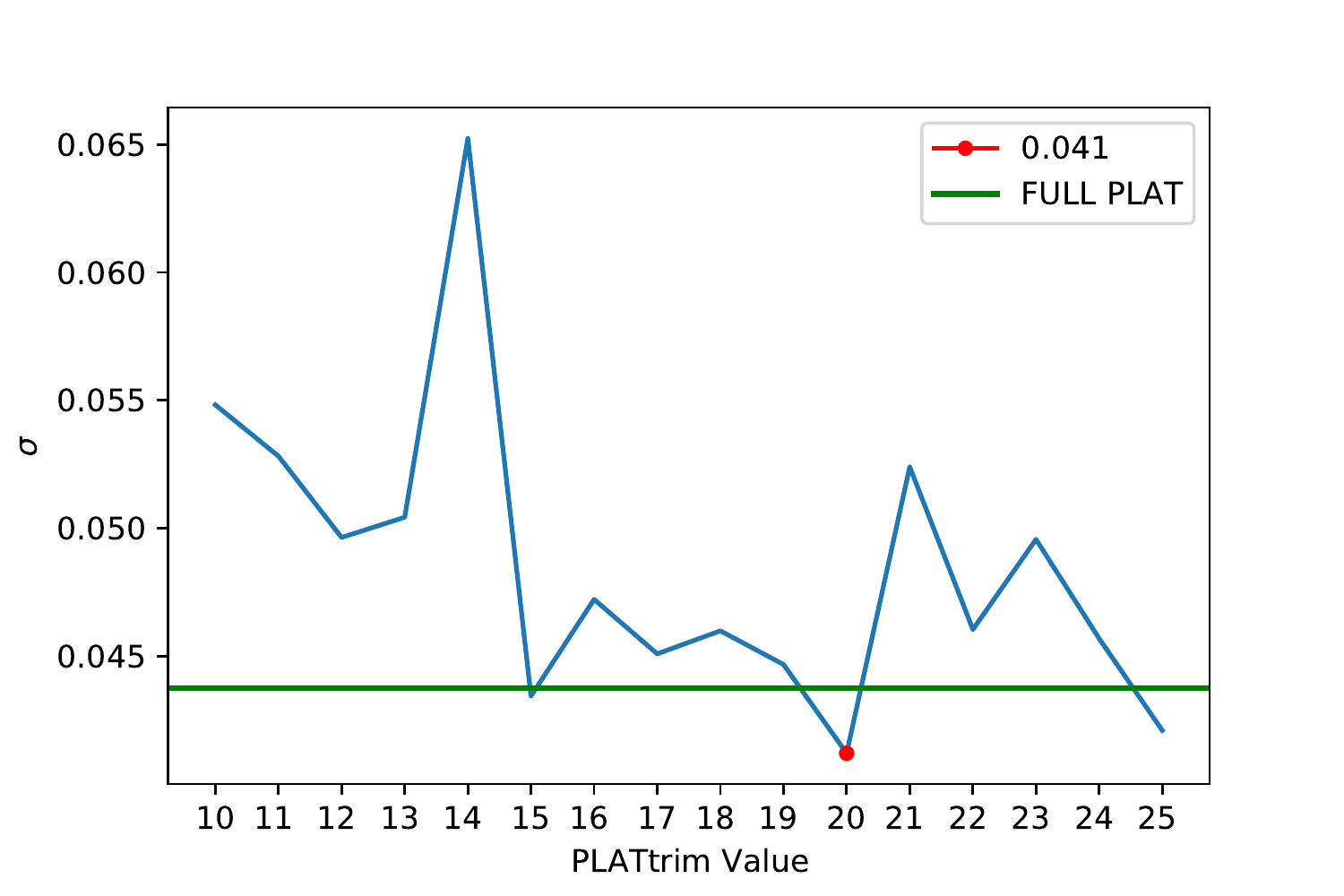} &
    \includegraphics[width=0.45\textwidth]{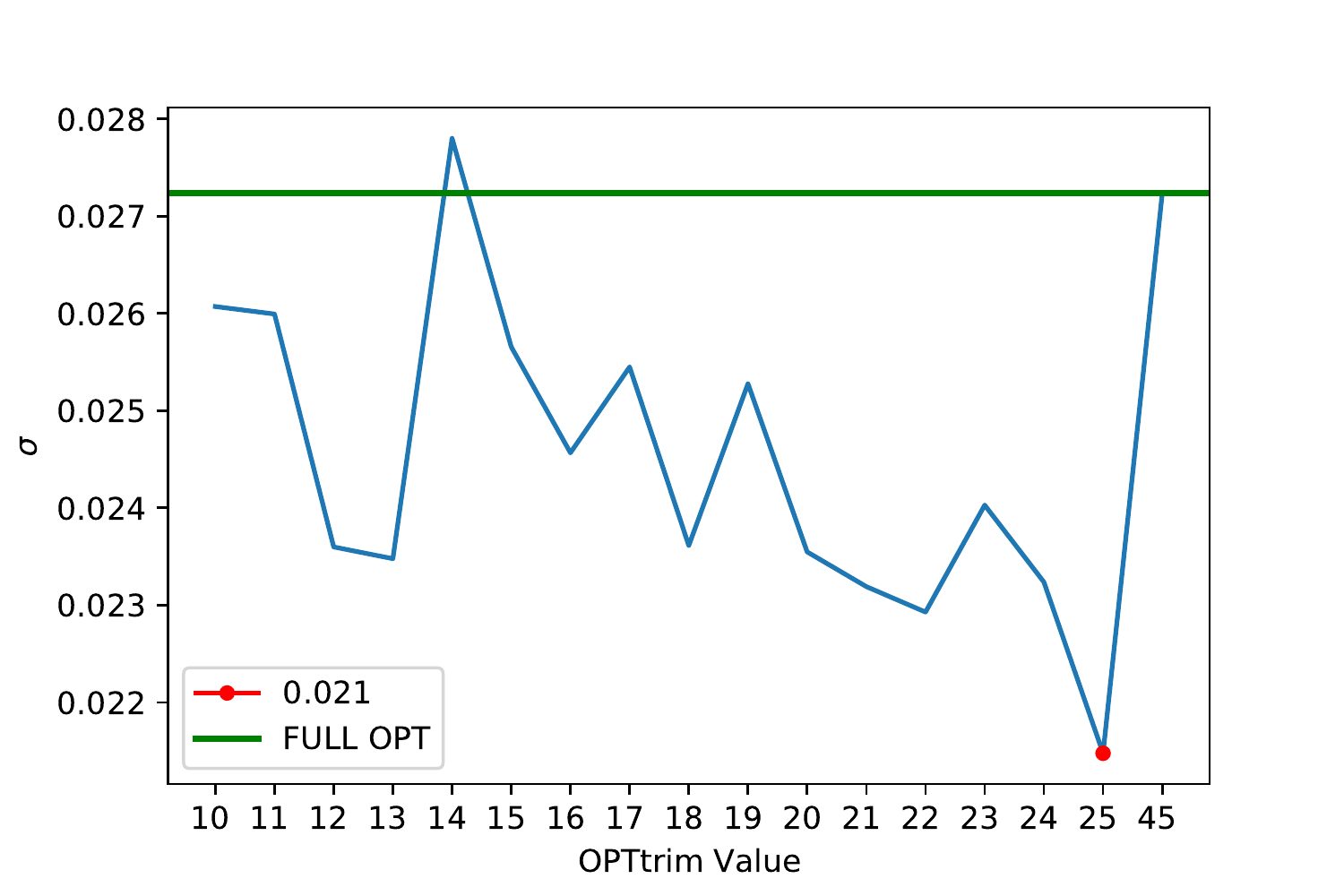} \\
    \includegraphics[width=0.45\textwidth]{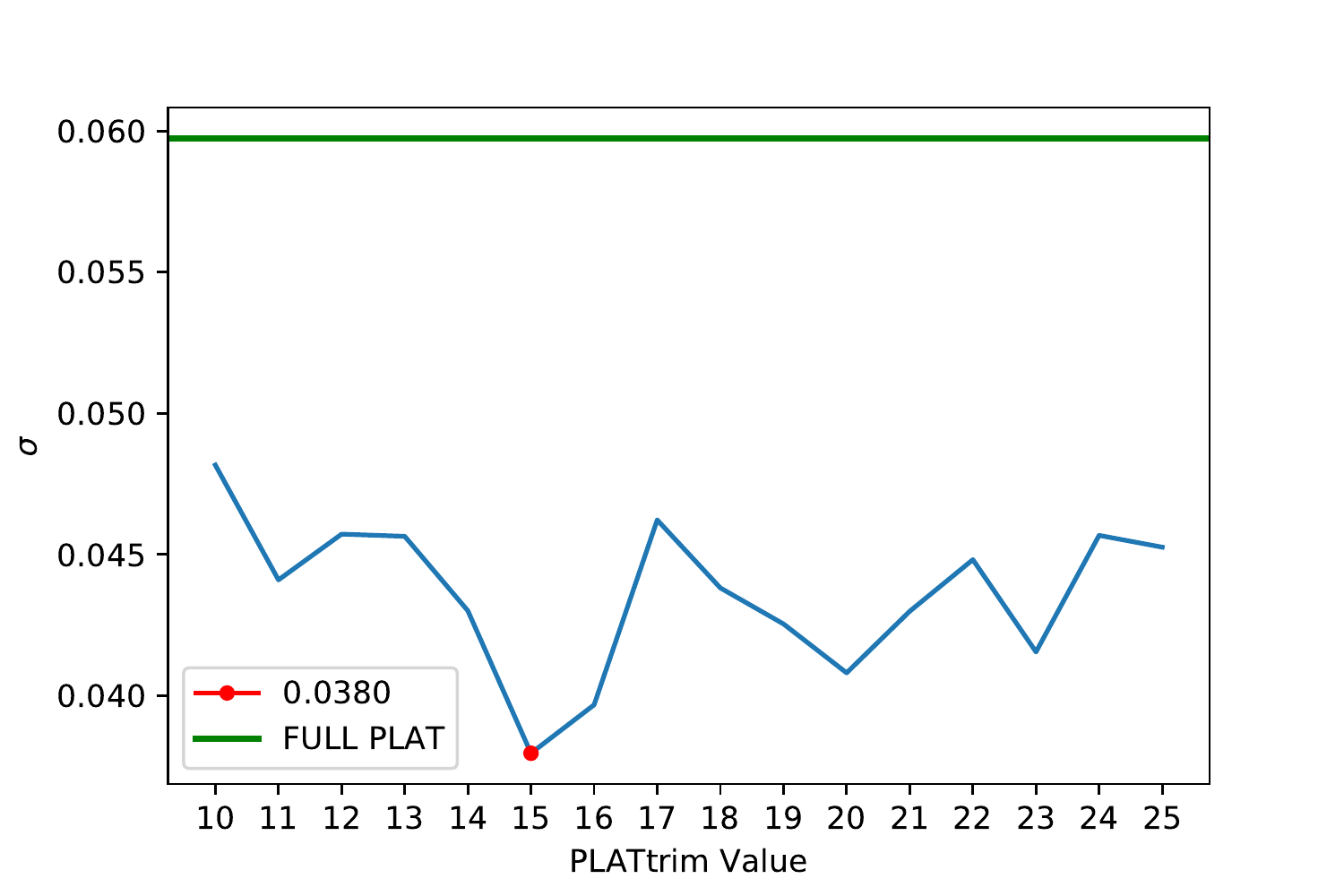} &
    \includegraphics[width=0.45\textwidth]{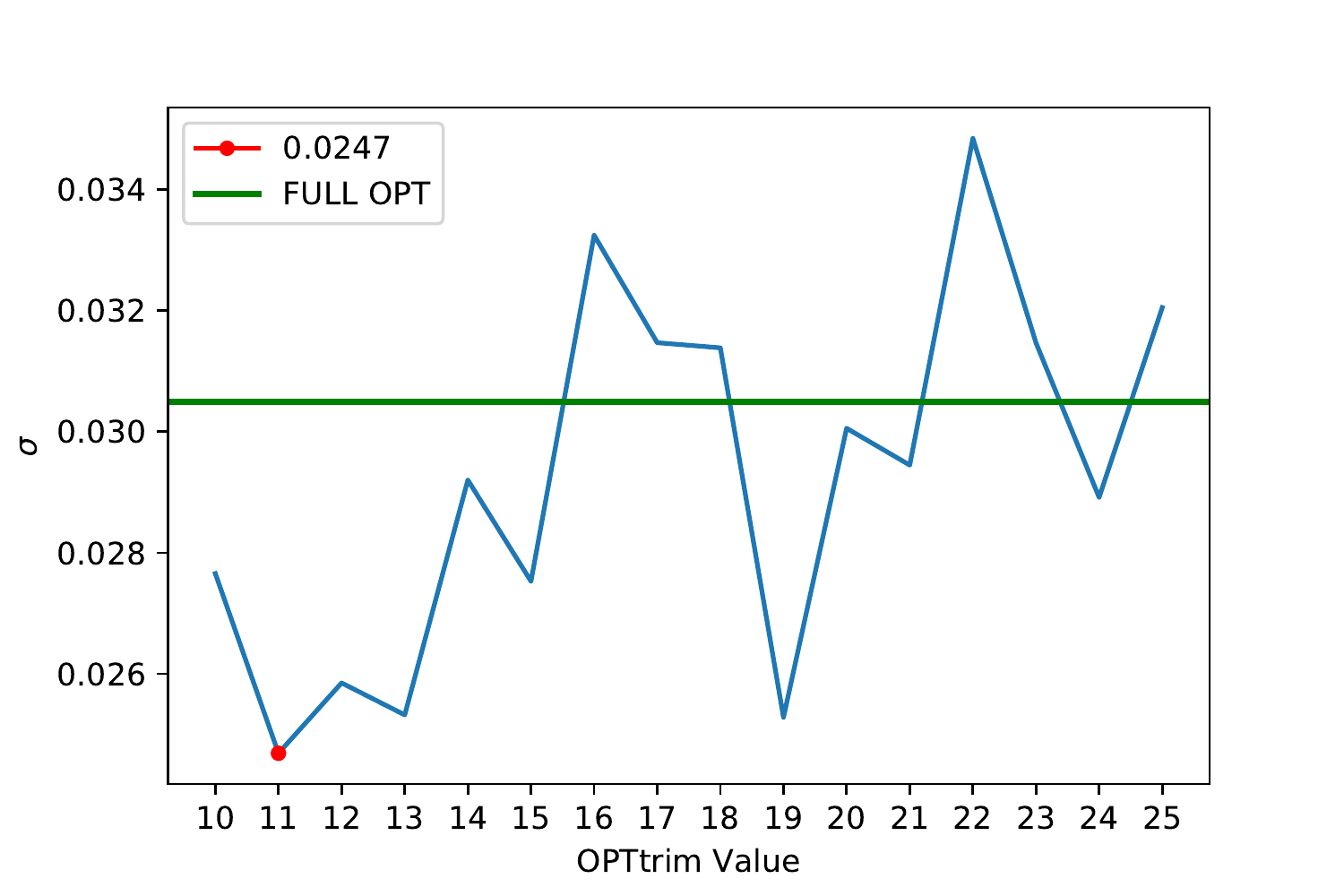} \\
    \end{tabular}
\caption{The leftmost plots in this Figure detail the trial simulations for varying PLATtrim values in the X-ray sample, whereas the rightmost plots do the same for varying OPTtrim in the Optical sample. The two upper panels do not consider redshift evolutionary effects; the bottom panels make corrections for both selection and redshift evolution biases.}\label{fig:retro picks}
\end{figure*}

 In this Appendix section, we detail the selection of the a posteriori sample trim number of GRBs. For a number of observed GRBs in the interval between 10 and 25, we ran test simulations with 2300 simulated GRBs to determine the trend of the standard deviation on $\Omega_{\text{M}}$ for gradually increasing $\sigmaint$ on the trimmed fundamental plane. This was performed for both optical and X-ray data, and with and without considering redshift evolution corrections. Fig. \ref{fig:retro picks} shows the general trend for the PLAT (left panels) and OPT samples (right panels); each minimum shown both without (upper panels) and with (lower panels) accounting for redshift evolution is the number on which our a posteriori trim is based upon. Our reasoning here is that running the simulations for a wide range of trim choices would allow for a better choice of the number of GRBs to be used as the base for simulations, rather than determining this value a priori (before simulating). This idea is confirmed in Fig. \ref{fig:retro picks} as we see that all of our a posteriori trims (minima) have significantly reduced uncertainties when simulated from our original choice of 10 GRBs indicated with the green solid line in  all the panels determined by $\sigmaint$ on their respective fundamental planes.

\end{appendix}

\end{document}